%% file: main.tex
\documentclass[twocolumn,twocolappendix]{aastex631}

\usepackage{CJK}
\usepackage[utf8]{inputenc}
\def\msun{\hbox{M$_{\odot}$}}
\def\msunyr{\mbox{\,${\rm M_{\odot}\, yr^{-1}}$}}
\def\EE#1{\times 10^{#1}}
\def\cm{\mbox{\,cm}}
\def\cm3{\mbox{\,cm$^{-3}$}}
\def\kms{\mbox{\,km~s$^{-1}$}}

\newcommand\R[1]{{\color{orange} \bf #1}} 

\shorttitle{Early Discovery of SN~2022xkq}
\shortauthors{Pearson et al.}

\begin{document}
\title{Strong Carbon Features and a Red Early Color in the Underluminous Type Ia SN~2022xkq}

\correspondingauthor{J. Pearson}
\email{jenivevepearson@arizona.edu}

\input{affiliation}
\input{authors}

\begin{abstract}
We present optical, infrared, ultraviolet, and radio observations of SN~2022xkq, an underluminous fast-declining type Ia supernova (SN Ia) in NGC~1784 ($\mathrm{D}\approx31$ Mpc), from $<1$ to 180 days after explosion.  
The high-cadence observations of SN~2022xkq, a photometrically transitional and spectroscopically 91bg-like SN Ia, cover the first days and weeks following explosion which are critical to distinguishing between explosion scenarios. 
The early light curve of SN~2022xkq has a red early color and exhibits a flux excess which is more prominent in redder bands; this is the first time such a feature has been seen in a transitional/91bg-like SN Ia. 
We also present 92 optical and 19 near-infrared (NIR) spectra,
beginning 0.4 days after explosion in the optical and 2.6 days after explosion in the NIR.
SN~2022xkq exhibits a long-lived \ion{C}{1} 1.0693~$\mu$m feature which persists until 5 days post-maximum. We also detect \ion{C}{2} $\lambda$6580 in the pre-maximum optical spectra. 
These lines are evidence for unburnt carbon that is difficult to reconcile with the double detonation of a sub-Chandrasekhar mass white dwarf. 
No existing explosion model can fully explain the photometric and spectroscopic dataset of SN~2022xkq, but the considerable breadth of the observations is ideal for furthering our understanding of the processes which produce faint SNe~Ia. 

\end{abstract}


\section{Introduction} \label{sec:intro}

Type Ia supernovae (SNe Ia) result from the thermonuclear explosions of carbon-oxygen white dwarfs \citep{Hoyle60}, and are a critical tool for measuring the expansion history of the universe, as they are standardizable candles \citep{Phillips93, Hamuy96, Guy07, Jha07}.  Despite their importance, the exact nature of the explosion mechanisms and progenitor systems of SNe Ia are still being actively investigated \citep[e.g., see][for a recent review]{Jha19}. 

There are several critical observational probes of SNe Ia that shed light on the possible explosion and progenitor scenarios.  Amongst the most promising are very early multiband and high cadence light curves, which in a handful of well-observed cases have exhibited color-dependent excesses over a smooth power law rise \citep[e.g.][]{Marion16,Hosseinzadeh17,Hosseinzadeh22,23bee,Jiang17,Jiang21,Miller18,Dimitriadis19,Shappee19,Miller20,Burke21,Ni22,Sai22, Wang23}.  Recent samples indicate that $\sim$10-30\% of all SNe Ia, if observed early enough and with high cadence, would exhibit such excesses \citep{Bulla20,Miller20_sample,Burke22_ztf,Burke22_dlt40,Deckers22}. 
Notably, such early light curve features were predicted for the single degenerate scenario due to the collision between the SN ejecta and a non-degenerate companion star, manifesting as a UV/blue excess which is visible for only a few days for favorable viewing angles \citep{Kasen10}.  Similar ``bump" features in the early light curve may also be attributable to an unusual distribution of radioactive $^{56}$Ni \citep[e.g.,][]{Magee20_Nidist, Magee20_Niclump}, rapid velocity evolution of the \ion{Ca}{2} H\&K feature \citep{Ashall22}, circumstellar medium (CSM)-ejecta interaction \citep[e.g.,][]{Piro16, Jiang21}, or a natural consequence of a sub-Chandrasekhar mass, double detonation explosion \citep[e.g.,][]{Polin19}.  Even well-observed examples that do not exhibit an early light curve excess are important for understanding SN Ia light curve demographics \citep[e.g.,][]{Olling15}, and the early light curves of many sub-types, including transitional and 91bg-like SNe Ia, have not yet been explored. 

Another key measurement in SNe Ia is the incidence of carbon in the early spectra. Depending on the explosion mechanism some carbon can escape being burned during the runaway carbon fusion in a C/O white dwarf, which makes carbon a direct probe of the explosion.  Several studies have focused on the \ion{C}{2} $\lambda$6580\,{\AA} feature, which is seen on the red shoulder of the ubiquitous \ion{Si}{2} $\lambda$6355\,{\AA} absorption feature and is detectable in $\sim$20-40\% of spectra taken before maximum light \citep{Thomas11,Parrent11,Blondin12, Folatelli12,Silverman12,Maguire14,15bp}. There is mounting evidence that the near-infrared (NIR) \ion{C}{1} $\lambda$1.0693~$\mu$m line is particularly strong in transitional and/or 91bg-like SN Ia \citep{Hsiao15,15bp, 12ij}. 

The amount of carbon (or lack thereof) depends on the explosion physics and the degree of mixing. The pure deflagration W7 model \citep{Nomoto84} leaves significant carbon behind in the outer ejecta, as does the violent merger model \citep{pak12} and the pulsational delayed-detonation models \citep{hoflich95}. Meanwhile, delayed-detonation models have nearly complete carbon burning at normal SN Ia luminosities \citep{Khokhlov91, Kasen09}, with increasing amounts of unburnt carbon remaining for fainter events \citep{99by_ir}. 
Notably, sub-Chandrasekhar double-detonation models, which has been suggested to be the dominate explosion mechanism for faint SNe Ia \citep{Blondin17,Goldstein18}, leave little to no carbon behind \citep[e.g.,][]{Polin19}. This indicates that the presence of unburnt carbon may be a distinguishing observational signature of the explosion mechanism in faint SNe Ia.

In this paper we present high cadence, multi-wavelength photometric and spectroscopic observations of the faint SN Ia SN~2022xkq. We discuss the discovery of SN~2022xkq in Section \ref{sec:disc}, observational details in Section \ref{sec:obs}, and the proper classification for SN~2022xkq in Section \ref{sec:class}. In Section \ref{sec:phot} and Section \ref{sec:spec}, we analyze the photometric and spectroscopic observations of SN~2022xkq and compare them to models of several different explosion scenarios. In Section \ref{sec:radio}, we examine radio observations and their constraints on the progenitor system. We discuss the possible origins of the observed color and the implications of spectral features in Section \ref{sec:discuss} and conclude in Section \ref{sec:conclude}. 

\section{Discovery}\label{sec:disc}

\begin{figure}
    \centering
    \includegraphics[width=\hsize]{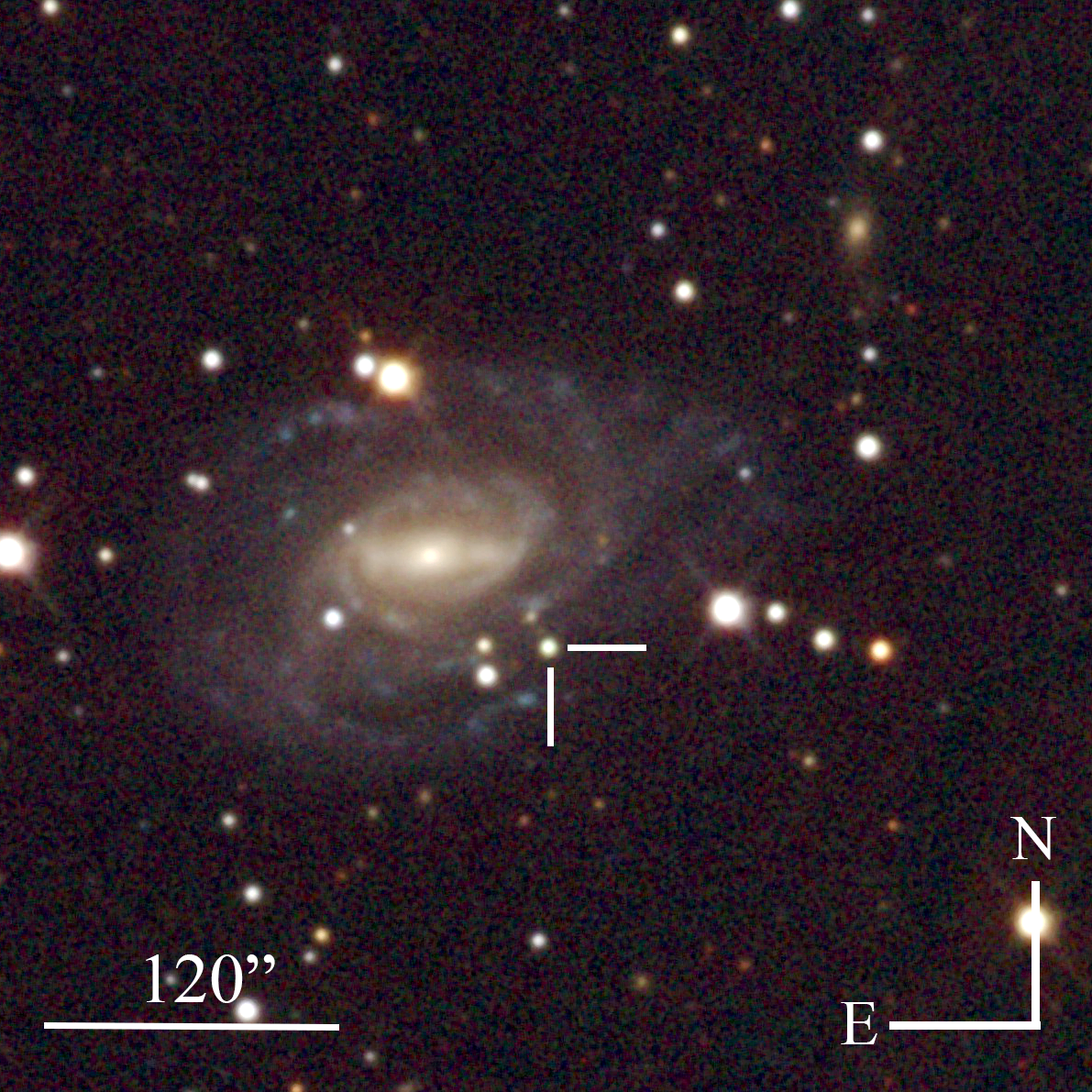}
    \caption{Composite $g,r,i$ image of SN~2022xkq in NGC~1784 obtained by Las Cumbres Observatory on 2022 Nov 14, +18 days after $B$-band peak.}
    \label{fig:Loc}
\end{figure}

SN~2022xkq, also known as DLT22r, was discovered at $\mathrm{RA(J2000)} = 05\textsuperscript{h}05\textsuperscript{m}23\fs710$, $\mathrm{Dec(J2000)} = -11\degr52'56\farcs10$ by the Distance Less Than 40 Mpc survey \citep[DLT40, for survey description see][]{Tartaglia2018} on 2022 October 13, 06:43:35 UTC \citep[59,865.28 MJD;][]{22xkqdis}. The last non-detection was on 2022 October 12, 11:47:37 UTC (2,459,864.99 JD) by the Asteroid Terrestrial-impact Last Alert System (ATLAS) to a $>2\sigma$ limit of $o>19.47$  \citep{Atlas_forced, Atlas_trans, Atlas_web}. 
SN~2022xkq was initially classified as a type I SN \citep{SNIclass}, then reclassified as a SN Ic \citep{SNIcclass}. It was eventually classified as a SN~1991bg-like SN Ia \citep{22xkqclass}; the exact classification will be discussed further in Section \ref{sec:class}.

As shown in Figure \ref{fig:Loc}\footnote{Image produced using https://afterglow.skynet.unc.edu}, SN~2022xkq is located in NGC~1784, a barred spiral SB(r)c galaxy \citep{deVaucouleurs91, NGC1784} and SN~2022xkq is the first recorded SN in the galaxy. 
Underluminous SNe Ia preferentially explode in older galaxy populations and are most likely to be found in elliptical and occasionally in early-type spiral galaxies \citep{Hamuy00, Howell01, Gallagher05, Sullivan06, Li11, Nugent23}. However, late-type spiral galaxies also possess old stellar populations, and while underluminous SNe Ia in late-type hosts are uncommon, there have been a few. The most notable is SN~1999by, which occurred in the SAb galaxy NGC~2841 \citep{99by_opt, 99by_ir}.

NGC~1784 has a redshift $z=0.007735$ \citep{redshift}. 
The Virgo infall distance to NGC~1784 is $31 \pm 2$~Mpc \citep[$\mu = 32.46 \pm 0.15$~mag;][]{VirgoInfall} for H$_0 = 70$~km~s$^{-1}$~Mpc$^{-1}$, which we use as the distance to SN~2022xkq, and is listed in Table \ref{tab:results}. This measurement is in general agreement with the Tully-Fisher distance of 28.07~Mpc \citep{TF} for H$_0 = 70$~km~s$^{-1}$~Mpc$^{-1}$ as well as the distance modulus calculated based off the multiband light curve by \texttt{SNooPy} \citep[see Section \ref{sec:phot};][]{snoopy} of $\mu=32.9 \pm 0.1$ mag for H$_0 = 72$~km~s$^{-1}$~Mpc$^{-1}$.

\section{Observations and Data Reduction}\label{sec:obs}

\subsection{Photometry}\label{sec:obs_phot}

\begin{figure*}
    \centering
    \includegraphics[width=\hsize]{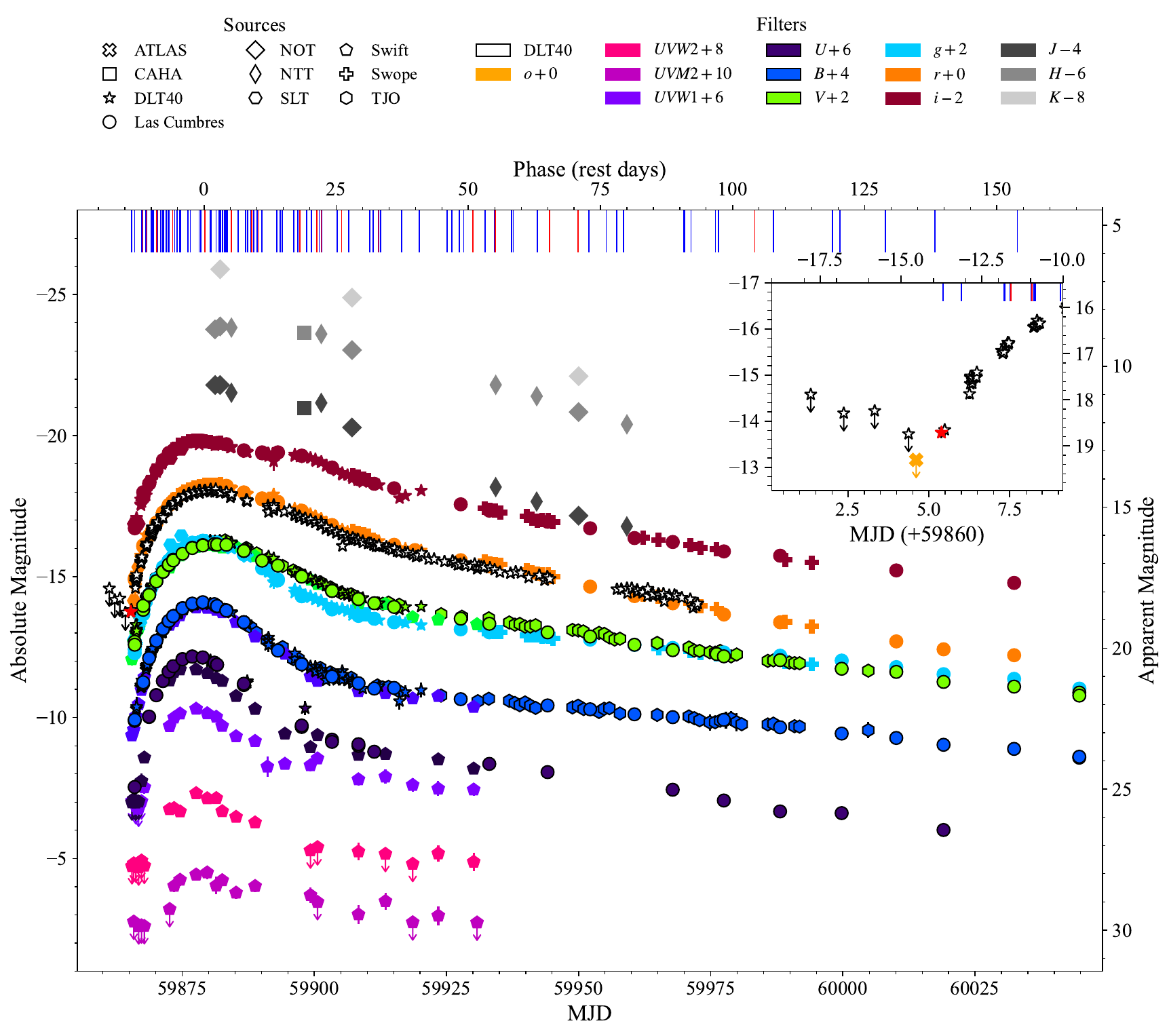}
    \caption{SN~2022xkq light curves from DLT40 (stars), Las Cumbres Observatory (circles), SLT (side up hexagons), Swift (pentagons), Swope (plus), and TJO (vertex up hexagons). NIR photometry from NTT, NOT, and CAHA are also included (quadrilaterals). Main figure: light curves with offsets in absolute and extinction-corrected apparent magnitudes. Swift observations are plotted as non-detections when the measurement error is $>3\sigma$. Upper right: zoom in of the $DLT40$ light curve immediately before and after discovery without offsets. The DLT40 discovery is marked by the red star and the last non-detection by ATLAS is marked by the yellow-orange x. Epochs of optical (blue) and infrared (red) spectra are displayed as lines along the upper x-axis. Phase is relative to the time of $B_{\mathrm{max}}$ (MJD 59,879.03). (The data behind this figure are available.)}
    \label{fig:phot}
\end{figure*}

Following the discovery of SN~2022xkq by the DLT40 survey, continued photometric monitoring was done by two of DLT40's discovery telescopes, the PROMPT5 0.4m telescope at the Cerro Tololo Inter-American Observatory and the PROMPT-USASK 0.4m telescope at Sleaford Observatory. Observations taken by these telescopes using a Clear or Open filter (labeled DLT40 in Figure \ref{fig:phot}) are calibrated to the SDSS \textit{r} band, as described in \citet{Tartaglia2018}. These ``DLT40"-band observations were taken with a cadence of $\sim$9 images per day until maximum light, $\sim$3 per day for the next 40 days, and $\sim$1 per day for the following 53 days. Additionally, multiband \textit{BVgri} photometry was taken at the Prompt5 telescope at a cadence of $\sim$6 observations per day until maximum light and $\sim$2 per day for the next 40 days. The \textit{BVgri} light curves were created using aperture photometry and calibrating to the APASS (\textit{BVgri}) catalog.

Further \textit{UBVgri} photometry of SN~2022xkq was obtained using the Sinistro cameras on Las Cumbres Observatory's robotic 1m telescopes \citep{LasCumbres}, located at the Siding Spring Observatory, the South African Astronomical Observatory, and the Cerro Tololo Inter-American Observatory as part of the Global Supernova Project (GSP) collaboration. 
The photometric data from Las Cumbres Observatory were reduced using \texttt{lcogtsnpipe} \citep{Valenti2016}, a PyRAF-based image reduction pipeline. \texttt{lcogtsnpipe} utilizes a low-order polynomial fit and a standard point-spread function (PSF) fitting technique to remove the background and calculate instrumental magnitudes. Apparent magnitudes were calibrated to the APASS (\textit{BVgri}) catalog and a Landolt (\textit{U}) standard field that was observed on 2022 November 24 alongside observations of SN~2022xkq.

We also include $BV$ photometry obtained with the 0.8m Telescopi Joan Or\'o (TJO) at the Montsec Observatory. 
Instrumental magnitudes were measured using \texttt{AutoPhOT} \citep{autophot}, and calibrated to $BV$ APASS tabulated magnitudes.

We also collect images with the 0.4 m Ritchey-Chr\'etien Super Light Telescope (SLT) at the Lulin Observatory, Taiwan, as a part of the Kinder project \citep{Chen21}, and use Photutils \citep{photutils} to perform aperture photometry on the images.

Additionally, we include photometric observations in $gri$ bands taken with the Swope 1-m optical telescope at Las Campanas Observatory, Chile taken as part of the Swope Supernova Survey. 
All Swope photometry was processed using biases and flat-fields in the same instrumental configuration as described in \citet{Kilpatrick18}, using the \texttt{photpipe} imaging and photometry package \citep{Rest05}, including bias-subtraction, flat-fielding, image stitching, and photometric calibration. Observations were calibrated using standard sources from the Pan-STARRS DR1 catalog \citep{Flewelling20}. Using the Pan-STARRS1 3$\pi$ images as templates, we performed image subtraction using {\tt hotpants} \citep{hotpants}, and forced photometry at the position of SN 2022xkq to obtain the reported Swope photometric measurements.

We include $JHK$ photometry obtained with SOFI mounted on the European Southern Observatory (ESO) 3.5m New Technology Telescope (NTT) at La Silla Observatory; Omega2000 at the 3.5m Calar Alto telescope (CAHA); and NOTCam at the 2.5m Nordic Optical Telescope (NOT). Photometry was measured with \texttt{AutoPhOT} and calibrated to 2MASS star catalogued photometry.

We also include optical and ultraviolet (UV) photometry from our high-cadence photometric follow-up campaign with the Ultraviolet/Optical Telescope \citep[UVOT;][]{Roming2005} on the Neil Gehrels Swift Observatory \citep{Gehrels2004}. The aperture photometry from the UVOT images was measured using the High-Energy Astrophysics software \citep[HEA-Soft;][]{HEA-Soft2014} and a 3\arcsec\ aperture centered at the position of SN~2022xkq. The background was measured from a region devoid of stars. Zero-points for the UVOT data were taken from \citet{Breeveld2011} with time-dependent sensitivity corrections updated in 2020.

The complete Milky Way and host extinction corrected (see Section \ref{sec:extinct} for details) multiband light curves of SN~2022xkq are shown in Figure \ref{fig:phot}.

\subsection{Spectroscopy}

\begin{figure*}
    \centering
    \includegraphics[width=\hsize]{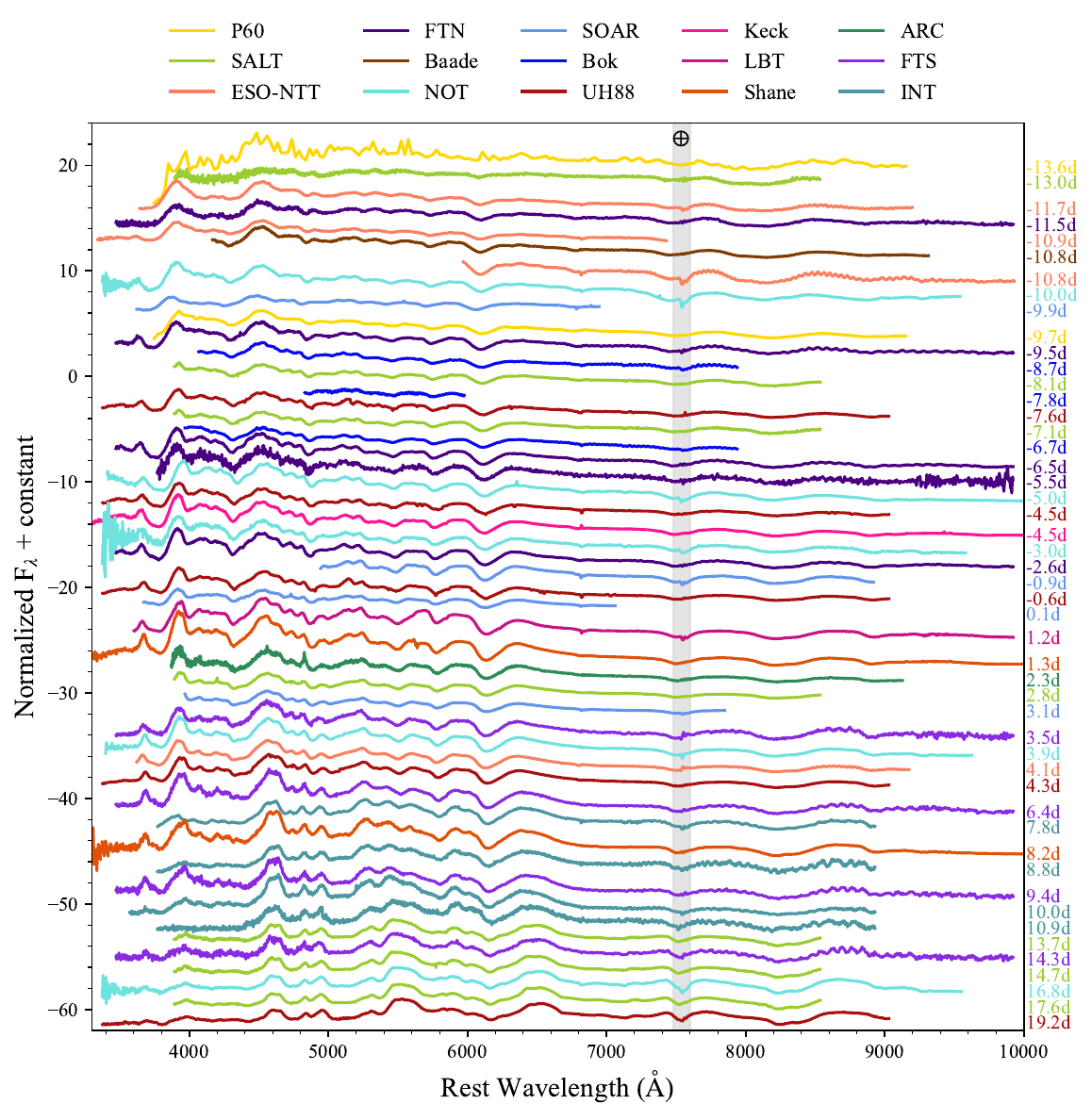}
    \caption{Optical spectroscopic evolution of SN~2022xkq until 20 days post-maximum light, corrected for $E(B-V)_\mathrm{tot} = 0.12$. SN~2022xkq clearly exhibits the characteristic 91bg-type \ion{Ti}{2} band from 4000 to 4500~{\AA} (see Figure \ref{fig:peak} for comparison to other underluminous SN Ia). Spectra are color-coded to denote telescope (see Section \ref{sec:obs} for more details). Labels are relative to the time of $B_{\mathrm{max}}$.}
    \label{fig:optspec1}
\end{figure*}

\begin{figure*}
    \centering
    \includegraphics[width=\hsize]{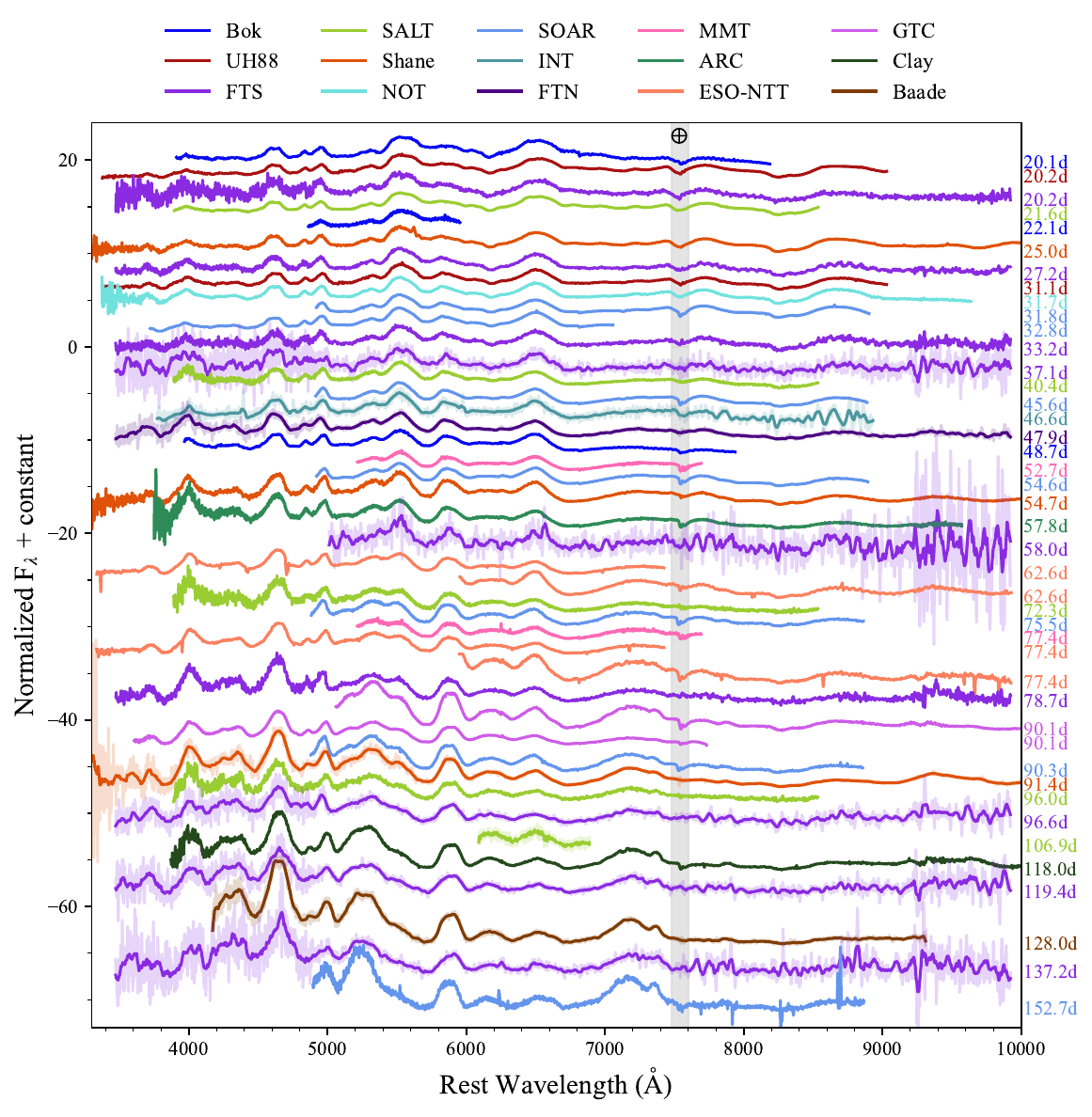}
    \caption{ Same as Figure \ref{fig:optspec1} but for epochs after 20 days post-maximum light. Some spectra have been rebinned for clarity, the unbinned spectra are displayed at lower opacity.}
    \label{fig:optspec2}
\end{figure*}

\begin{figure*}
    \centering
    \includegraphics[width=\hsize]{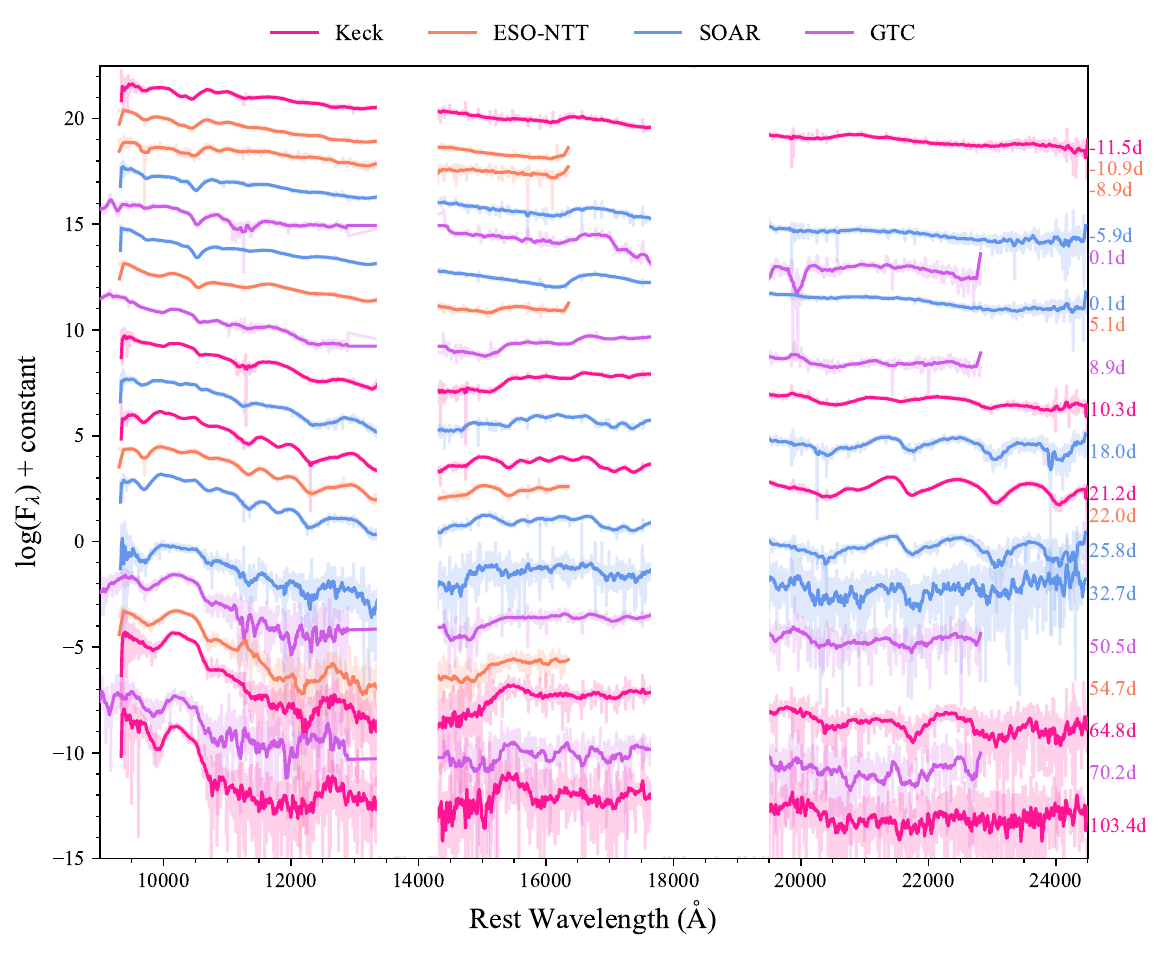}
    \caption{NIR spectroscopic evolution of SN~2022xkq, corrected for $E(B-V)_\mathrm{tot} = 0.12$. Labels are relative to the time of $B_{\mathrm{max}}$. Spectra have been rebinned and telluric bands have been removed for clarity; the unbinned spectra are displayed at lower opacity.}
    \label{fig:irspec}
\end{figure*}

We present 92 optical spectra from the first 153 days following the explosion. 
These spectra where observed using the Robert Stobie Spectrograph (RSS) on the Southern African Large Telescope \citep[SALT,][]{SALT}; the ESO Faint Object Spectrograph and Camera \citep[EFOSC2,][]{EFOSC2} on the NTT (ESO-NTT) as part of the ePESSTO+
survey \citep{PESSTO}; the FLOYDS spectrographs \citep{LasCumbres} on the Las Cumbres Observatory's 2m Faulkes Telescopes North and South (FTN/FTS) as part of the GSP collaboration; the Inamori-Magellan Areal Camera and Spectrograph \citep[IMACS,][]{IMACS} on the Magellan Baade Telescope; the Alhambra Faint Object Spectrograph and Camera (ALFOSC) on the Nordic Optical Telescope (NOT); the Goodman High Throughput Spectrograph \citep[GHTS;][]{GHTS} on the Southern Astrophysical Research Telescope (SOAR); the Boller and Chivens Spectrograph (B\&C) on the University of Arizona's Bok 2.3m telescope located at Kitt Peak Observatory; 
the University of Hawaii 2.2m telescope (UH88) on Mauna Kea using the Supernova Integral Field Spectrograph \citep[SNIFS,][]{SNIFS} as part of the Spectroscopic Classification of Astronomical Transients survey \citep[SCAT, ][]{Tucker22}
one of the Multi-Object Double Spectrographs \citep[MODS1,][]{MODS} on the Large Binocular Telescope (LBT);
the Kitt Peak National Observatory Ohio State Multi-Object Spectrograph \citep[KOSMOS;][]{KOSMOS1, KOSMOS2} instrument at the Astrophysical Research Consortium 3.5m telescope (ARC) at Apache Point Observatory; the Intermediate Dispersion Spectrograph (IDS) on the Isaac Newton Telescope (INT); the Binospec instrument on the MMT \citep{Binospec}; the Optical System for Imaging and
low-Resolution Integrated Spectroscopy \citep[OSIRIS,][]{OSIRIS1, OSIRIS2} at Gran Telescopio Canarias (GTC); the Kast dual-beam spectrograph \citep{KAST} on the Lick Shane 3m telescope; the Low-Resolution Imaging Spectrograph \citep[LRIS;][]{LRIS} on the 10m Keck~I telescope as part of the Young Supernova Experiment \citep[YSE;][]{Jones21, Aleo23}; and the Low Dispersion Survey Spectrograph \citep[LDSS-3,][]{ldss3} on the Magellan Clay Telescope.
We also include the two classification spectra \citep{SNIclass, 22xkqclass} from the Transient Name Server that were not obtained by our team. These spectra were taken by the Zwicky Transient Facility \citep[ZTF;][]{ZTF} using the Spectral Energy Distribution Machine \citep[SEDM;][]{SEDM} instrument on the Palomar 60-inch telescope (P60). 
Optical spectra were reduced using the FLOYDS pipeline \citep{FLOYDS}, the PyNOT-redux reduction pipeline \footnote{https://pypi.org/project/PyNOT-redux/}, the PESSTO pipeline \citep{PESSTO}, PySALT \citep{SALTpipe}, the Goodman pipeline \citep{Torres17}, the IDSRED package \citep{Bravo23}, SNIFS reduction pipeline \citep{Tucker22}, the Binospec IDL pipeline \citep{Kansky19}, the modsCCDRed package \citep{Pogge19}, the UCSC Spectral Pipeline \footnote{\url{https://github.com/msiebert1/UCSC\_spectral\_pipeline}} \citep{siebert19}, and standard IRAF and PyRAF routines \citep{iraf, pyraf}. 

We also present 19 NIR spectra taken using the Near-Infrared Echellette Spectrometer \citep[NIRES,][]{NIRES} on Keck II as part of the Keck Infrared Transient Survey (KITS) collaboration \citep{Tinyanont23}; the Son of ISAAC spectrograph \citep[SOFI,][]{SOFI} on ESO-NTT; the Espectr\'{o}grafo Multiobjeto Infra-Rojo \citep[EMIR,][]{EMIR} on GTC; and TripleSpec \citep{TripleSpec} on the Southern Astrophysical Research Telescope (SOAR). The NIRES data were reduced using the PypeIt package \citep{pypeit_arXiv, pypeit_zenodo}. SOFI data were reduced using the PESSTO pipeline \citep{PESSTO}. EMIR observations were reduced using PyEmir \citep{Pascual10, Cardiel19}. The Triplespec data were reduced using a modified version of the Spextool package \citep{Cushing04}, and were corrected for telluric absorption with the software and prescription of \citet{Vacca03}.

All spectra are corrected for Milky Way extinction and the host galaxy redshift ($z=0.007735$). Optical spectra are plotted in Figure \ref{fig:optspec1} and \ref{fig:optspec2} and infrared spectra are plotted in Figure \ref{fig:irspec}. All spectra are logged in Table \ref{tab:specInst}.

\subsection{Radio observations}\label{sec:radobs}

SN~2022xkq was observed with the Australia Telescope Compact Array (ATCA) 
at 5.5 and 9.0 GHz on 2022 15 Oct (12 days before $B_{\mathrm{max}}$) between 19:55-22:50 UT using a bandwidth of 2 GHz  
\citep{ryder22}. The observations and reductions were done in a similar way to what was described
for SN~2018gv in \citet{plund20}. Since the ATCA primary flux calibrator was below the horizon throughout, PKS B0823-500 was used as the flux and bandpass calibrator, while PKS B0514-161 was observed every 15 minutes for gain and phase calibration. No radio emission was detected to 3$\sigma$ upper limits on the flux
density of 0.07 mJy and 0.04 mJy at 5.5 GHz and 9.0 GHz, respectively.
For a host galaxy distance of 31 Mpc, this implies upper limits on the luminosity of 
$8.05~(4.60)\EE{25}$ erg s$^{-1}$ Hz$^{-1}$ for 5.5~(9.0) GHz. 

\subsection{Extinction}\label{sec:extinct}

The equivalent widths of \ion{Na}{1} D absorption lines correlate with interstellar dust extinction \citep{Richmond1994, Munari1997}. \ion{Na}{1} D equivalent widths can be converted to $E(B-V)$ values using Eq 9. of \citet{Poznanski2012} with an additional normalization factor of 0.86 from \citet{Schlafly2010}. However, this relation saturates at an equivalent width of $\sim$0.2\,{\AA}.  Milky Way \ion{Na}{1} D lines are clearly apparent in the spectra of SN~2022xkq, however the equivalent widths of these lines are $\ge0.5$\,{\AA} for all observed spectra. We instead use the \citet{Schlafly2011} value of $E(B-V)_\mathrm{MW} = 0.1155 \pm 0.0023$ mag. 
There is no evidence of host \ion{Na}{1} D in any of the spectra of SN~2022xkq, so the host extinction is likely negligible. 

We also estimate the host extinction by using the slope of the $B-V$ color from 30-90 days after $V$-band maximum \citep{Lira, Phillips1999}. We employ this method, known as the Lira Law, by using the procedure described in \citet{Phillips1999} and the equation:
\begin{equation}
    (B-V)_0 = E(B-V)_\mathrm{host} + 0.725 - 0.0118(t_V-60) 
    \label{eq:Lira}
\end{equation}

\begin{figure}
    \centering
    \includegraphics[width=\hsize]{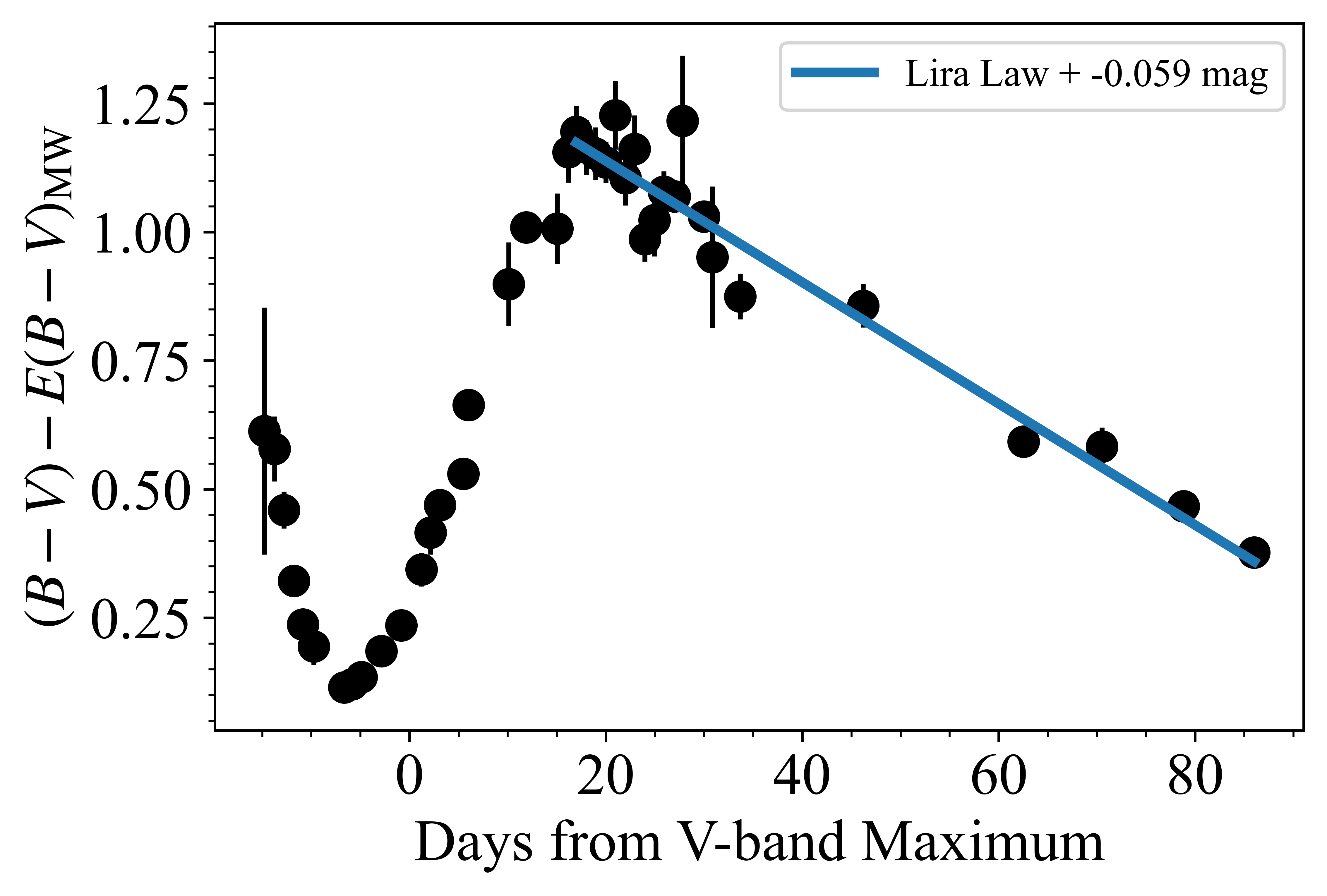}
    \caption{Milky Way extinction-corrected $B-V$ color curve of SN~2022xkq. The color on the radioactive tail is offset from the \citet{Lira} Law by $E(B-V)_\mathrm{host}=-0.059\pm 0.009$ mag, which is consistent with zero given the known scatter of the Lira law. 
    We discuss the color curve of SN~2022xkq and compare to other SNe Ia in Section \ref{sec:phot} and Figure \ref{fig:color}.}
    \label{fig:Lira}
\end{figure}

\noindent In Figure \ref{fig:Lira}, we perform a weighted least-squares fit to Eq. \ref{eq:Lira}, where $(B-V)_0$ is the $B-V$ color curve corrected for only Milky Way extinction and $t_V$ is the phase with respect to V-band maximum. 
From this we find an unphysical $E(B-V)_\mathrm{host} = -0.059 \pm 0.009$ mag, assuming R$_\mathrm{V}$ of 3.1, which is consistent with zero given the known scatter of the Lira Law \citep[$\sim0.05$ mag;][]{Phillips1999}.
Given the location of the SN in the host (see Figure~\ref{fig:Loc}), minimal host extinction is expected.  We therefore assume negligible host extinction and adopt $E(B-V)_\mathrm{tot} =$ $E(B-V)_\mathrm{MW} = 0.116 \pm 0.002$ mag.

\section{Classification}\label{sec:class}

\begin{figure*}
    \centering
    \includegraphics[width=\hsize]{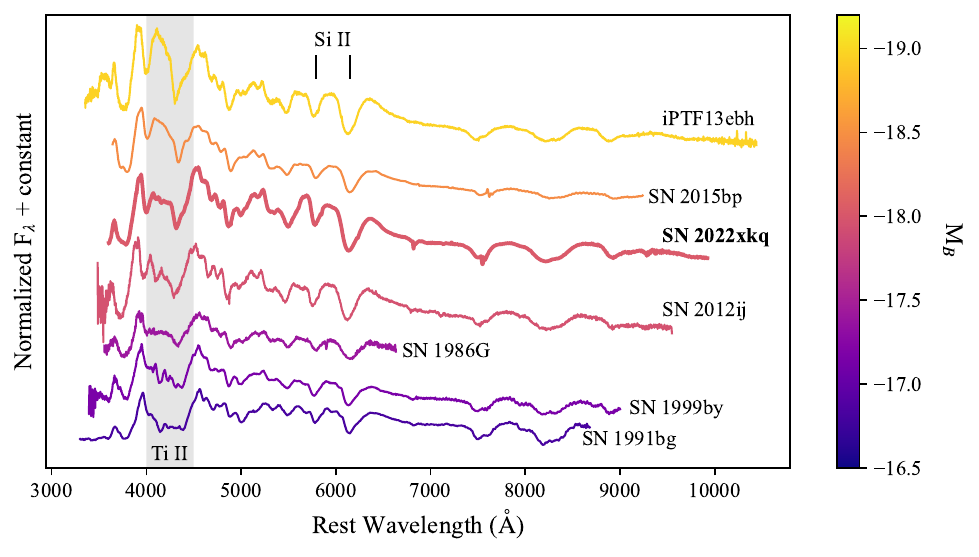}
    \caption{Comparison of spectra taken near $B_{\mathrm{max}}$ for transitional and 91bg-like SNe Ia. SN~2022xkq is underluminous ($B_{\mathrm{max}}=-18.01$) and exhibits a ``titanium trough" (highlighted), characteristics of 91bg-like SNe Ia. Note that the blue part of the titanium trough is actually \ion{Sc}{2}.}
    \label{fig:peak}
\end{figure*}

Underluminous SNe Ia come in a variety of sub-types, often with overlapping definitions. This is made more complex by the fact that some sub-type classifications are based on spectroscopy and others are based on photometry. 
The most common of the underluminous sub-types is 91bg-like, which make up $\sim$5--20\% of all SNe Ia \citep{Graur17, Desai23}. 91bg-like is a spectroscopic classification, characterized by strong \ion{Ti}{2} and and \ion{Sc}{2} features which create a distinctive ``titanium trough" \citep{Liu23_review}.

In contrast, ``subluminous" and ``transitional" are photometrically classified underluminous sub-types. 
Subluminous SNe have fast-declining light curves, weak or non-existent secondary NIR maxima, and NIR peaks which occur after the $B$-band peak \citep{99by_opt, Phillips2012, Hsiao15}. 
\citet{Hsiao15} suggest the existence of a photometrically-classified ``transitional" sub-type, which bridges the gap in the \citet{Phillips93} relationship between between subluminous 91bg-like and normal SNe Ia. Transitional SNe Ia have fast-declining light curves similar to subluminous SNe, but have a primary NIR ($iYJHK$) maximum before $B$-band maximum like normal SNe Ia. 

The majority of 91bg-like SNe are subluminous (e.g., SN~1991bg and SN~1999by), and the terms 91bg-like and subluminous are often mistakenly used interchangeably.
Conversely, many transitional SNe Ia (e.g., iPTF13ebh and SN~2015bp), do not possess a \ion{Ti}{2} trough and are not spectroscopically 91bg-like (see Figure \ref{fig:peak}). However, since transitional is a photometric classification and 91bg-like is a spectroscopic classification, these classes are not entirely separate populations.
 
Objects with both 91bg-like and transitional characteristics are not uncommon. Further investigation into the infrared properties of 91bg-like SNe reveal that this population is bimodal in that some 91bg-like SNe display the NIR properties of subluminous SNe while others display those of transitional SNe \citep{03gs, Folatelli2010, Phillips2012, Hsiao15, Dhawan17}. These SNe tend to be slightly brighter than subluminous 91bg-like SNe and they have weaker \ion{Ti}{2} features, though the feature is still a prominent trough-shape \citep[see Figure \ref{fig:peak}]{Dhawan17}. Notable examples include SN~1986G \citep{86G}, SN~2003gs \citep{03gs}, and SN~2012ij \citep{12ij}, all of which would be spectroscopically classified as 91bg-like and photometrically classified as transitional. 

As shown in Figure \ref{fig:peak}, SN~2022xkq has a strong \ion{Ti}{2} trough (4000-4500 {\AA}) at $B_{\mathrm{max}}$, leading to the classification of SN~2022xkq as a 91bg-like SN Ia. The shape of this titanium trough is strikingly similar to SN~1986G, which is notably photometrically transitional \citep{Ashall2016}. 
Similarly, SN~2022xkq exhibits all the photometric characteristics of transitional SNe (see Section \ref{sec:phot}). 
With a $B_{\mathrm{max}} = -18.01$ mag, SN~2022xkq is somewhat faint but still within the expected luminosity range of transitional SN~Ia. Furthermore, SN~2022xkq is a fast-decliner, ${\Delta} m_{15}(B) = 1.65 \pm 0.03$, similar to other SNe Ia with comparable peak magnitudes. 

\begin{figure}
    \centering
    \includegraphics[width=\hsize]{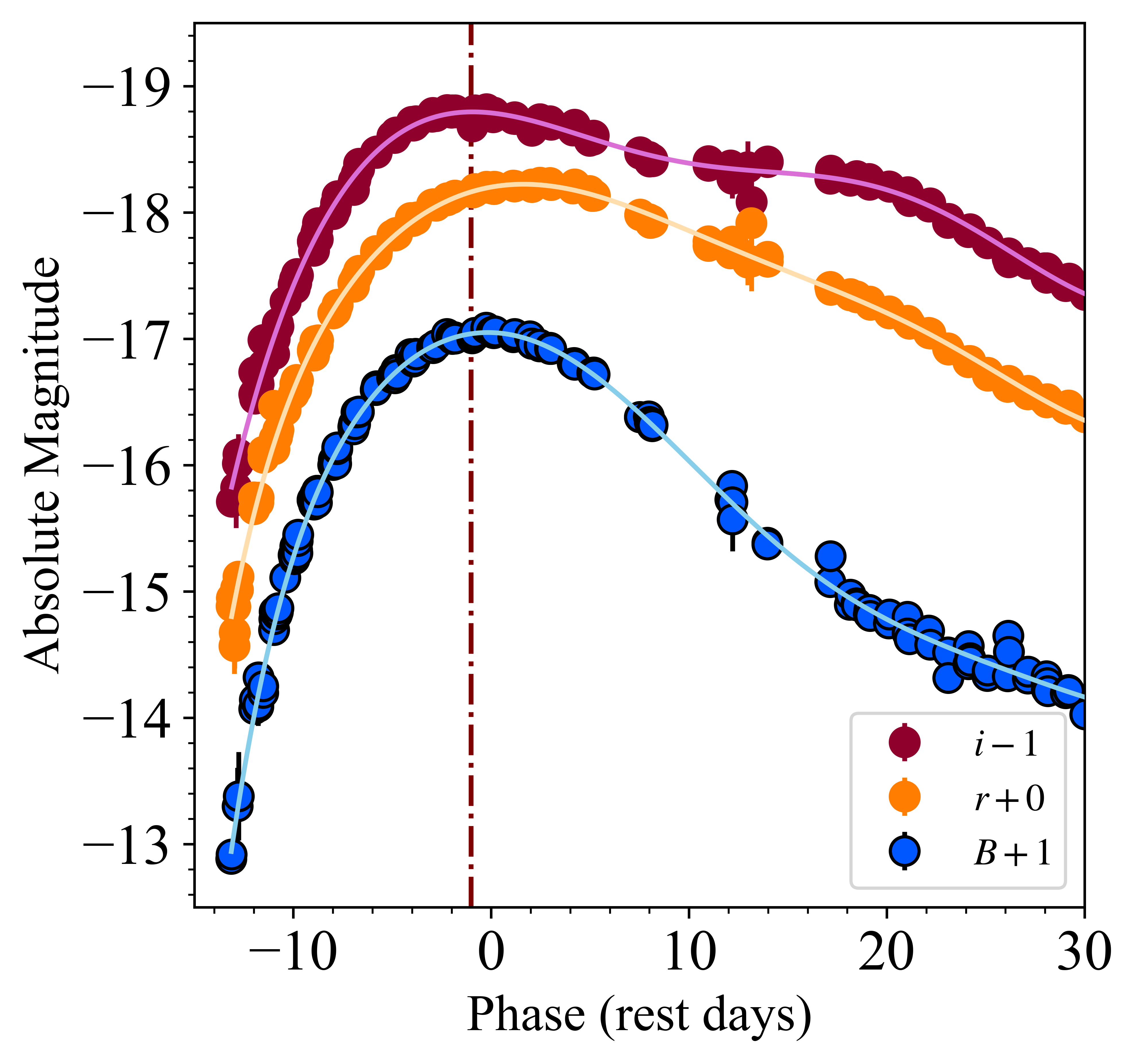}
    \caption{Plot of the $B$-, $r$-, and $i$-band DLT40 and Las Cumbres Observatory light curves of SN~2022xkq. The solid lines are polynomial fits to each band of the light curve. The maroon vertical dashed-dotted line marks the time of $i$-band max. The $i$-band peaks at $\sim{-1}$ day relative to $B_{\mathrm{max}}$ and also exhibits a clear secondary maximum. The secondary maximum is also present in $r$-band though much weaker. These light curve attributes are typical of transitional SNe Ia.
    }
    \label{fig:Bir}
\end{figure}

To determine if SN~2022xkq has the NIR properties of the transitional subclass, we fit the $i$-band light curve. We exclude the $JHK$ observations from this analysis since these observations began after NIR maximum. We find that the DLT40 and Las Cumbres Observatory $i$-band light curves peak at $-1.2 \pm 0.1$ and $-1.0 \pm 0.2$ days with respect to B-band maximum, respectively. Given that these measurements are consistent with one another, the $i$-band peak occurs $\sim$1~day before the $B$-band peak, as shown in Figure \ref{fig:Bir}. Further, the $i$-band light curve displays a clear secondary peak (the $r$-band also displays a weak secondary peak as seen Figure \ref{fig:Bir}). Indeed, the early NIR maximum and clear secondary bump in the $i$-band classify SN~2022xkq as a transitional SN Ia. 

Given the strong titanium trough and early NIR maximum, 
we classify SN~2022xkq as spectroscopically 91bg-like and photometrically transitional. SN~2022xkq adds to the growing sample of SNe with such properties, which link purely transitional SNe and subluminous 91bg-like SNe. The photometric and spectroscopic dataset presented in this work makes SN~2022xkq the best-observed SN of this type and can provide insight into the different progenitor scenarios for both 91bg-like and transitional SNe.  

\section{Photometric Properties} \label{sec:phot}
Optical and Swift UV light curves (shown in Figure \ref{fig:phot}) were obtained for SN~2022xkq beginning immediately following the discovery of the SN (MJD 59,865.28; 14 days before $B_{\mathrm{max}}$). 

Light curve properties were fit using the \texttt{SNooPy} software package \citep[with $H_0=72$ km s$^{-1}$ Mpc$^{-1}$;][]{snoopy}. Currently, \texttt{SNooPy} lacks UV templates for SNe Ia fitting, thus, we exclude any UV data from these fits and only include DLT40 and Las Cumbres Observatory $BVgri$ data.
The results of these processes are tabulated in Table \ref{tab:results}. SN~2022xkq reached $B_{\mathrm{max}} = -18.01 \pm 0.15$ mag (m$_B=14.45$ mag) on MJD $59,879.03 \pm 0.34$. 
Further, SNooPy measures a decline rate of ${\Delta} m_{15} = 1.62 \pm 0.06$ mag and a color-stretch parameter of s$_{BV} = 0.63 \pm 0.03$. The color-stretch parameter is a better characterization of the light curves of fast declining SNe Ia. The $B$-band declines very quickly in many faint SNe, resulting in a diverging M$_B$ versus ${\Delta} m_{15}$ relation \citep{Burns2014}. In contrast, using the s$_{BV}$ color-stretch parameter allows these fast-declining events to appear as part of the tail end of the normal SN Ia population \citep{Burns2018, Gall2018}. 
We also measure ${\Delta} m_{15}$($B$) directly from the DLT40 and Las Cumbres Observatory $B$-band light curve by fitting an eight-degree polynomial to the data between maximum light at 15 days post-maximum light using a least-squares method; this returns a value of $1.65 \pm 0.03$ (reported in Table \ref{tab:results}).
The measured ${\Delta} m_{15}$ and s$_{BV}$ values for SN~2022xkq are consistent with other 91bg-like and transitional SNe Ia with similar maximum M$_B$ magnitudes. 

At maximum light, the bolometric luminosity of a SN Ia is related to the amount of $^{56}$Ni produced in the explosion; this relation is known as Arnett's Rule \citep{Arnett82, Arnett85}. To determine the $^{56}$Ni mass we use Equation 7 of \citet{Stritzinger05}, which gives the peak luminosity of a SN Ia with 1.0 M$_{\odot}$ of $^{56}$Ni as 
\begin{equation}
    \mathrm{L_{max}}=(2.0\pm0.3)\times10^{43}\,  \frac{\mathrm{M_{Ni}}}{\mathrm{M_{\odot}}}\,\mathrm{erg\,s}^{-1}
\end{equation}
where M$_\mathrm{Ni}$ is the nickel mass and L$_\mathrm{max}$ is the maximum bolometric luminosity. The complete bolometric light curve, which includes the $JHK$ photometry, has a peak luminosity of $4.3\times10^{42}$ erg s$^{-1}$, resulting in a $^{56}$Ni mass of $0.22\pm0.03$ M$_{\odot}$.  
To ensure this mass is reasonable, we also determine the $^{56}$Ni mass from the maximum of the pseudo-bolometric light curve which we calculated from the $UBVgri$ Las Cumbres Observatory light curve. To account for the $\sim10\%$ of light at maximum which is emitted outside of the optical \citep{Suntzeff96}, we follow the procedure in \citet{Stritzinger06}, which multiplies the resulting mass by a factor of $1.1$. This method results in a pseudo-bolometric luminosity of $3.9\times10^{42}$ erg s$^{-1}$ and a $^{56}$Ni mass of $0.21\pm0.03$ M$_{\odot}$. This mass is consistent with the value from the full bolometric light curve so we assume that $^{56}$Ni mass and list it in Table \ref{tab:results}.
\begin{table}
 \centering
 \caption{Summary of SN~2022xkq Properties} \label{tab:results}
 \begin{tabular}{ l c l}
    \hline
    Parameter & & Value \\
    \hline
    Last Non-Detection & & MJD 59,864.49\\ 
    First Detection & & MJD 59,865.28\\
    Explosion Epoch\footnote{from power law fit} & & MJD $59,865.0\pm 0.3$\\
    Redshift $z$ & & 0.007735 \\
    Distance & & $31 \pm 2$ Mpc\\
    Distance modulus ($\mu$) & & $32.46 \pm 0.15$ mag\\
    $E(B-V)_\mathrm{tot}$ & & $0.1155 \pm 0.0023$ mag\\
    Peak Magnitude ($B_{\mathrm{max}}$) & & $-18.01 \pm 0.15$ mag\\
    Time of $B_{\mathrm{max}}$ & & MJD $59,879.03 \pm 0.34$\\
    $s_{BV}$ & & $0.63 \pm 0.03$\\
    $\Delta m_{15}(B)$  & & $1.65 \pm 0.03$ mag \\
    $^{56}\mathrm{Ni}$ mass& & $0.22\pm0.03$ M$_{\odot}$ \\
    \ion{Si}{2} velocity (-11.6 d) & & $11,920 \pm 350$ km s$^{-1}$\\
    \ion{Si}{2} velocity (1.2 d) & & $9,980 \pm 200$ km s$^{-1}$\\ 
    \ion{Si}{2} $\lambda$5972 EW (1.2 d) & & $38 \pm 1$~{\AA}\\
    \ion{Si}{2} $\lambda$6355 EW (1.2 d) & & $129 \pm 7$~{\AA}\\
    H$\alpha$ luminosity & & $<8.8\times10^{36}$ erg s$^{-1}$\\
    \ion{He}{1} $\lambda$5875 luminosity & & $<4.0\times10^{37}$ erg s$^{-1}$\\
    \ion{He}{1} $\lambda$6678 luminosity & & $<8.8\times10^{36}$ erg s$^{-1}$\\ 
    Stripped H mass (Boty\'anszki) & & $\lesssim 2\times 10^{-4}$ M$_{\odot}$\\
    Stripped H mass (Dessart) & & $\lesssim 7\times 10^{-5}$ M$_{\odot}$\\
    Stripped He mass (Boty\'anszki) & & $\lesssim 7\times 10^{-4}$ M$_{\odot}$\\
    \hline
 \end{tabular}
\end{table}

\begin{figure*}
    \centering
    \includegraphics[width=\hsize]{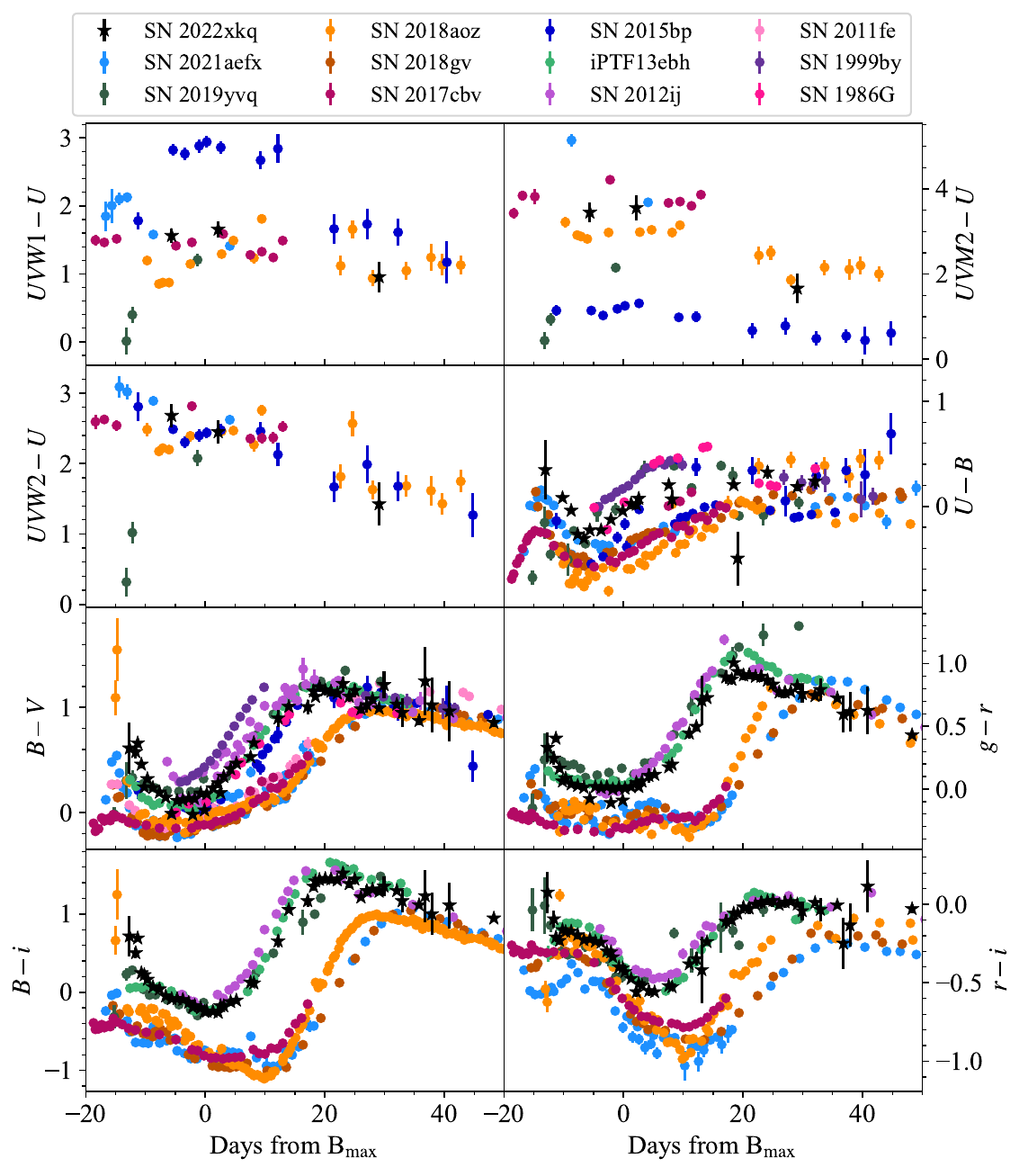}
    \caption{Color evolution of SN~2022xkq compared to other notable SNe Ia, including SN~1986G \citep{86G}, SN~1999by \citep{99by_opt},  SN~2011fe \citep{11fe_lc}, SN~2012ij \citep{12ij}, iPTF13ebh \citep{Hsiao15}, SN~2018aoz \citep{Ni22}, and SN~2021aefx \citep{Hosseinzadeh22}. The color curve of SN~2022xkq is remarkably similar to that of SN~1986G. SN~2022xkq exhibits an early red color, which is most visible in ($B-i$). This early red color before maximum light is most similar to that observed in SN~2018aoz, which has a bump that coincides with its early red evolution. However, SN~2022xkq is generally redder than SN~2018aoz and has a different color evolution at later times.}
    \label{fig:color}
\end{figure*}

In Figure \ref{fig:color}, we compare the color evolution of SN~2022xkq with that of notable and relevant SNe Ia: SN~1986G \citep{86G}, SN~1999by \citep{99by_opt}, SN~2011fe \citep{11fe_lc}, SN~2012ij \citep{12ij}, iPTF13ebh \citep{Hsiao15}, SN~2015bp \citep{Srivastav17}, SN~2017cbv \citep{Hosseinzadeh17}, SN~2018gv \citep{18gv}, SN~2018aoz \citep{Ni22}, SN~2019yvq \citep{Burke21}, and SN~2021aefx \citep{Hosseinzadeh22}. The color evolution of SN~2022xkq is very similar to the transitional SNe Ia iPTF13ebh and SN~2015bp but also to that of SN~2012ij and SN~1986G. SN~1986G is heavily extincted and we use $E(B-V)_\mathrm{tot} = 0.88$ from \citet{Ashall2018}, though $E(B-V)$ values as high as $1.1$ \citep{86G_high} and as low as $0.63$ \citep{86G_low} have also been suggested. If we assume SN~1986G has a slightly lower extinction ($\sim0.78$), the color evolution of SN~1986G is almost identical to that of SN~2022xkq. The similarity in color between SN~2022xkq and SN~1986G is particularly of note because these two SNe are also remarkably similar spectroscopically.

A striking aspect of SN~2022xkq's color evolution is its very red color immediately following explosion (until $\sim-10$ days before the time of $B_{\mathrm{max}}$), which is especially notable in the ($B-i$) color curve, though it is present in all bands. The strong red color is somewhat similar to that observed in SN~2018aoz \citep{Ni22, Ni23_18aoz}. However, SN~2022xkq is generally redder than SN~2018aoz and there is no bright bump in SN~2022xkq as was observed in the very early light curve of SN~2018aoz. 
The flux excess of SN~2022xkq is similar to the normal SNe Ia ZTF18aayjvve and ZTF18abdfwur \citep[see Figure~6 in][]{Deckers22} which have been identified as having early red colors. 
Underluminous SNe Ia are intrinsically redder than normal SNe Ia \citep{Taubenberger08} and to our knowledge, there is no 91bg-like SN Ia with multiband photometry as early as for SN~2022xkq. Therefore, this very early (within $\sim5$ days of explosion) red color could be typical for the subclass. The possible causes of this color will be further discussed in Section \ref{sec:red}. 

\subsection{Power Law Fit}
Densely sampled early light curves, like those obtained for SN~2022xkq, provide useful constraints on the progenitor system and/or explosion mechanism. In particular, bumps and excesses in the early light curve can be distinguishing factors among various explosion mechanisms or progenitor scenarios. Some light curve excesses and smaller bumps can be difficult to distinguish by eye but these features are notable when the observed light curves are compared to a power law rise \citep{Dimitriadis19}.

\begin{figure}
    \centering
    \includegraphics[width=\hsize]{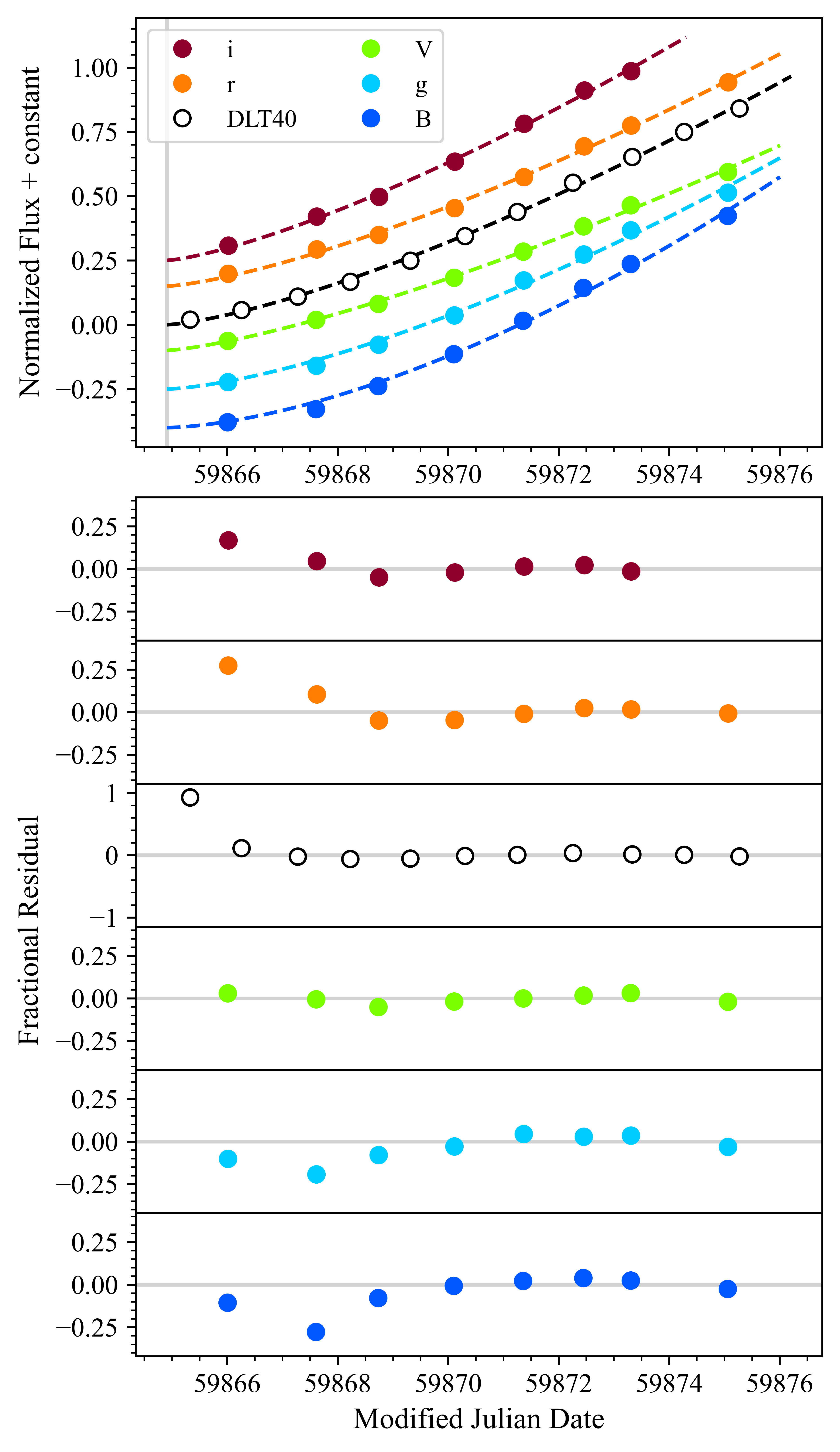} 
    \caption{The binned LCO multiband and DLT40 open band light curves fit to a power law, with $t_{\rm exp}=59865.0 \pm 0.3$ MJD (marked by grey line in top panel). Residuals for each band are plotted as fractional residuals, i.e. residual divided by the power law model flux. Immediately following explosion the power law fit over predicts flux in the blue filters and under predicts flux in the red, this indicates the presence of a slight red flux excess.
    }
    \label{fig:powerlaw} 
\end{figure}

Using the non-linear least squares method, we fit the first 10 days of the Las Cumbres Observatory multiband and the $DLT40$-band (see Section \ref{sec:obs_phot} for more information) light curve to a power law of the form:
\begin{equation}
    F = c(t-t_{exp})^a
\end{equation}
where $t_{\rm exp}$ is the time of explosion, $F$ is the flux, and $c$ and $a$ are free parameters. Photometry is binned to 0.3 days to reduce the impact of the different observational sites. When we limit the fit to only when the light curve reaches half maximum flux \citep[the procedure used in][]{Firth15} the unconstrained fit returns an explosion epoch which is inconsistent with the observations.
Alternatively, we constrain $t_{\rm exp}$ to vary only between the last non-detection and discovery epoch, though an unconstrained fit of the first 10 days returns a weighted mean $t_{\rm exp}$ value consistent with the observed limits. When each individual filter is fit separately, a weighted mean of $t_{\rm exp} = 59864.9 \pm 0.2$ MJD is returned, shown in Figure \ref{fig:powerlaw}. We use this value as the explosion epoch for SN~2022xkq.  

The power law model favors $a = 1.6\pm0.2$ in $B$-band and $a \sim 1.4$ in all others. This differs from the $a = 2$ of the fireball model, which is commonly used to describe the rising light curves of SNe Ia \citep{Riess99, Goldhaber01}. Fixing $a = 2$ results in $t_{\rm exp}$ almost 2 days before the last non-detection and is unable to reproduce the shape of the first few days of the rising light curve. 
Surveys have found $a=2$ to be roughly consistent with normal SNe Ia \citep{Conley06, Hayden10, Ganeshalingam11, Zheng17, Miller20_sample, Fausnaugh23}; however, those with low $s_{BV}$ ($s_{BV}<$0.8) have been observed to have systematically lower $a$ values than normal SNe Ia \citep{Gonzalez12}, so a value $<2$ is expected for SN~2022xkq. 

There is a trend in the residuals of the power law fit, such trends have been linked to early time light curve excesses \citep{Ni22, Burke22_dlt40, Ni23}. 
To best highlight the flux excess, we plot the residual as a fraction of the power law models flux at each epoch (see lower panels of Figure \ref{fig:powerlaw}).
The observed residual trends indicate that the power law fit over predicts flux in the blue filters and under predicts flux in the red. This red flux excess is coincident with the early red colors of SN~2022xkq.  

\subsection{Comparison to Models}\label{sec:lcmodels}

The trend in the residual of the power law fit is an indicator of an excess of flux in the early light curve ($\lesssim3$ days after inferred explosion). Here we compare to specific models which display early bumps and excesses to examine the possible source of this flux. 

\subsubsection{Companion-Shocking}\label{sec:companion}

\begin{deluxetable*}{lLlllLl}
\tablecaption{Companion-shocking Model Parameters\label{tab:params}}
\tablehead{\colhead{Parameter} & \colhead{Variable} & \colhead{Prior Shape} & \twocolhead{Prior Parameters\tablenotemark{a}} & \colhead{Best-fit Value\tablenotemark{b}} & \colhead{Units}}
\tablecolumns{7}
\startdata
\cutinhead{Companion-shocking Model \citep{Kasen10}}
Explosion time & t_0 & Uniform & 59864.49 & 59865.28 & 59864.7^{+0.3}_{-0.2} & MJD \\
Binary separation & a & Uniform & 0 & 0.1 & 0.04^{+0.02}_{-0.01} & $10^{13}$ cm \\
Viewing angle & \theta & Uniform & 0 & 180 & 50 \pm 30 & $\degr$ \\
\cutinhead{SiFTO Model \citep{sifto}}
Time of $B$ maximum & t_\mathrm{max} & Uniform & 59877.14 & 59881.14 & 59879.26 \pm 0.07 & MJD \\
Stretch & s & Log-uniform & 0.5 & 2 & 0.858 \pm 0.008 & dimensionless \\
Time shift in $U$ & $\Delta t_U$ & Gaussian & 0 & 1 & +0.14 \pm 0.07 & d \\
Time shift in $i$ & $\Delta t_i$ & Gaussian & 0 & 1 & +0.46 \pm 0.06 & d \\
\cutinhead{Combined Model}
Intrinsic scatter & \sigma & Half-Gaussian & 0 & 1 & 7.3 \pm 0.3 & dimensionless
\enddata
\tablenotetext{a}{The ``Prior Parameters'' column lists the minimum and maximum for a uniform distribution, and the mean and standard deviation for a Gaussian distribution.}
\tablenotetext{b}{The ``Best-fit Value'' column is determined from the 16th, 50th, and 84th percentiles of the posterior distribution, i.e., $\mathrm{median} \pm 1\sigma$.}
\end{deluxetable*}\vspace{-24pt}

\begin{figure*}
    \centering
    \includegraphics[width=0.495\hsize]{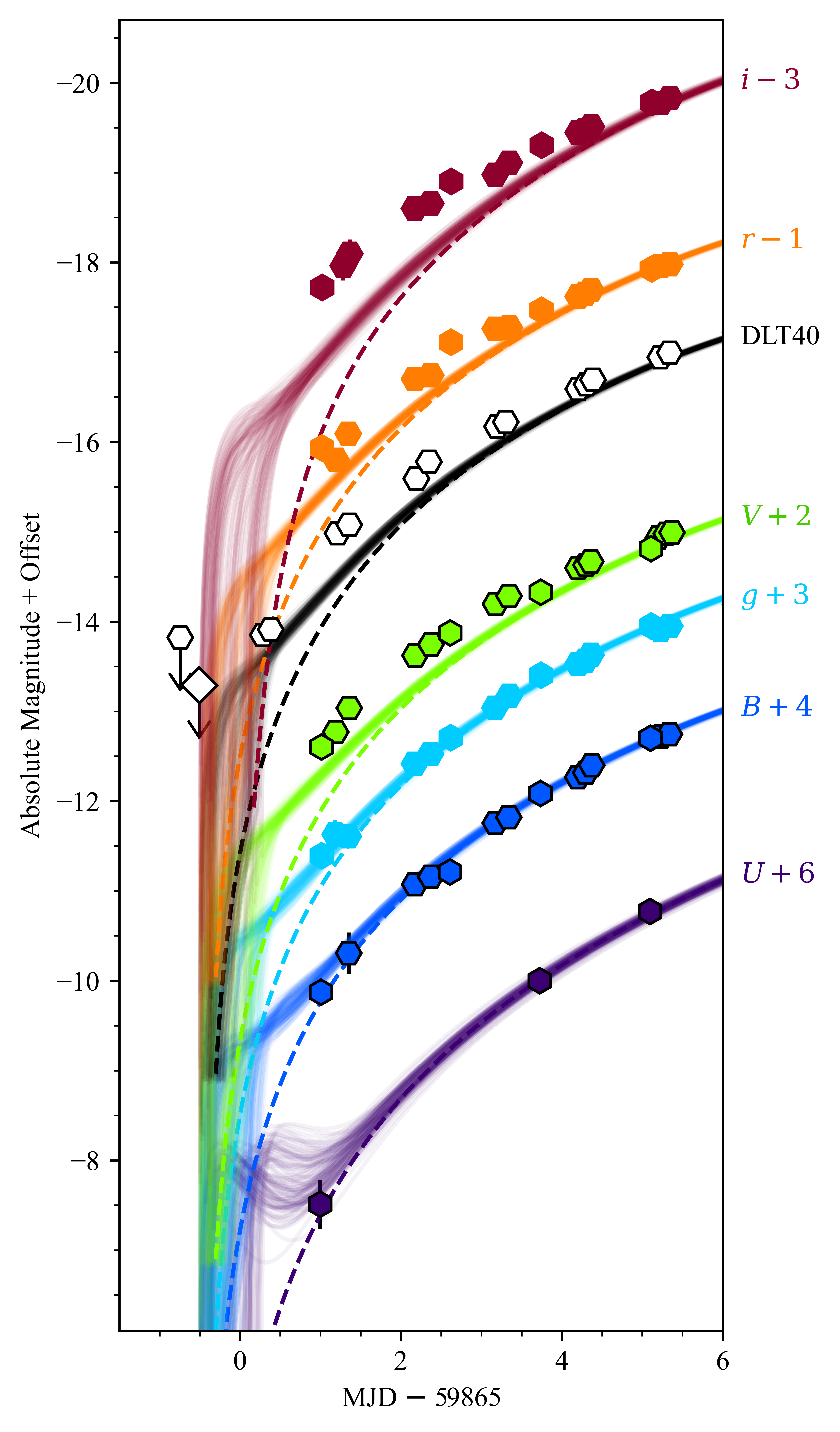} 
    \includegraphics[width=0.495\hsize]{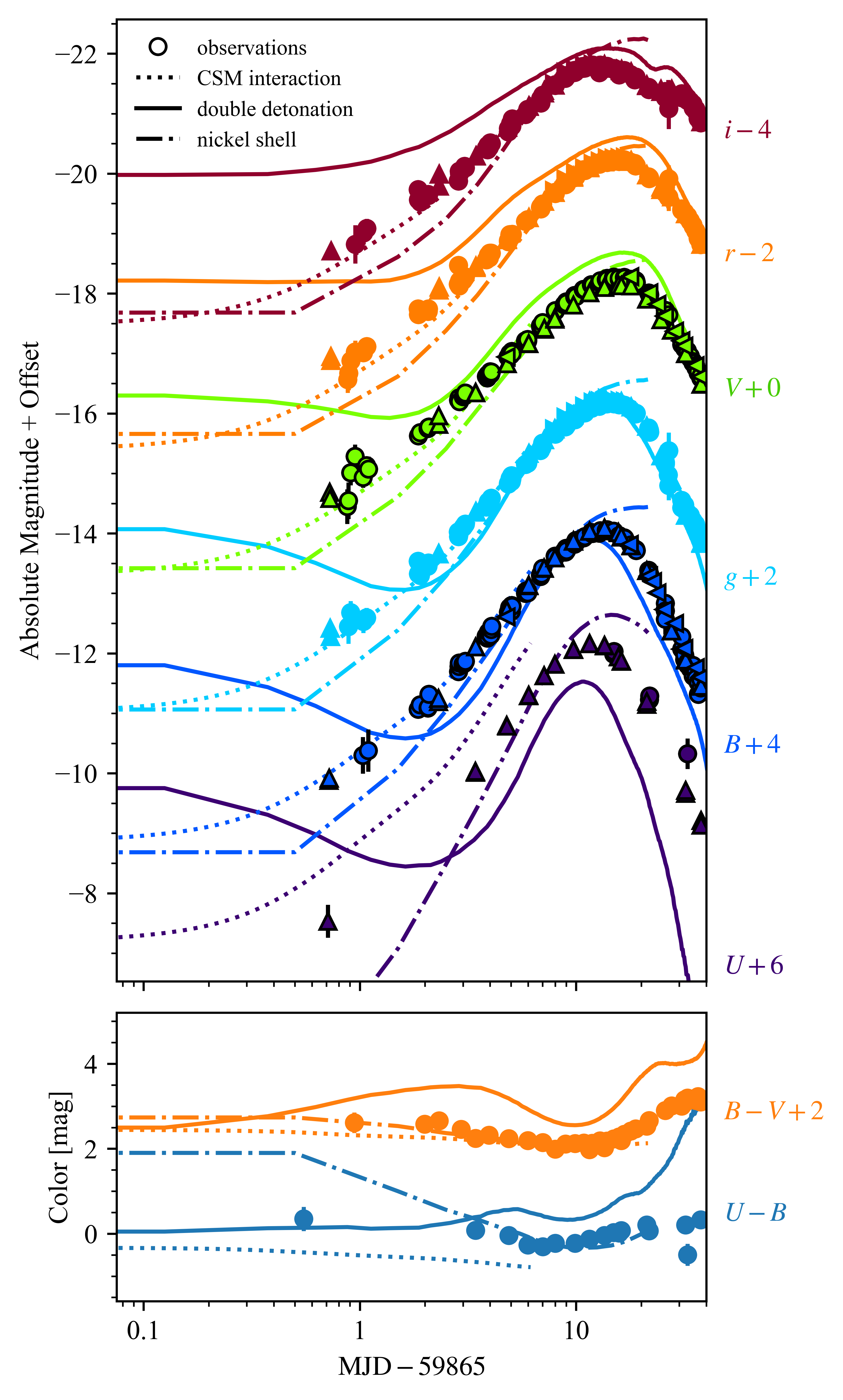} 
    \caption{Left: SiFTO templates (dashed) and the combination of the SiFTO templates and \citet{Kasen10} companion-shocking component (solid lines) fit using the Light Curve Fitting package. A companion-shocking component is unable to reproduce the light curve in the redder bands without introducing a large blue excess which is not observed. In combination with the residuals from the power law fit (see Figure \ref{fig:powerlaw}), this suggests the existence of a red excess in the early light curve.
    Right: The double detonation model (solid lines) from \citet{Polin19}, nickel distribution model (dash-dotted lines) from \citet{Magee20_Nidist}, and CSM interaction model (dotted lines) from \citet{Piro16} which best match the observed photometry of SN~2022xkq. No model is a particularly good description of the observed light curve, though the nickel distribution and CSM interaction models do somewhat reproduce the rising light curve, particularly in the redder bands.}
    \label{fig:earlyLC} 
\end{figure*}

Early light curve excesses may arise from SN ejecta colliding with a nondegenerate companion resulting in the SN becoming briefly bluer and more luminous \citep{Kasen10}. We fit a combination of the \citet{Kasen10} companion-shocking component and SiFTO templates \citep{sifto} using an Markov-Chain Monte Carlo (MCMC) routine implemented in the Light Curve Fitting \citep{lightcurvefitting, lightcurvefitting2} package using the same parameters and procedure as described in \citet{23bee}. 
The SiFTO models are scaled to each band so that they match the observed maximum for SN~2022xkq without preserving the colors of the original SiFTO template.  The SiFTO models for $U$- and $i$-band have also been slightly shifted in time when necessary to best match the observed light curve. We only include the data up to the time of $B_\mathrm{max}$ (59879.14 MJD) in the fit since the SiFTO template does not match the observed post-maximum light well, especially the secondary infrared maximum. 
All SN~2022xkq optical photometry is binned by 0.1 days.
The SiFTO template does not include the UV so we exclude the Swift photometry. Finally, we include a $\mathrm{\sigma}$ term which allows for proper handling of the intrinsic scatter in the photometry by effectively inflating the photometric uncertainties by a factor of $\sqrt{1+\mathrm{\sigma}^2}$.
The prior and posterior distributions of the fit parameters are given in Table \ref{tab:params} and the best-fit model and SiFTO templates are compared to the observed light curve in Figure \ref{fig:earlyLC} (left).

The possibility of an early excess is further justified by a comparison to a SiFTO SN Ia template \citep{sifto}, which fits $U$, $B$, $V$, and $g$ well but underestimates the light curve in the redder bands, most notably in the $i$-band. 
The companion-shocking component fits excesses which exist primarily in bluer bands. The red color of the excess in SN~2022xkq is not well explained by these models. 
However, the redder color of SN~2022xkq's light curve does not necessarily exclude the possibility of companion interaction 
\citep{Deckers22}. 
Further modeling is needed to determine if companion interaction can produce a red color without producing a detectable bump in bluer optical bands. 

\subsubsection{Double Detonation}

Another scenario that could result in early excess flux is the double detonation of a sub-Chandrasekhar mass white dwarf. Double detonation occurs when a shell of surface helium detonates before the white dwarf core ignites. This process has been suggested as an explosion mechanism for producing underluminous SNe Ia \citep{Polin19, Gronow21}. Further, double detonation models are linked to early red excesses in SNe Ia \citep[see Section \ref{sec:red} for further discussion]{Jiang17, Ni23_18aoz}. 
We compare the \citet{Polin19} double-detonation models to the observations of SN~2022xkq. To account for uncertainties in explosion time, we apply a time offset from 0 to 3 days at 0.5 day intervals. We calculate the reduced $\chi^2$ ($\bar\chi^2 \equiv \chi^2/N$, where N is the number of photometric points which overlap the model) for each model in the grid. The model with the lowest $\bar\chi^2$ (see right panel of Figure \ref{fig:earlyLC}) is of a 0.9 M$_{\odot}$ white dwarf with a 0.05 M$_{\odot}$ helium shell offset by 3.0 days. 

The double detonation model can somewhat reproduce the photometry near and after maximum light in all bands except $U$, where the model under-predicts the luminosity. However, it fails to reproduce the color and early light curve and requires a large time shift which is inconsistent with observed non-detections. 
To determine if this was a result of comparing to the photometry rather than the color evolution, we also run the same process to find the model with light curves closest to the observed $U-B$ and $B-V$ color. The model best matched to the color is a 1.0 M$_{\odot}$ white dwarf with a 0.1 M$_{\odot}$ helium shell offset by 3.0 days. This model is overluminous in all bands while still requiring a time shift which conflicts with non-detections. 

The \citet{Polin19} double detonation models were not designed for SN~2022xkq, adjusting any number of parameters in the model could result in a double detonation model which is a better match to SN~2022xkq. Specifically, we note that the next best match to the photometry is a model of a 0.9 M$_{\odot}$ white dwarf with a 0.01 M$_{\odot}$ helium shell. The available model grid does not include any models of 0.9 M$_{\odot}$ white dwarfs with helium shell masses between 0.05 and 0.01 M$_{\odot}$ which may produce a better match to SN~2022xkq. 
Further, double detonation is an asymmetric process and the viewing angle can have a significant impact on the light curve, particularly in the bluer bands \citep{Shen21}. The \citet{Polin19} models are 1D and cannot account for viewing angle effects. 
We are therefore unable to rule out the possibility of a double detonation scenario using the early light curves alone.

\subsubsection{Distribution of Nickel}

Another possible explanation for the red color in the early light curve of SN~2022xkq is shells or clumps of $^{56}$Ni near the surface of the ejecta, providing extra photons in the rising light curve as a result of radioactive decay. We compare the observed color curves of SN~2022xkq to nickel models from \citet{Magee20_Nidist} and \citet{Magee20_Niclump} which were produced to explain the early excesses in SN~2017cbv and SN~2018oh (see dashed lines in right panel of Figure \ref{fig:earlyLC}). Following the same process as for the double detonation models, we find the best-matched nickel model to be 0.4 M$_{\odot}$ of $^{56}$Ni distributed according to an exponential density profile with a scaling parameter of 9.7 and a kinetic energy of $5.04\times10^{50}$ erg; the model is offset by 3.0 days. 
This model is a good match to the observed $B-V$ color and the $U-B$ color after the first few days. Overall, the nickel models are a much better match to the observed light curve than the best double detonation model, though the nickel model underpredicts the rising light curve and slightly overpredicts the peak luminosity in all bands. Similar to the double detonation models, the nickel models to which we compare were not designed for SN~2022xkq and presumably could be improved to produce a better match to the observations.

\subsubsection{Interaction with Circumstellar Medium}

Accretion or merging proceses are integral to the production of SN Ia and these processes could result in CSM near an exploding white dwarf. This ejecta-CSM interaction can result in excesses in the early light curve similar to those observed in companion shocking models. 
We compare the photometry of SN~2022xkq with models of interaction with CSM from \citet{Piro16} using the same procedure as for the double detonation and the nickel models.  
The best matching model is of a 1.25 M$_{\odot}$ C/O white dwarf with 0.5 M$_{\odot}$ of $^{56}$Ni which has been distributed using boxcar averaging of width 0.1 M$_{\odot}$ (for details see \citealt{Piro16}) interacting with extended CSM material of radius 10$^{11}$ cm and a mass of 0.3 M$_{\odot}$. 
As shown in the right panel of Figure \ref{fig:earlyLC}, this model provides the best match to photometry of any considered model in the $BgVri$ bands and is a great match to the observed $B-V$ color.
However, the CSM interaction model significantly exceeds the observed $U$-band flux, resulting in a poor match to the $U-B$ color. The model also slightly undershooting the $r$- and $i$-bands in the epochs where the red early color is most obvious ($\sim3$ days after explosion). 
The models presented in \citet{Piro16} primarily produce light curves with blue early colors and flux excesses. However, these models do not consider radiative transfer and opacity effects in detail and are not expected to describe real SNe. While CSM signatures may be present in the early light curves of SNe Ia, the primary observational signature of this interaction is the presence of hydrogen in nebular spectra (see Section \ref{sec:neb}). A more detailed modeling of ejecta-CSM interaction in SN~2022xkq, particular once late time ($>300$ days) nebular spectra are available, may be able to better describe the rising light curve and $U-B$ color.

\subsubsection{Other Possible Scenarios}

Additionally, we compare the photometry of SN~2022xkq with models of pulsational-delayed-detonations from \citet{Dessart14} using the previously discussed procedure. The pulsational-delayed-detonation model set produces the worst match to the light curve of SN~2022xkq of any of the previously discussed models and we refrain from discussing it further. 

We further compare to the light curves of the delayed-detonation of a Chandrasekhar white dwarf, the violent merger of two sub-Chandrasekhar white dwarfs, the detonation of a bare sub-Chandrasekhar white dwarf, 
and two pure deflagration models presented in \citet{Noebauer17}. The majority of these models have 
red colors at early times (merger, sub-Chandrasekhar detonation, and delayed-detonation). While the shape of the early light curve is somewhat similar to the violent merger model, none of these models produce a $B-V$ color evolution similar to what was observed in SN~2022xkq. 
\\
\\
Here we briefly summarize our observations of the early light curve, and our model exploration.  The $B-V$ color of SN~2022xkq reveals an early color which is generally redder than most of the SNe Ia to which we compare. This color evolution is somewhat similar to the evolution observed for SN~2018aoz which has an identified red excess.
Further, the early light curve of SN~2022xkq exhibits a red excess in the residual when the light curve is fit using a simple power law rise. 
The power law residuals combined with the SN's early color suggests the presence of a red early excess in SN~2022xkq. 
To determine the mechanism responsible for this excess we compare to several models of SN Ia explosions. 
However, the rising light curve and color evolution of SN~2022xkq is not well described by any of the models we consider here, though we are unable to rule out any of the mechanisms to which we compare, as tailored models may produce better results. 

\section{Spectroscopic Properties} \label{sec:spec}

\subsection{Optical Spectra}\label{sec:optspec}

As is typical of SNe Ia, the optical spectra of SN~2022xkq lack hydrogen and helium features while exhibiting characteristic strong absorption features of intermediate-mass and iron-peak elements such as \ion{Si}{2}, \ion{Ca}{2}, and \ion{Fe}{2}, as seen in Figures \ref{fig:optspec1} and \ref{fig:optspec2}. 

In Figure \ref{fig:peak}, we compare an optical spectrum of SN~2022xkq taken near $B_{\mathrm{max}}$ to the peak spectra of other notable low luminosity SNe Ia: the transitional iPTF13ebh \citep{Hsiao15}, SN~2015bp \citep{15bp}, the 91bg-like SN~1991bg \citep{91bg_92, 91bg_93, 91bg_96} and SN~1999by \citep{99by_opt}, the transitional/91bg-like SN~2012ij \citep{12ij} and SN~1986G \citep{86G}. 
SN~2022xkq has a strong \ion{Ti}{2} feature at maximum light, indicating a lower photospheric temperature than found in typical SNe Ia \citep{Mazzali97}.
This \ion{Ti}{2} feature is similar to those seen in the spectra of the 91bg-like SNe, confirming the classification of SN~2022xkq. 
At peak SN~2022xkq is most spectroscopically analogous to SN~1986G, a highly reddened SN which is spectroscopically 91bg-like but has photometric properties not typical of 91bg-like SNe (for more in-depth discussion see Section \ref{sec:class}). 

\begin{figure}
    \centering
    \includegraphics[width=\hsize]{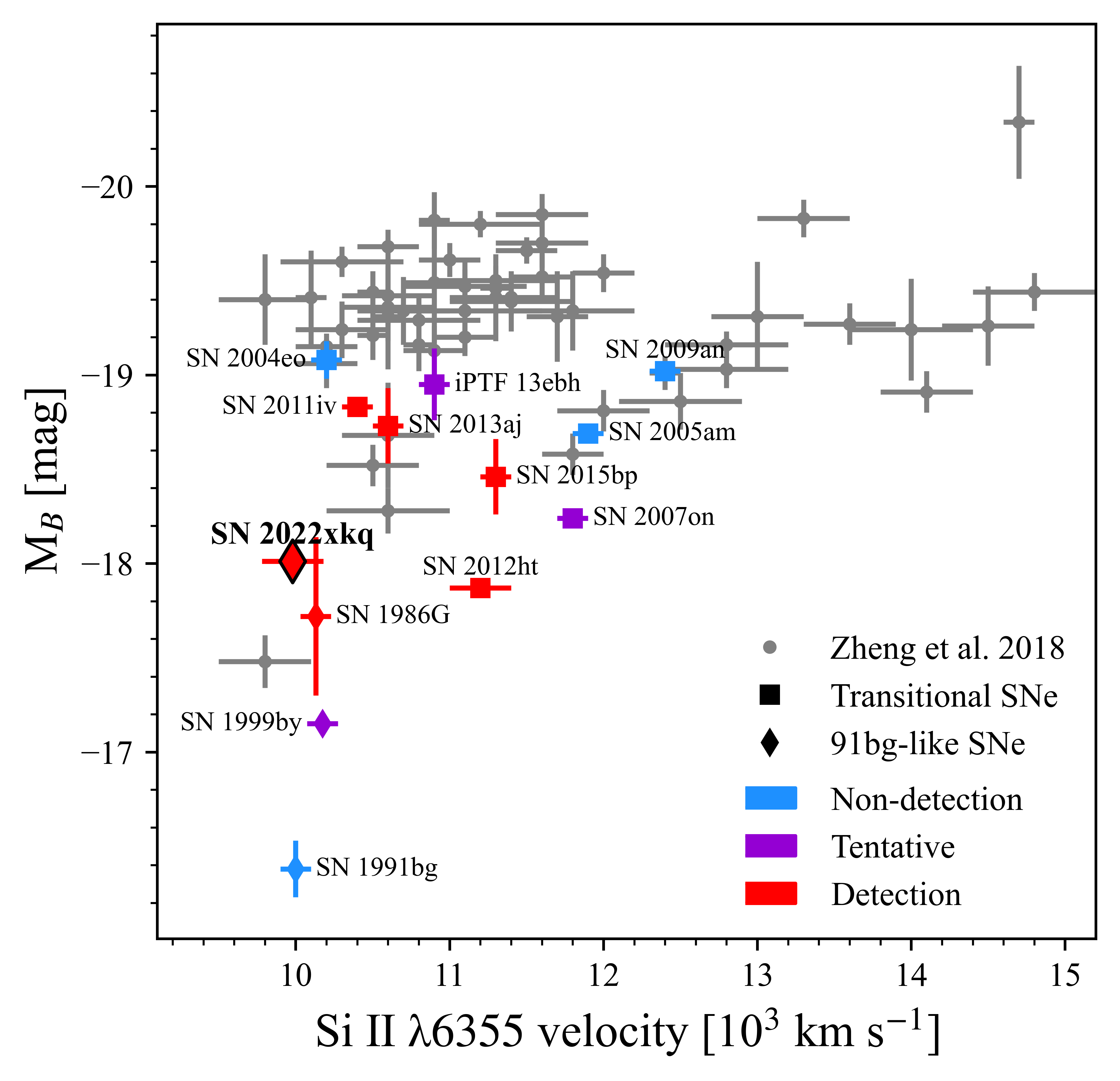}
    \caption{Plot of the $B$-band peak absolute magnitude versus the \ion{Si}{2} $\lambda$6355 velocity at $B_{\mathrm{max}}$ of SN~2022xkq compared to SNe Ia reported in \citet{Zheng2018},  the transitional SN Ia sample (squares) presented in \citet[][see their table 7]{15bp}, and a few 91bg-like SNe Ia (stars): SN~1986G, SN~1991bg and SN~1999by. Transitional and 91bg-like SNe Ia points are colored based on whether \ion{C}{2} $\lambda$6580 is observed in their early data. A tentative detection is illustrated by a ``flat" profile in the \ion{C}{2} region. \ion{C}{2} classifications are taken from \citet{15bp} for transitional objects and \citet{Ashall2016} for SN~1986G. In addition to SN~2022xkq, we classify SN~1991bg and SN~1999by.}
    \label{fig:SiIIvsBmax}
\end{figure}

The \ion{Si}{2} $\lambda$5972 and $\lambda$6355 absorption features are ubiquitous in SNe Ia. In SN~2022xkq, both of these \ion{Si}{2} lines are asymmetric starting $\sim10$ days before peak and persisting to $\sim3$ days post-peak, though \ion{Si}{2} $\lambda$6355 remains asymmetric until its disappearance. The \ion{Si}{2} lines over these epochs are not well fit with a single Gaussian. When the \ion{Si}{2} lines at $-10$ days relative to $B_\mathrm{max}$ are fit using a two-component Gaussian, we find a $13,000\pm500$ and a $10,000\pm500$ km s$^{-1}$ component. We note this is a small difference ($\sim3000$ km s$^{-1}$) and it is therefore difficult to determine if there are truly two separate features \citep{Silverman2015}. 91bg-like SNe often exhibit asymmetric \ion{Si}{2} absorption features (see Figure \ref{fig:peak}), possibly the result of a spectral-forming region located at higher velocities than for normal SNe Ia \citep{Blondin18} or of the blending of \ion{Fe}{2} and \ion{Si}{2} lines \citep{Galbany19}. 
So while the observed asymmetry could indicate the presence of two velocity components, we refrain from making a definitive statement. 
We instead measure the velocities and equivalent widths (EW) of \ion{Si}{2} lines by fitting both with a fourth degree spline. Using the location of the absorption trough, we find a \ion{Si}{2} velocity of 11,920 $\pm$ 350 km s$^{-1}$ in an early spectrum ($-11.6$~days) and of 9,980 $\pm$ 200 km s$^{-1}$ near $B_{\mathrm{max}}$ (1.2 days). Near B-band maximum (1.2 days), we measure a EW of 38 $\pm$ 1~{\AA} and 129 $\pm$ 7~{\AA} for \ion{Si}{2} $\lambda$5972 and $\lambda$6355, respectively. These results are tabulated in Table \ref{tab:results}.

SN~2022xkq's \ion{Si}{2} velocities and EWs at $B_{\mathrm{max}}$ are similar to SN~1986G and close to other 91bg-like SNe.
\citet{Polin19} introduced the use of a plot of \ion{Si}{2} $\lambda$6355 velocity and peak absolute $B$-band magnitude to identify SNe Ia which may be the result of sub-Chandrasekhar mass double detonation explosions. \citet{Burrow20} identifies three subgroups of SNe Ia in this plot: ``main", which may result from near-Chandrasekhar mass progenitors, ``fast" and ``dim", both of which may have sub-Chandrasekhar mass progenitors with thin He shells \citep{Polin19}. 
As shown in Figure \ref{fig:SiIIvsBmax}, SN~2022xkq is a dim SNe, consistent with other underluminous SNe Ia. We note that the frequency of \ion{C}{2} $\lambda$6580 detection in dim SNe has raised questions about the mass of their progenitors (see Section \ref{sec:carbon} and \ref{sec:subch}). 
The EWs of SN~2022xkq's \ion{Si}{2} lines at peak are also similar to other 91bg-like SNe, which categorizes SN~2022xkq as ``cool" in the Branch classification scheme \citep{Branch06, Burrow20}.  
Further, the \ion{Si}{2} velocities of SN~2022xkq following explosion and at peak are consistent other 91bg-like SNe \citep{Wang2009}. We note that \citet{Wang2009} excludes 91bg-like SNe from the normal-velocity (NV) and high-velocity  (HV) classification scheme, but that they roughly coincide with NV SNe.

\subsection{Infrared Spectra}\label{sec:irspec}

In addition to optical spectra we also collected an infrared spectral sequence (see Figure \ref{fig:irspec}). 
The pre-maximum light NIR evolution of SN~2022xkq is very similar to both SN~1999by and iPTF13ebh.
However, post-maximum light the NIR evolution is much closer to that of iPTF13ebh than SN~1999by, which has more distinct absorption features in the $J$-band than either SN~2022xkq or iPTF13ebh.

There is a NIR feature around 2~$\mu$m which appears, both in emission and absorption, in some of the NIR spectra. Given the lack of consistent evolution and the proximity of the nearby telluric region, we do not consider this feature real and do not mention it further.  

Prominent iron-peak emission in the $H$-band begins to appear around 5 days post $B_\mathrm{max}$. The blue-edge velocity, $\mathrm{v_{edge}}$, of this feature can indicate the outer location of the Ni distribution. 
Assuming the iron-peak emission begins at 1.57~$\mu$m \citep[value used in][]{Ashall19}, we measure $\mathrm{v_{edge}}$ of $\sim11000$ km s$^{-1}$  at 10 days post $B_\mathrm{max}$ and $\sim5500$ km s$^{-1}$  at 21 days post $B_\mathrm{max}$. This is in line with transitional SNe Ia like SN~2012ij \citep{12ij}, SN~2015bp, and iPTF13ebh \citep{Ashall19} but is significantly larger than 91bg-like SN Ia SN~1999by \citep{Ashall19}. This behavior may indicate a Ni distribution similar to that of transitional SNe Ia, though we note that SN~1999by is the only other 91bg-like SN Ia for which robust NIR observations exist. 

\subsection{Optical and Infrared Carbon Detection}\label{sec:carbon}

\begin{figure*}
    \centering
    \includegraphics[width=\hsize]{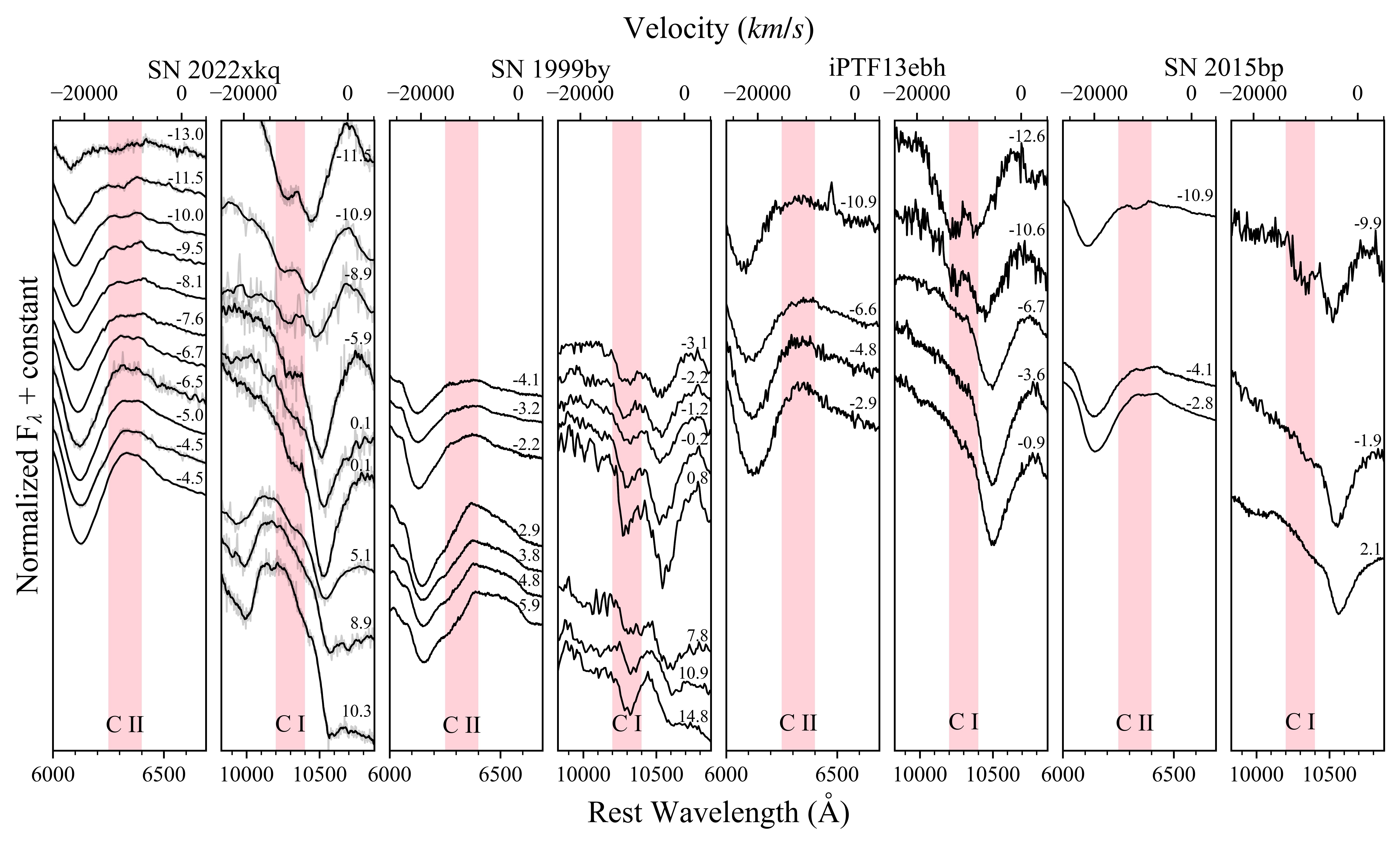}
    \caption{The optical \ion{C}{2} $\lambda$6580 and NIR \ion{C}{1} $\lambda$10693 evolution of SN~2022xkq compared to transitional/91bg-like SNe Ia: SN~1999by \citep{99by_opt, 99by_ir}, SN~2015bp \citep{15bp}, and iPTF13ebh \citep{Hsiao15}. The \ion{C}{2} evolution of SN~2022xkq is similar to that of SN~2015bp, starting with a strong \ion{C}{2} line which evolves to a ``flat" feature before disappearing near $B_\mathrm{max}$. 
    SN~2022xkq also has a strong NIR \ion{C}{1} $\lambda$10693 feature which persists until $\sim5$ days post-$B_\mathrm{max}$, which is somewhat comparable to the NIR \ion{C}{1} evolution in SN~1999by.
    } 
    \label{fig:carbon}
\end{figure*}

The presence of C lines in the early spectra can provide important constraints on the explosion mechanisms of SNe Ia. SN~2022xkq displays both optical \ion{C}{2} $\lambda$6580 and NIR \ion{C}{1} $\lambda$10693 in the pre-$B_\mathrm{max}$ spectra (see Figure \ref{fig:carbon}). We compare the evolution of the carbon features in SN~2022xkq to that of other transitional/91bg-like SNe Ia which also exhibit strong carbon features: SN~1999by, iPTF13ebh, and SN~2015bp.

The optical carbon absorption feature in SN~2022xkq is clear in the earliest spectra before weakening to a ``flat" feature by $-7$ days post $B_\mathrm{max}$ and disappearing soon after. 
This evolution, though more rapid, is similar to what was observed in SN~2015bp \citep{15bp}. In contrast, iPTF13ebh, despite its strong NIR carbon feature, has a flat \ion{C}{2} $\lambda$6580 profile in the earliest epochs which quickly fades. The optical carbon observed in SN~1999by also has a flat \ion{C}{2} pre-maximum, though this feature persists much closer to peak than observed in SN~2022xkq. 

Based on the classification scheme of \citet{Folatelli12}, the clear presence of \ion{C}{2} $\lambda$6580 in the $-13$ to $-9$ day spectra of SN~2022xkq can be classified as ``A" (i.e. absorption), for a clear \ion{C}{2} $\lambda$6580 feature in the earliest spectrum. This is also the classification assigned to SN~2015bp. In contrast, iPTF13ebh was classified as ``F", named for the flat profile on the red side of \ion{Si}{2} $\lambda$6355 which indicates a marginal detection of \ion{C}{2} $\lambda$6580. SN~1999by was also classified as F, though its earliest optical spectrum is from a significantly later epoch ($-$3 days relative to the time of $B_\mathrm{max}$) than SN~2022xkq or iPTF13ebh which prevents the detection of an earlier, and perhaps stronger, carbon feature. No \ion{C}{2} $\lambda$6580 (classified as ``N") was observed in SN~1991bg, though the lack of early spectra likely precludes its detection. Most transitional/91bg-like SNe with early optical spectra show evidence for the presence of \ion{C}{2} $\lambda$6580 in some capacity (see Figure \ref{fig:SiIIvsBmax}), including SN~2022xkq. We note that this feature has been attributed to \ion{Fe}{2} $\lambda$6516 in some of the 91bg-like SNe \citep{Galbany19}, however we do not observe any strong \ion{Fe}{2} lines (i.e. $\lambda$$\lambda$6456,6516,7308,7462,7711) within this wavelength region in SN~2022xkq.

The earliest NIR spectrum of SN~2022xkq was taken only 3 days after the initial detection of the SN. 
This spectrum, and several subsequent spectra, exhibits a feature at 1.03~$\mu$m on the blue side of the \ion{Mg}{2} 1.09~$\mu$m multiplet. This feature is analogous to those observed in the early NIR spectra of several other transitional/91bg-like SNe (e.g., SN~1999by, iPTF13ebh and SN~2015bp) where it was identified as \ion{C}{1} $\lambda$10693. Further, as shown in Figure \ref{fig:carbon}, this feature has a similar velocity evolution to \ion{C}{2} $\lambda$6580. We therefore identify this line as \ion{C}{1} $\lambda$10693. We note that this feature has also been  suggested to be \ion{He}{1} $\lambda$10830 \citep{Boyle2017, Collins23} and we explore this possibility in Section \ref{sec:helium}.

The line we identify as \ion{C}{1} $\lambda$10693 persists past $B_\mathrm{max}$ and is still present at $+5$ days with respect to $B$-band maximum (see Figure \ref{fig:carbon}). In the earliest spectra, SN~2022xkq has a \ion{C}{1} $\lambda$10693 feature similar in shape to that of iPTF13ebh \citep{Hsiao15} which displays a strong feature in the $-13$ and $-11$ day spectra but the feature has weakened significantly by $-7$ days. The persistence of the \ion{C}{1} $\lambda$10693 feature in SN~2022xkq is most analogous to that of SN~1999by which displays a strong \ion{C}{1} $\lambda$10693 feature post-$B_\mathrm{max}$ as well. However, the \ion{C}{1} $\lambda$10693 feature in SN~1999by is still present at 15 days post-$B_\mathrm{max}$. \citet{99by_ir} suggested that there may have also been a strong \ion{C}{1} $\lambda$9406 feature in SN~1999by given the strength of the \ion{C}{1} $\lambda$10693 line. However, we do not detect \ion{C}{1} $\lambda$9406 in the spectra of SN~2022xkq. The \ion{C}{1} $\lambda$10693 in SN~1999by persists for much longer than observed in SN~2022xkq, where the feature has mostly disappeared after 5 days post-$B_\mathrm{max}$. We note that in the 9 and 10 day NIR spectra of SN~2022xkq there is a shoulder in the \ion{Mg}{2} $\lambda$10927 feature, this shoulder has been linked to fading \ion{C}{1} $\lambda$10693 \citep{Hsiao15} so the NIR carbon may be present even past 5 days after the time of $B_\mathrm{max}$. 

SN~2022xkq exhibits more persistent carbon lines, both in the optical and NIR, than those observed in either iPTF13ebh or SN~2015bp. Of the three transitional/91bg-like SNe Ia with clear NIR carbon features to which we compare, perhaps only SN~1999by displays evidence of stronger carbon than observed in SN~2022xkq. With the addition of SN~2022xkq, the sample of transitional/91bg-like SNe Ia with carbon features indicates that these lines may be commonplace in underluminous SNe. 
This is perhaps not a surprising result as several authors \citep{Thomas11, Maguire14} have noted a correlation between narrow light curves, denoted by low s$_{BV}$ values, and the presence of carbon. 
The detection of carbon in SN~2022xkq may have significant implications for the explosion mechanism of the SN, this is further discussed in Section \ref{sec:expmech}.

\subsection{H and He Constraints from Nebular Spectroscopy}\label{sec:neb}

To place further constraints on the progenitor of SN~2022xkq we use the late-time, low-resolution (R$\approx$650) high signal-to-noise spectrum taken using the LDSS-3 spectrograph at the Magellan Clay telescope +118 days post-$B_\mathrm{max}$. 
If the progenitor system of SN~2022xkq has a nondegenerate companion, models indicate the SN ejecta will ablate part of its envelope, resulting in narrow hydrogen or helium emission lines $\gtrsim100$ days after explosion \citep[most recently][]{Boehner17, Botyanszki18, Dessart20}. Models predict $\gtrsim0.1$ M$_{\odot}$ of stripped hydrogen \citep{Botyanszki18, Dessart20}, however there is diversity in the predicted shape and strength of the H$\alpha$ emission line based on the details of the SN explosion, the companion, and treatment of non-LTE effects. 

There are no hydrogen or helium emission features visible in the LDSS-3 spectrum (see Figure \ref{fig:optspec2}). Using the methodology of \citet{Sand18, Sand19,Sand21}, we set limits on narrow H$\alpha$ and \ion{He}{1} ($\lambda5875$, $\lambda6678$) emission. We note the NIR spectrum taken at +103d is too low signal-to-noise to perform a meaningful analysis on the \ion{He}{1} $\lambda10830$ line. The extinction-corrected and flux-calibrated spectrum is binned to the native resolution and smoothed on scales larger than the expected emission (FWHM $\approx1000$ km s$^{-1}$) using a second-order \citet{Savitzky64} filter with a width of 132~\AA. Hydrogen and helium emission features with widths similar to those expected from a single-degenerate scenario would be apparent when comparing the unsmoothed and smoothed versions of the spectrum.  

We insert emission lines with width 1000 km s$^{-1}$ and a peak flux five times the root-mean-square of the residual spectrum into our data in order to set quantitative limits on the H$\alpha$ and \ion{He}{1} nondetections. As shown in Table \ref{tab:results}, the resulting flux and luminosity limits (assuming D=31 Mpc) are: 1) H$\alpha$ of 7.6$\times$10$^{-17}$ erg s$^{-1}$ cm$^{-2}$ and 8.8$\times$10$^{36}$ erg s$^{-1}$; 2) \ion{He}{1} $\lambda5875$ of 3.4$\times$10$^{-16}$ erg s$^{-1}$ cm$^{-2}$  and 4.0$\times$10$^{37}$  erg s$^{-1}$; and 3) \ion{He}{1} $\lambda6678$ of 7.6$\times$10$^{-17}$ erg s$^{-1}$ cm$^{-2}$ and 8.8$\times$10$^{36}$ erg s$^{-1}$. For reference, the H$\alpha$ luminosity limit for SN~2022xkq is between $\sim$4--10 times fainter than the H$\alpha$ detections seen in three recent low luminosity, fast declining SNe Ia with similar photometric properties to SN~2022xkq \citep{Kollmeier19, Vallely:2019, Prieto20, Elias21}, indicating that we would detect such features if they were present.  

First we use the radiation transport model sets from \citet{Botyanszki18} to translate the hydrogen and helium luminosity limits to masses. The 3D radiation transport results of \citet{Botyanszki18} present simulated SN Ia spectra 200 days after explosion based off the ejecta-companion interaction simulations of \citet{Boehner17}, see \citet{Botyanszki18} for further details. The simulated spectra presented in \citet{Botyanszki18} have $L_{\mathrm{H}\alpha} \approx (4.5-15.7)\times 10^{39}$ erg s$^{-1}$ ($M_\mathrm{strip} \sim$ 0.2-0.4 $M_{\odot}$) for the main sequence, subgiant, and red giant companion star models. This is roughly 500--1800 times brighter than the detection limit for SN~2022xkq. In order to quantify the stripped mass limit for SN~2022xkq, we adopt the quadratic fitting relation between H$\alpha$ (and \ion{He}{1} $\lambda5875$ and $\lambda6678$) luminosity and stripped hydrogen(/helium) mass (Equation 1 in \citealt{Botyanszki18} as updated by \citealt{Sand18}) and correct for the epoch of our observations (118 days after $B_\mathrm{max}$ or 133 days after explosion; see \citealt{Botyanszki18}). This yields a stripped hydrogen mass limit of $M_{\mathrm{H, strip}} \lesssim 2\times$10$^{-4} \mathrm{M}_{\odot}$ and a helium mass limit of $M_{\mathrm{He, strip}} \lesssim 2\times$10$^{-3} \mathrm{M}_{\odot}$ and $M_{\mathrm{He, strip}} \lesssim 7\times$10$^{-4} \mathrm{M}_{\odot}$ for the \ion{He}{1} $\lambda5875$ and $\lambda6678$ lines, respectively.

We also constrain the amount of stripped hydrogen based on 1D non-LTE radiative transfer calculations 
of several delayed-detonation models \citep[DDC0, DDC15, and DDC25]{Blondin13} presented in \citet{Dessart20}. We interpolate the available model grid, 100--300 days post-explosion, to the explosion epoch of SN~2022xkq. This gives a stripped hydrogen mass of M$_{\mathrm{H, strip}} \lesssim 7\times$10$^{-5}$ $\mathrm{M}_{\odot}$ 
with very little variation between the models. \citet{Dessart20} does not present a luminosity to mass conversion for helium so we do not derive a stripped helium mass for this model. 

Limits on hydrogen and helium luminosity, and stripped mass, are listed in Table \ref{tab:results}. 
Other narrow emission lines, such as [\ion{O}{1}] $\lambda6300$ and \ion{Ca}{2} $\lambda7291$ and $\lambda7324$, may be sensitive to helium star companions in particular \citep[e.g.,][]{Lundqvist13}. The LDSS-3 spectrum of SN~2022xkq constrains the luminosity of [\ion{O}{1}] $\lambda6300$ to be L$_{\mathrm{[O~I]}} < 8.8\times10^{36}$~erg s$^{-1}$. Further collection of nebular spectra of SN~2022xkq is needed to search for signatures of a single-degenerate scenario at later times. This is especially important given the detection of H$\alpha$ in several similar low-luminosity SNe Ia \citep{Kollmeier19, Vallely:2019, Prieto20, Elias21}.

\section{Radio Nondetection}\label{sec:radio}

\begin{figure}
    \centering 
    \includegraphics[width=\hsize]{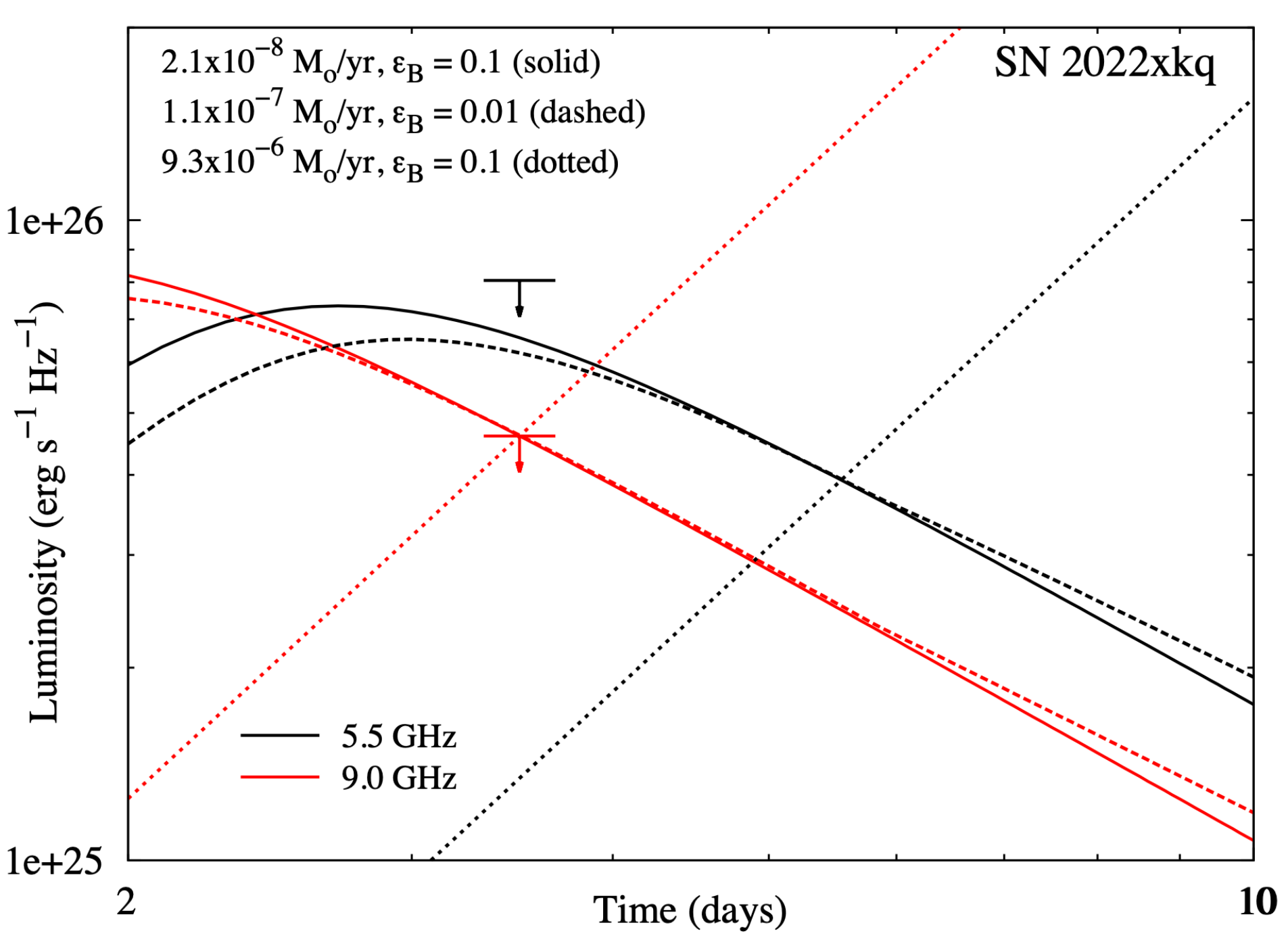}
    \caption{
    Radio data for SN~2022xkq \citep{ryder22} together
    with models at 5.5 GHz and 9.0 GHz for a $\rho(r) \propto r^{-2}$ wind. Models use 
    $\dot{M} = 2.1~(11.0) \EE{-8} \msunyr~(v_w/100 \kms)$
    for $\epsilon_{\rm B} = 0.1~(0.01)$, with solid (dashed) lines being for
    $\epsilon_{\rm B} = 0.1~(0.01)$. Common parameters in both models
    are $\epsilon_{\rm e} = 0.1$, $T_{\rm bright} = 5\EE{10}~{\rm K}$,
    $n=13$ and $v_w = 100~\kms$. The most constraining observation is that at 9.0 GHz on 
    day 3.5. As there is only one epoch of observations, there is also an upper limit
    on the derived mass-loss rate shown here for $\epsilon_{\rm B} = 0.1$ as dotted lines.
    See text for more details.
    }
    \label{fig:22xkq_fig}
\end{figure}

We have modeled any possible radio emission from SN 2022xkq in the same way as in \citet{plund20} and \citet{Hosseinzadeh22}---that is, 
we assume that any emission would arise as a result of circumstellar interaction. Behind the SN blast-wave, electrons are accelerated to 
relativistic speeds, magnetic fields are generated, and the relativistic electrons thereby radiate synchrotron emission. For a power-law distribution
of the electron energies, $dN/dE = N_0E^{-p}$, where $E=\gamma m_ec^2$ is the energy of the electrons and $\gamma$ is the Lorentz factor,
the intensity of optically thin synchrotron emission $\propto \nu^{-\alpha}$, where $\alpha = (p-1)/2$. We have adopted $p=3$ \citep{che06}. 
The structure of the circumstellar medium depends on the progenitor system. For a double-degenerate progenitor system, a constant density
medium may be expected (as might be expected for relatively undisturbed interstellar medium), whereas for the single-degenerate scenario, we might expect the circumstellar environment to be shaped by the companion star and binary processes. In this case, we assume wind-like circumstellar medium described by a constant mass-loss rate ($\dot M$), and a constant wind 
speed ($v_w$), implying a density decrease with radius as $\rho(r) = \dot M/(4\pi r^2 v_w)$. We assume a constant function of post-shock energy density from amplified magnetic fields, defined as $\epsilon_{\rm B} = u_{\rm B} / u_{\rm th}$, where $u_{\rm th}$ is the post-shock thermal energy density ($\sim \rho v_s^2$ where $v_s$ is the velocity of the shock), and  $u_{\rm B}$ is the magnetic field energy density.
We use the model N100 \citep{rop12,sei13} for the SN ejecta to test the single-degenerate scenario. To the fastest ejecta in the model ($\sim 2.8\times10^4 \kms$), we add even faster ejecta with a density profile $\rho_{ej} \propto r^{-n}$ with $n = 13$ \citep[see][for a discussion on $n$]{kun17}. 
In both models, we also assume a constant function of post-shock energy density from relativistic electrons, $\epsilon_{\rm rel} = 0.1$, where $\epsilon_{\rm rel} = u_{\rm rel} / u_{\rm th}$, and $u_{\rm rel}$ 
is the energy density in relativistic electrons. 
For the brightness temperature of the radio emission at the frequency at which the optical depth of synchrotron self-absorption is unity we adopt  $T_{\rm bright} = 5\EE{10}~{\rm K}$ \citep[see][for a discussion on $T_{\rm bright}$]{bjo14}. 
 
In Figure~\ref{fig:22xkq_fig} we show the predicted radio emission from our wind model. We have assumed that the explosion occurs 3.5 days prior to the epoch of radio data, i.e., one day before detection in the optical. The ratio of $\dot M/v_w$ has been tuned to give the highest luminosity possible that does not conflict with the observational radio limits. Solid (dashed) lines are for $\epsilon_{\rm B} = 0.1~(0.01)$. A higher wind density is needed to compensate for less efficient conversion to magnetic field 
energy density, and we obtain $\dot{M} \lesssim 2.1~(11.0) \EE{-8}\, \msunyr~(v_w/100 \kms)$ for $\epsilon_{\rm B} = 0.1~(0.01)$. The observed
upper limit at 9.0 GHz constrains the models the most, and for the $\epsilon_{\rm B} = 0.1~(0.01)$ model the circumstellar shock is 
at $\approx 2.4~(1.9)\EE{15}$~cm 3.5 days after explosion. The figure also shows a model with $\dot{M} = 9.3\EE{-6}\, \msunyr~(v_w/100 \kms)$ and 
 $\epsilon_{\rm rel} = 0.1$ (dotted lines). For higher wind densities than this, synchrotron self-absorption becomes so large that the modeled emission
is predicted to fall below the observed luminosity at 9.0 GHz. So in essence, our models can only rule out the interval  
$\dot{M} = (2.1\EE{-8} - 9.3\EE{-6}) \msunyr~(v_w/100 \kms)$ for $\epsilon_{\rm B} = 0.1$. However, since multi-epoch radio observations 
have limited $\dot{M}$ to a few $\EE{-8} \msunyr~(v_w/100 \kms)$ for many SNe~Ia \citep{Chomiuk16,plund20}, with the notable exception of SN Ia-CSM \citep{Kool23}, the $\dot{M}$ value highlighted by 
the dotted lines in Figure~\ref{fig:22xkq_fig} should only be considered as a formal limit.
 
To test the double-degenerate scenario we have used a constant density for the ambient medium. In this case the modeled 
radio flux increases with time \citep[e.g.,][]{Chomiuk12, Chomiuk16, kun17,plund20}. The progenitor model used is the violent merger model 
by \citet{pak12}, simulating the merger of two C/O degenerate stars with masses $1.1~\msun$ and $0.9~\msun$. The total mass and 
asymptotic kinetic energy of the ejecta for this model are $1.95~\msun$, and $1.7\EE{51}$ erg, respectively. We assume the same $n = 13$ density profile for this model as for the single degenerate model. Again, the 9.0 GHz data
are the most constraining, and we obtain an upper limit on the density of the ambient medium of $1450~(9800) \cm3$ for $\epsilon_{\rm B} = 0.1~(0.01)$.
For the $\epsilon_{\rm B} = 0.1~(0.01)$ model the circumstellar shock is at $\approx 2.6~(2.1)\EE{15}$~cm 3.5 days after explosion.

In summary, the radio data can be used to put a limit on the density of the ambient medium outside $\sim 2\EE{15}$ cm. In a single-degenerate 
scenario, the mass loss rate of the progenitor system is $\lesssim  2.1\EE{-8} \msunyr~(v_w/100 \kms)$ for $\epsilon_{\rm B} = 0.1$.  When we compare these limits to the mass-loss rate parameters of single-degenerate models defined by \citet{Chomiuk12}, we find that these limits are deep enough to rule out most symbiotic systems (red-giant companions). These systems are characterized by slow winds ($10-100$ km s$^{-1}$) and mass-loss rates of $10^{-6}-10^{-8}$ \msunyr \citep{Seaquist90}. Radio upper limits have ruled out red-giant companions for the majority of SNe Ia \citep{Chomiuk12, Horesh12, Perez-Torres14, Chomiuk16, Lundqvist20, Pellegrino20, Burke21, Hosseinzadeh22, 23bee}. However, many models involving main-sequence companions still lie within the radio limits for SN~2022xkq. 
For the 
double-degenerate scenario, or the single-degenerate scenario with spun-up/spun-down super-Chandrasekhar mass white 
dwarfs \citep{dis11,jus11}, where mass transfer no longer occurs at the time of explosion, a near uniform density of the ambient medium is 
expected. However, the early radio data here are not ideal to test this since the likely number density is of order 1\,cm$^{-3}$, and our upper limit on the 
density is 1450\,cm$^{-3}$ for $\epsilon_{\rm B} = 0.1$. The limits on mass-loss rates and ambient medium density is a factor of $\approx 5-7$ larger for 
$\epsilon_{\rm B} = 0.01$. Given that the radio flux is expected to increase with time in the case of interaction with uniform medium \citep{Chomiuk16}, we encourage further monitoring of this SN in order to obtain more stringent limits.

\section{Discussion}\label{sec:discuss}

\subsection{Origin of the Red Excess}\label{sec:red}
Several SNe Ia exhibit early flux excesses in the UV and bluer optical bands \citep{Hosseinzadeh17, Miller20, Burke21, Hosseinzadeh22, Srivastav23, 23bee, Wang23}. Population studies of SNe Ia have suggested that these ``blue bumps" may be a common occurrence \citep{Magee22, Deckers22, Burke22_dlt40, Burke22_ztf}.  
However, red flux excesses are much less common. Only a handful of SNe Ia have been suggested to have an early time ``red excess", notably MUSSES1604D \citep{Jiang17}, SN~2018aoz \citep{Ni22}, and the more peculiar SN~2018byg \citep{De19} and SN~2022joj \citep{Gonzalez23, Liu23}. 
\citet{Ni23} suggest that SN~2021aefx may also have had a red flux excess, although it was not characterized this way by other authors \citep{Ashall22, Hosseinzadeh22}.

\citet{Jiang17} attribute the red excess in MUSSES1604D to double detonation, where the He-shell detonation produces radioactive isotopes in the outer ejecta layers, creating an early flux excess. The He-shell detonation also creates iron-group and intermediate mass elements, like Ti and Ca, which primarily absorb in the blue optical bands, thus blocking the blue flux and producing an overall red $B-V$ color. The early bump in the light curve of MUSSES1604D was thus attributed to a flash produced by decaying radioactive isotopes. \citet{Ni23_18aoz} give three possible explanations for the red flux excess in SN~2018aoz: 1) surface radioactive isotopes; 2) interaction with a binary companion; and 3) interaction with surrounding circumstellar material. They note that, similar to MUSSES1604D, some Fe-group line blanketing is required to produce the observed color evolution, therefore any scenario will likely include the production of radioactive materials in the outer ejecta. 
Similarly, the red excesses in SN~2018byg and SN~2022joj have also been linked to double-detonation and iron-group element absorption in the blue. However, double-detonation models can not reproduce the observed \ion{C}{2} features in SN~2022joj \citep{Gonzalez23, Liu23}.
For SN~2021aefx, double detonation has been used to explain the UV suppression and the early redder color \citep{Ni23}. However, \citet{Hosseinzadeh22} note that there are observational features of SN~2021aefx that do not support a double detonation scenario, including the presence of carbon lines at pre-maximum epochs and the lack of strong [\ion{Ca}{2}] emission in the nebular spectra. 

Other origins for an early red color in SNe Ia have also been suggested. ZTF18aayjvve and ZTF18abdfwur were identified as possibly having red excesses, but neither were well explained by a double-detonation scenario and were better explained as the result of the presence of $^{56}$Ni in the outer layers or companion/CSM interaction \citep{Deckers22}.  

As we discuss further below, we do not favor a double detonation mechanism for SN~2022xkq, but we cannot rule it out. A wide variety of explosion parameters can significantly change the color evolution of double detonation explosions including: the exact mass of the white dwarf progenitor and the He-shell \citep{Shen10, Kromer10, Sim2012, Polin19}; the degree of mixing \citep{Polin19, Gronow20}; non-LTE effects \citep{Shen21_LTE, Dong22}; and viewing angle effects \citep{Kromer10, Sim2012, Gronow20, Shen21}. Viewing angle effects in particular have been shown to have a significant impact on the color, light curve morphology, and spectral evolution of double detonation SNe \citep{Sim2012, Gronow20, Shen21}.  
We also cannot discount impacts from other processes which might result in bluer early colors in more luminous SNe Ia, like companion and CSM interaction. The strong Ti absorption features that denote 91bg-like SNe dominate in $U$- and $B$-bands and may suppress any blue bumps, resulting in a red early color. 
Further modeling is needed to understand the origin of the early red color excess in SN~2022xkq. 

\subsection{Further Explosion Constraints}\label{sec:expmech}

The progenitors and explosion mechanisms of transitional/91bg-like SNe Ia are the subject of some debate.  
Studies have long suggested double detonation in sub-Chandrasekhar mass white dwarfs as the origin for faint SNe Ia \citep{Goldstein18, Polin19, Shen21, Collins23}. 
However, the analyses of a few transitional and 91bg-like SNe suggest that many of these objects are best described by the delayed-detonation of near-Chandrasekhar mass white dwarfs \citep{99by_ir, Ashall2016, Ashall2018}. 
One of the differentiators between these two scenarios is the amount of unburnt carbon in the ejecta. Here we discuss the presence of carbon in sub-Chandrasekhar mass models, both pure and double detonation, and compare the spectra of SN~2022xkq with models of a Chandrasekhar mass delay-detonation and pure sub-Chandrasekhar mass detonation to try to shed further light on the origins of this SN.
Finally, it has been suggested that the double detonation scenario may produce a \ion{He}{1} line which could be misidentified as \ion{C}{1} \citep{Boyle2017, Collins23}; we discuss this possibility in Section \ref{sec:helium}.

\subsubsection{Carbon in sub-Chandrasekhar Mass Models}\label{sec:subch}

The carbon lines observed in SN~2022xkq, and other transitional/91bg-like SNe Ia, can be difficult to reconcile with sub-Chandrasekhar mass progenitors. 
For underluminous SNe Ia both pure and double detonation sub-Chandrasekhar mass models have higher carbon burning efficiencies, resulting in less unburnt carbon than Chandrasekhar mass models with similar $^{56}$Ni masses and therefore weaker or nonexistent carbon features. However, we note that the actual amount of unburnt carbon is model dependent. 

In the double detonation sub-Chandrasekhar models presented by \citet{Polin19}, the amount of unburnt carbon left following detonation would not produce optical carbon features in the spectra of a SN Ia. However, the sub-Chandrasekhar mass models presented in \citet{Blondin17} have NIR carbon in the spectra pre-maximum light. We note that the \citet{Blondin17} models are of pure central detonation of a bare CO white dwarf and do not account for any He-shell effects which could potentially result in less unburnt carbon. Neither \citet{Polin19} nor \citet{Blondin17} predict the presence of optical carbon in the spectra of a SN Ia produced by a sub-Chandrasekhar mass progenitor. The \citet{Blondin17} models produce a feature on the red wing of \ion{Si}{2} $\lambda$6355, though this feature is due to the \ion{Mg}{2} $\lambda$6347 doublet rather than \ion{C}{2} $\lambda$6580. Given that the amount of unburnt carbon following the detonation (whether pure or double) of a sub-Chandrasekhar white dwarf is dependent on the physical parameters of the model, we caution against using the presence of carbon as the sole indicator of the progenitor.

\subsubsection{Sub-Chandrasekhar vs Delayed-Detonation}

\begin{figure*}
    \centering
    \includegraphics[width=\hsize]{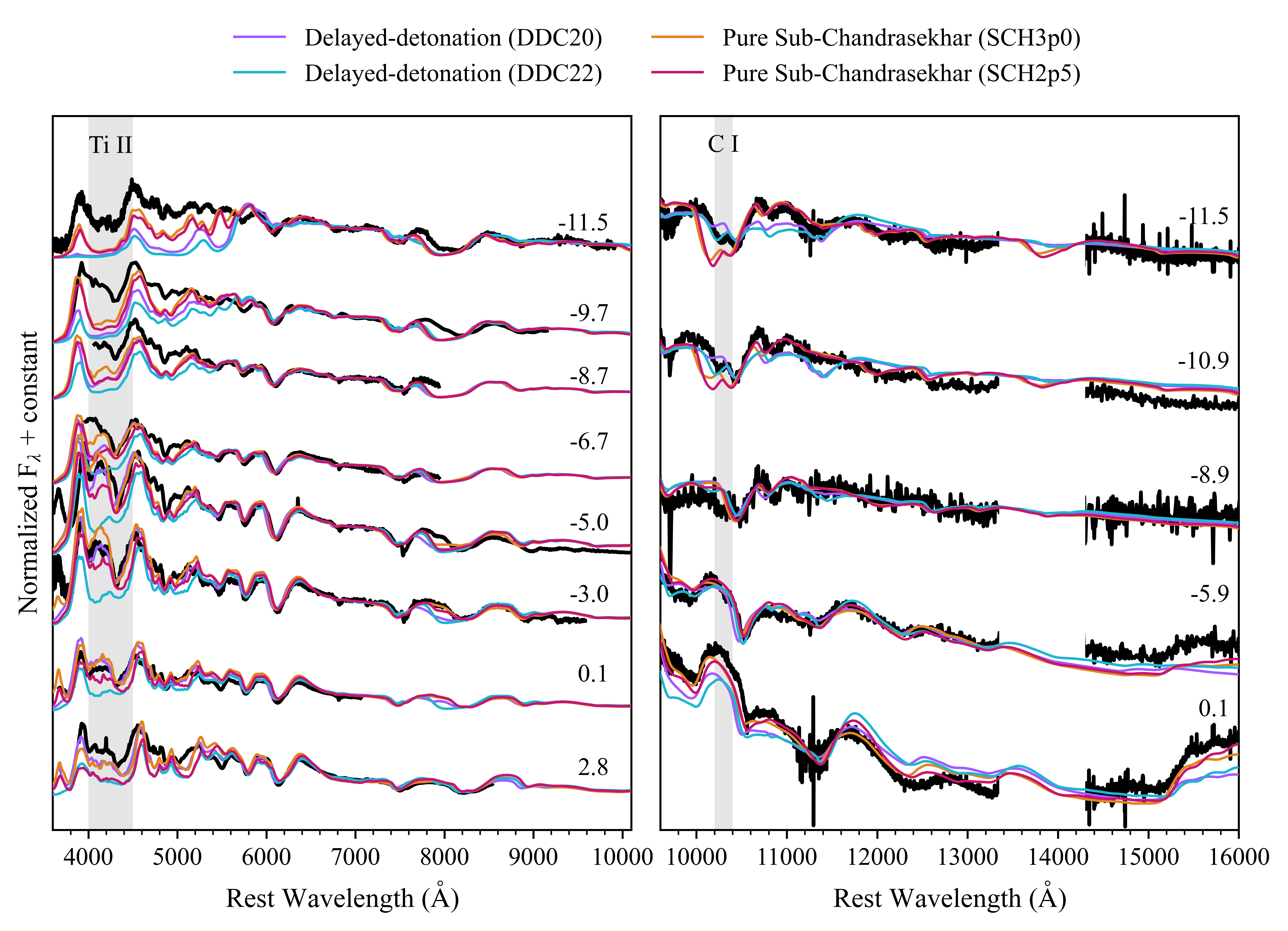}
    \caption{Optical and NIR spectra of SN~2022xkq compared to models of delayed-detonation and pure detonation sub-Chandrasekhar mass explosions from \citet{Blondin17}. Optical models are scaled to the weighted average flux from 6000-7500\,\AA\ of the spectrum to which they are compared and NIR models are scaled to the wavelength weighted average flux from 10000-13000\,\AA. The spectral evolution of SN~2022xkq is closer to the models of the pure detonation of a sub-Chandrasekhar mass white dwarf. Phases are relative to the time of $B_{\mathrm{max}}$. The input models and synthetic spectra shown in this figure are publicly available on Zenodo \citep{zenodo}.} 
    \label{fig:Blondin_models} 
\end{figure*}

\citet{Blondin17} modeled the explosion of a transitional SN Ia using two types of explosion models: Chandrasekhar mass delayed-detonation models with nickel masses of 0.3 and 0.21 M$_\odot$, respectively denoted as DDC20 and DDC22; and models of the pure central detonation of a sub-Chandrasekhar 0.95 and 0.93 M$_{\odot}$ white dwarf ($^{56}$Ni masses of 0.23 and 0.17 M$_{\odot}$) labeled as SCH3p0 and SCH2p5, respectively.
We compare these four models to the spectra of SN~2022xkq in Figure \ref{fig:Blondin_models}. 
For the optical spectra (left panel), we scale all models to the wavelength weighted average flux from 6000-7500 {\AA} of the comparison spectra of the same epoch, this region was chosen because it is primarily continuum absorption. In the NIR (right panel), models were similarly scaled to the spectral region from 10000-13000 {\AA}, a region with little influence from telluric lines.

We find that immediately following explosion ($<-7$ days from time of $B_{\mathrm{max}}$) the pure sub-Chandrasekhar mass models are a somewhat better reproduction of the shape of the optical spectra of SN~2022xkq compared to the delayed-detonation models; this is most evident on the blue end of the optical spectra (see Figure \ref{fig:Blondin_models} left). We note however that the pure sub-Chandrasekhar mass models are still too red to map the 4000-6000 {\AA} region at early times.
A week before $B_{\mathrm{max}}$ and beyond, of the sub-Chandrasekhar mass models SCH3p0 is a slightly better representation, though at peak SCH2p5 matches the optical data remarkably well. Despite being not blue enough at early times, the DDC20 delayed-detonation explosion also matches the observations near and after peak light. 

In the infrared, the pure detonation sub-Chandrasekhar mass models have features with higher velocities than observed in SN~2022xkq at early times. At peak light, the sub-Chandrasekhar mass models are a slightly better match to the data than the delayed-detonation models. However, in the early spectra the delayed-detonation models reproduce the NIR \ion{C}{1} feature better than the pure detonation sub-Chandrasekhar mass models. The pure sub-Chandrasekhar mass models have more \ion{C}{1} $\lambda$10693 than the delayed-detonation models immediately following explosion since the line-forming region is located at higher velocities where the carbon abundance is significantly higher. Conversely, neither the sub-Chandrasekhar mass nor the delayed-detonation Chandrasekhar mass models display \ion{C}{2} $\lambda$6580 since its line formation region is located in the ejecta layers where the carbon abundance is too low. 
All of the models reproduce the evolution of the NIR \ion{C}{1} line in the spectra, strongest immediately following explosion and then eventually weakening and disappearing. However, none of the models to which we compare have carbon post-maximum light.

Given the less-pronounced \ion{Sc}{2}/\ion{Ti}{2} absorption trough around 4000--4200\,\AA\ and the decent match to the optical and infrared spectral evolution near maximum light, the observations of SN~2022xkq are closer to the pure sub-Chandrasekhar mass white dwarf model. However, we note that pure central detonation is only an approximation and the detonation mechanism will have significant impacts on the spectral evolution. Further,
modelling uncertainties in nucleosynthesis and burning front propagation can have significant impacts on carbon abundance in delayed-detonation models. Other delayed-detonation models have shown considerable promise in describing the general spectral evolution, including the \ion{C}{2} $\lambda$6580 and NIR \ion{C}{1} features, of other SNe similar to SN~2022xkq, including SN~1986G \citep{99by_ir, Ashall2016, Ashall2018}.
Ultimately, neither of the models to which we compare was designed for SN~2022xkq, which has an inferred $^{56}$Ni mass which falls between these models, and dedicated work may result in models which more closely follow the evolution of the SN for either scenario.

\subsubsection{Helium in Double Detonation Models}\label{sec:helium}

\begin{figure}
    \centering
    \includegraphics[width=\hsize]{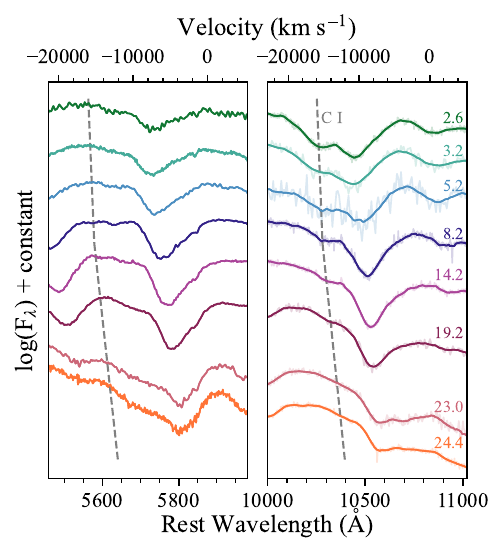}
    \caption{The evolution of the region where the \ion{He}{1} $\lambda$5875 and $\lambda$10830 lines are expected to appear in the infrared spectra. Labeled epochs are the time since explosion as determined by a power-law fit. The feature we identify as \ion{C}{1} $\lambda$10693 is marked in the right panel with the dashed line. If this feature was \ion{He}{1} $\lambda$10830 rather than \ion{C}{1} there should be a corresponding \ion{He}{1} $\lambda$5875 feature in the earliest spectra (marked by the dashed line in the left panel). There is no notable absorption feature from \ion{He}{1} $\lambda$5875, indicating that the feature on the blue wing of \ion{Mg}{2} is likely \ion{C}{1} $\lambda$10693 rather than \ion{He}{1} $\lambda$10830.} 
    \label{fig:HeI}
\end{figure}

In double detonation models, the surface detonation of helium can result in observable trace amounts of unburnt helium in the outer ejecta of the SN. These helium features are difficult to model since their optical depth depends on the treatment of non-LTE effects, the evolution of the velocity gradient, the ejecta density, variations in radiation temperature, the viewing angle, and so on \citep{Kromer10, Boyle2017, Shen21, Collins23}. 
\citet{Boyle2017} present the expected evolution of NIR \ion{He}{1} $\lambda$1.0830 and 2.0581~$\mu$m in both ``high-mass" (1.025 M$_{\odot}$ CO core mass) and ``low-mass" (0.58 M$_{\odot}$ CO core mass) models. The high-mass model corresponds to normal-luminosity SNe Ia \citep{Fink2010}. The low-mass model is designed to describe underluminous, peculiar thermonuclear events \citep{Sim2012}. SN~2022xkq is an intermediate case, though closer in luminosity to the low-mass model. 
The helium absorption lines are much stronger in the low-mass model where more helium remains in the ejecta. However, both low- and high-mass models have prominent \ion{He}{1} 1.0830~$\mu$m features which somewhat resemble the absorption features we identify as \ion{C}{1} 1.0693~$\mu$m in Figure \ref{fig:carbon}. 

In both the high- and low-mass models, helium is visible at maximum light and grows stronger past maximum \citep{Boyle2017}. In contrast, the observed feature in SN~2022xkq gets weaker post-maximum rather than stronger.
However, \citet{Collins23} recently presented a sub-Chandrasekhar mass double detonation model in which helium is strongest immediately following explosion and fades near peak luminosity. This model plausibly reproduces the \ion{C}{1} feature in iPTF13ebh with just \ion{He}{1} 1.0830~$\mu$m. Therefore, the evolution of the observed feature in SN~2022xkq could be explained by either helium or carbon. 

The model presented in \citet{Collins23} does reproduce the flat profile of \ion{C}{2} $\lambda$6580 observed in iPTF13ebh so optical carbon features alone cannot be used to confirm the presence of NIR \ion{C}{1}. Instead, we use other \ion{He}{1} features to determine if the \ion{C}{1} 1.0693~$\mu$m feature is actually due to \ion{He}{1} 1.0830~$\mu$m. We cannot properly assess if there is a \ion{He}{1} 2.0581~$\mu$m absorption feature given its proximity to the nearby telluric region. The next strongest \ion{He}{1} line in the \citet{Collins23} models is \ion{He}{1} $\lambda$5875. This line is blended with other lines in the region and is completely gone by $\sim$9 days after explosion but is detectable in earlier spectra. The earliest infrared spectrum of SN~2022xkq was taken just 2.6 days after explosion. This is well within the period in which we would expect to observe \ion{He}{1} $\lambda$5875 given the strength of the potential \ion{He}{1} 1.0830~$\mu$m line. As shown in Figure \ref{fig:HeI}, there is no clear detection of \ion{He}{1} $\lambda$5875 at the expected \ion{He}{1} velocity. Therefore, there is no evidence that the NIR feature at 1.03~$\mu$m is \ion{He}{1} 1.0830~$\mu$m.
\\
\\
Based on current models, we are unable to suggest either the delayed-detonation or the double detonation scenario as the explosion mechanism of SN~2022xkq at this time. More tailored modeling would be needed to determine the origin of this SN.

\section{Summary and Conclusions}\label{sec:conclude}

SN~2022xkq is one of the best observed underluminous SNe Ia and is by far the most well studied photometrically transitional and spectroscopically 91bg-like SNe to date. 
The high-cadence observations of SN~2022xkq offer a detailed view of its photometric and spectroscopic evolution, in particular the early light curve and carbon features.
The early multiband light curve of SN~2022xkq reveals a possible flux excess which is most prominent in the redder bands. To our knowledge, SN~2022xkq is the first transitional/91bg-like SN Ia with multiband observations soon enough after explosion to detect the presence of an early excess. Multiband photometry within days of explosion of more transitional/91bg-like SNe is crucial to understanding if the red color in SN~2022xkq is unique or typical of this subclass of SNe Ia.

SN~2022xkq is another member of the increasing sample of transitional/91bg-like SNe Ia which exhibit strong carbon features, both in the optical and infrared. The spectroscopic data set presented in this paper offers near complete coverage of the evolution of the \ion{C}{2} $\lambda$6580 and \ion{C}{1} $\lambda$10693 lines. The NIR data set collected for this SN is the most complete \ion{C}{1} evolution observed to date and will be invaluable for understanding the processes which produce carbon-rich SNe Ia.

No current model of any explosion scenario predicts a red excess and early light curve like those exhibited in SN~2022xkq.
Red excesses are often linked to double detonation mechanisms. However, the overall light curve shape and color, as well as the strong C lines of SN~2022xkq do not agree with existing double detonation models. Other scenarios predicted to produce early flux excesses, including companion shocking, non-typical nickel distribution, and mergers, are unable to describe both the color and shape of the pre-maximum light curve as well. 

Spectroscopically, SN~2022xkq is not particularly well described by the models we consider either. The spectra are closer to modeled spectra of the detonation of a pure sub-Chandrasekhar mass white dwarf when compared to models of delayed-detonation. However, models currently suggest that sub-Chandrasekhar mass progenitors, for both pure and double detonation, do not produce the strong and persistent carbon features observed in SN~2022xkq. 
The evolution of the line we identify as \ion{C}{1} could be produced by \ion{He}{1} in a sub-Chandrasekhar double detonation scenario \citep{Collins23}, though we do not detect any other expected \ion{He}{1} lines despite obtaining spectra less than 48 hours after explosion.
In contrast, delayed-detonation models have been shown to better explain the carbon evolution, but do not explain the weaker \ion{Sc}{2}/\ion{Ti}{2} absorption trough at 4000--4200\,\AA\ in the early spectra of SN~2022xkq. 

No current publicly-available model can explain the early color and light curve, the spectroscopic evolution, and the strength of the carbon features of SN~2022xkq all at once. Given the breadth of the observational record available, this SN offers the unique opportunity to rigorously test progenitor scenarios. SN~2022xkq will be an important reference point in our efforts to understand the progenitors and explosions of underluminous SNe Ia.

\section*{Acknowledgments}

Time-domain research by the University of Arizona team, J.P. and D.J.S.\ is supported by NSF grants AST-1821987, 1813466, 1908972, 2108032, and 2308181, and by the Heising-Simons Foundation under grant \#2020-1864.

PL acknowledges support from the Swedish Research Council.

L.G. acknowledges financial support from the Spanish Ministerio de Ciencia e Innovaci\'on (MCIN), the Agencia Estatal de Investigaci\'on (AEI) 10.13039/501100011033, and the European Social Fund (ESF) ``Investing in your future" under the 2019 Ram\'on y Cajal program RYC2019-027683-I and the PID2020-115253GA-I00 HOSTFLOWS project, from Centro Superior de Investigaciones Cient\'ificas (CSIC) under the PIE project 20215AT016, and the program Unidad de Excelencia Mar\'ia de Maeztu CEX2020-001058-M.

This publication was made possible through the support of an LSSTC Catalyst Fellowship to K.A.B., funded through Grant 62192 from the John Templeton Foundation to LSST Corporation. The opinions expressed in this publication are those of the authors and do not necessarily reflect the views of LSSTC or the John Templeton Foundation.

Research by Y.D., S.V., N.M.R, E.H., and D.M. is supported by NSF grant AST-2008108. 


The Keck Infrared Transient Survey was executed primarily by members of the UC Santa Cruz transients team, who were supported in part by NASA grants NNG17PX03C, 80NSSC21K2076, 80NSSC22K1513, 80NSSC22K1518; NSF grant AST--1911206; and by fellowships from the Alfred P.\ Sloan Foundation and the David and Lucile Packard Foundation to R.J.F.  KITS was directly supported by NASA grant 80NSSC23K0301.

S.B.'s work was supported by the `Programme National de Physique Stellaire' (PNPS) of CNRS/INSU co-funded by CEA and CNES, and has made use of computing facilities operated by CeSAM data centre at LAM, Marseille, France.

M.N. is supported by the European Research Council (ERC) under the European Union's Horizon 2020 research and innovation program (grant agreement No.~948381) and by UK Space Agency Grant No.~ST/Y000692/1.

This work was funded in part by ANID, Millennium Science Initiative, ICN12\_009

J.S. acknowledges support from the Packard Foundation.


Y.-Z. Cai is supported by the National Natural Science Foundation of China (NSFC, Grant No. 12303054) and the International Centre of Supernovae, Yunnan Key Laboratory (No. 202302AN360001).

IS is supported by funding from the Italian Ministry of Education, University and Research (MIUR), PRIN 2017 (grant 20179ZF5KS) and PRIN-INAF 2022 project ``Shedding light on the nature of gap transients: from the observations to the models" and acknowledges the support of the doctoral grant funded by Istituto Nazionale di Astrofisica via the University of Padova and MIUR.

The SALT data presented here were obtained via Rutgers University program 2022-1-MLT-004 (PI: S.W.J.). 

L.A.K. acknowledges support by NASA FINESST fellowship 80NSSC22K1599.


C.L. acknowledges support from the National Science Foundation Graduate Research Fellowship under grant No. DGE-2233066.

M.M. acknowledges support in part from ADAP program grant No. 80NSSC22K0486, from the NSF AST-2206657 and from the HST GO program HST-GO-16656.

L.C. is grateful for support from NSF grants AST-2107070 and AST-2205628.

This work was performed in part at the Aspen Center for Physics, which is supported by National Science Foundation grant PHY-2210452.

The Australia Telescope Compact Array is part of the Australia Telescope National Facility\footnote{https://ror.org/05qajvd42} which is funded 
by the Australian Government for operation as a National Facility managed by CSIRO. 
We acknowledge the Gomeroi people as the traditional owners of the Observatory site.
The ATCA data reported here were obtained under Program C1473 (P.I. S. Ryder). 

The Las Cumbres Observatory group is supported by NSF grants AST-1911151 and AST-1911225.

This research has made use of the NASA Astrophysics Data System (ADS) Bibliographic Services, 
and the NASA/IPAC Extragalactic Database (NED), which is operated by the Jet Propulsion Laboratory, 
California Institute of Technology, under contract with the National Aeronautics and Space Administration.

This paper uses observations made with the MuSCAT3 instrument, developed by the Astrobiology Center and under financial supports by JSPS KAKENHI (JP18H05439) and JST PRESTO (JPMJPR1775), at Faulkes Telescope North on Maui, HI, operated by the Las Cumbres Observatory. 

Some observations reported here were obtained at the MMT Observatory, a joint facility of the University of Arizona and the Smithsonian Institution.

This work includes observations obtained at the Southern Astrophysical Research (SOAR) telescope, which is a joint project of the Minist\'{e}rio da Ci\^{e}ncia, Tecnologia e Inova\c{c}\~{o}es (MCTI/LNA) do Brasil, the US National Science Foundation's NOIRLab, the University of North Carolina at Chapel Hill (UNC), and Michigan State University (MSU).

The LBT is an international collaboration among institutions in the United States, Italy and Germany. LBT Corporation Members are: The University of Arizona on behalf of the Arizona Board of Regents; Istituto Nazionale di Astrofisica, Italy; LBT Beteiligungsgesellschaft, Germany, representing the Max-Planck Society, The Leibniz Institute for Astrophysics Potsdam, and Heidelberg University; The Ohio State University, and The Research Corporation, on behalf of The University of Notre Dame, University of Minnesota and University of Virginia.
This paper made use of the modsCCDRed data reduction code developed in part with funds provided by NSF Grants 
AST-9987045 and AST-1108693. 

The data presented here were obtained in part with ALFOSC, which is provided by the Instituto de Astrofisica de Andalucia (IAA) under a joint agreement with the University of Copenhagen and NOT.

This work has made use of data from the Asteroid Terrestrial-impact Last Alert System (ATLAS) project. The Asteroid Terrestrial-impact Last Alert System (ATLAS) project is primarily funded to search for near earth asteroids through NASA grants NN12AR55G, 80NSSC18K0284, and 80NSSC18K1575; byproducts of the NEO search include images and catalogs from the survey area. This work was partially funded by Kepler/K2 grant J1944/80NSSC19K0112 and HST GO-15889, and STFC grants ST/T000198/1 and ST/S006109/1. The ATLAS science products have been made possible through the contributions of the University of Hawaii Institute for Astronomy, the Queen's University Belfast, the Space Telescope Science Institute, the South African Astronomical Observatory, and The Millennium Institute of Astrophysics (MAS), Chile.

This work includes data obtained with the Swope Telescope at Las Campanas Observatory, Chile, as part of the Swope Time Domain Key Project (PI: Piro, Co-Is: Coulter, Drout, Phillips, Holoien, French, Cowperthwaite, Burns, Madore, Foley, Kilpatrick, Rojas-Bravo, Dimitriadis, Hsiao). We thank Abdo Campillay and Yilin Kong-Riveros for observations on the Swope Telescope.

A major upgrade of the Kast spectrograph on the Shane 3m telescope at Lick Observatory was made possible through generous gifts from the Heising-Simons Foundation as well as William and Marina Kast. Research at Lick Observatory is partially supported by a generous gift from Google.

NASA Keck time is administered by the NASA Exoplanet Science Institute. Data presented herein were obtained at the W.\ M.\ Keck Observatory from telescope time allocated to the National Aeronautics and Space Administration through the agency's scientific partnership with the California Institute of Technology and the University of California. The Observatory was made possible by the generous financial support of the W.\ M.\ Keck Foundation. The authors wish to recognize and acknowledge the very significant cultural role and reverence that the summit of Maunakea has always had within the indigenous Hawaiian community.  We are most fortunate to have the opportunity to conduct observations from this mountain.

\facilities{ADS, ATCA, CTIO:PROMPT, LBT (MODS), Las Cumbres Observatory (Sinistro, FLOYDS), Keck I (LRIS), Keck II (NIRES), NED, SALT (RSS), SOAR (GHTS, TripleSpec), Lick Shane (Kast), WISeREP, USASK:PROMPT}

\software{  
AIPS \citep{Wells85}, astropy \citep{astropy:2013, astropy:2018}, \texttt{AutoPhOT} \citep{autophot}, Binospec pipeline \citep{Kansky19}, CASA \citep{mcm07}, CMFGEN \citep{HillierMiller1998,HillierDessart2012}, corner \citep{corner}, emcee \citep{emcee}, FLOYDS pipeline \citep{FLOYDS}, HEA-Soft \citep{HEA-Soft2014}, \texttt{HOTPANTS} \citep{hotpants}, IDSRED \citep{Bravo23}, \texttt{lcogtsnpipe} \citep{Valenti2016}, Light Curve Fitting \citep{lightcurvefitting}, MatPLOTLIB \citep{mpl}, MIRIAD \citep{sault95}, modsCCDRed \citep{Pogge19}, NumPy \citep{numpy}, PESSTO pipeline \citep{PESSTO}, \texttt{photpipe} \citep{Rest05}, Photutils \citep{photutils}, PyEmir \citep{Pascual10, Cardiel19}, PyNOT-redux (https://pypi.org/project/PyNOT-redux/), PypeIt \citep{pypeit_arXiv, pypeit_zenodo}, PySALT \citep{SALTpipe},  Scipy \citep{scipy}, SNIFS pipeline \citep{Tucker22}, \texttt{SNooPy} \citep{snoopy}, \texttt{Spextool} \citep{Cushing04}, {\tt UCSC Spectral Pipeline} \citep{siebert19}, WISeREP \citep{wiserep}, \texttt{YSE-PZ} \citep{Coulter2022, Coulter2023}
}

\appendix

Table~\ref{tab:specInst} shows the complete spectroscopic log, both optical and NIR, of observations of SN~2022xkq reported in this work. All spectra are available on WISeRep (\url{https://www.wiserep.org}).
\input{spectra_log}

\bibliography{sources}

\end{document}

%% file: affiliation.tex
\newcommand{\LCO}{\affiliation{Las Cumbres Observatory, 6740 Cortona Drive, Suite 102, Goleta, CA 93117-5575, USA}}
\newcommand{\UCSB}{\affiliation{Department of Physics, University of California, Santa Barbara, CA 93106-9530, USA}}
\newcommand{\KITP}{\affiliation{Kavli Institute for Theoretical Physics, University of California, Santa Barbara, CA 93106-4030, USA}}
\newcommand{\UCD}{\affiliation{Department of Physics and Astronomy, University of California, Davis, 1 Shields Avenue, Davis, CA 95616-5270, USA}}
\newcommand{\WIS}{\affiliation{Department of Particle Physics and Astrophysics, Weizmann Institute of Science, 76100 Rehovot, Israel}}
\newcommand{\OKC}{\affiliation{Oskar Klein Centre, Department of Astronomy, Stockholm University, Albanova University Centre, SE-106 91 Stockholm, Sweden}}
\newcommand{\OAPD}{\affiliation{INAF-Osservatorio Astronomico di Padova, Vicolo dell'Osservatorio 5, I-35122 Padova, Italy}}
\newcommand{\UniPd}{\affiliation{Dipartimento di Fisica e Astronomia ``G. Galilei'', Universit\`a degli studi di Padova Vicolo dell'Osservatorio 3, I-35122 Padova, Italy}}
\newcommand{\Caltech}{\affiliation{Cahill Center for Astronomy and Astrophysics, California Institute of Technology, Mail Code 249-17, Pasadena, CA 91125, USA}}
\newcommand{\GSFC}{\affiliation{Astrophysics Science Division, NASA Goddard Space Flight Center, Mail Code 661, Greenbelt, MD 20771, USA}}
\newcommand{\UMD}{\affiliation{Joint Space-Science Institute, University of Maryland, College Park, MD 20742, USA}}
\newcommand{\UCB}{\affiliation{Department of Astronomy, University of California, Berkeley, CA 94720-3411, USA}}
\newcommand{\TTU}{\affiliation{Department of Physics, Texas Tech University, Box 41051, Lubbock, TX 79409-1051, USA}}
\newcommand{\STScI}{\affiliation{Space Telescope Science Institute, 3700 San Martin Drive, Baltimore, MD 21218-2410, USA}}
\newcommand{\UT}{\affiliation{University of Texas at Austin, 1 University Station C1400, Austin, TX 78712-0259, USA}}
\newcommand{\IoA}{\affiliation{Institute of Astronomy, University of Cambridge, Madingley Road, Cambridge CB3 0HA, UK}}
\newcommand{\QUB}{\affiliation{Astrophysics Research Centre, School of Mathematics and Physics, Queen's University Belfast, Belfast BT7 1NN, UK}}
\newcommand{\IPAC}{\affiliation{Spitzer Science Center, California Institute of Technology, Pasadena, CA 91125, USA}}
\newcommand{\JPL}{\affiliation{Jet Propulsion Laboratory, California Institute of Technology, 4800 Oak Grove Dr, Pasadena, CA 91109, USA}}
\newcommand{\Southampton}{\affiliation{Department of Physics and Astronomy, University of Southampton, Southampton SO17 1BJ, UK}}
\newcommand{\LANL}{\affiliation{Space and Remote Sensing, MS B244, Los Alamos National Laboratory, Los Alamos, NM 87545, USA}}
\newcommand{\Tsinghua}{\affiliation{Physics Department and Tsinghua Center for Astrophysics, Tsinghua University, Beijing, 100084, People's Republic of China}}
\newcommand{\NAOC}{\affiliation{National Astronomical Observatory of China, Chinese Academy of Sciences, Beijing, 100012, People's Republic of China}}
\newcommand{\Itagaki}{\affiliation{Itagaki Astronomical Observatory, Yamagata 990-2492, Japan}}
\newcommand{\Einstein}{\altaffiliation{Einstein Fellow}}
\newcommand{\Hubble}{\altaffiliation{Hubble Fellow}}
\newcommand{\CfA}{\affiliation{Center for Astrophysics \textbar{} Harvard \& Smithsonian, 60 Garden Street, Cambridge, MA 02138-1516, USA}}
\newcommand{\UA}{\affiliation{Steward Observatory, University of Arizona, 933 North Cherry Avenue, Tucson, AZ 85721-0065, USA}}
\newcommand{\MPIA}{\affiliation{Max-Planck-Institut f\"ur Astrophysik, Karl-Schwarzschild-Stra\ss{}e 1, D-85748 Garching, Germany}}
\newcommand{\DSFP}{\altaffiliation{LSSTC Data Science Fellow}}
\newcommand{\HCO}{\affiliation{Harvard College Observatory, 60 Garden Street, Cambridge, MA 02138-1516, USA}}
\newcommand{\Carnegie}{\affiliation{Observatories of the Carnegie Institute for Science, 813 Santa Barbara Street, Pasadena, CA 91101-1232, USA}}
\newcommand{\TAU}{\affiliation{School of Physics and Astronomy, Tel Aviv University, Tel Aviv 69978, Israel}}
\newcommand{\Edinburgh}{\affiliation{Institute for Astronomy, University of Edinburgh, Royal Observatory, Blackford Hill EH9 3HJ, UK}}
\newcommand{\Birmingham}{\affiliation{Birmingham Institute for Gravitational Wave Astronomy and School of Physics and Astronomy, University of Birmingham, Birmingham B15 2TT, UK}}
\newcommand{\Bath}{\affiliation{Department of Physics, University of Bath, Claverton Down, Bath BA2 7AY, UK}}
\newcommand{\CTIO}{\affiliation{Cerro Tololo Inter-American Observatory, National Optical Astronomy Observatory, Casilla 603, La Serena, Chile}}
\newcommand{\Potsdam}{\affiliation{Institut f\"ur Physik und Astronomie, Universit\"at Potsdam, Haus 28, Karl-Liebknecht-Str. 24/25, D-14476 Potsdam-Golm, Germany}}
\newcommand{\INPE}{\affiliation{Instituto Nacional de Pesquisas Espaciais, Avenida dos Astronautas 1758, 12227-010, S\~ao Jos\'e dos Campos -- SP, Brazil}}
\newcommand{\UNC}{\affiliation{Department of Physics and Astronomy, University of North Carolina, 120 East Cameron Avenue, Chapel Hill, NC 27599, USA}}
\newcommand{\Ohio}{\affiliation{Astrophysical Institute, Department of Physics and Astronomy, 251B Clippinger Lab, Ohio University, Athens, OH 45701-2942, USA}}
\newcommand{\AAS}{\affiliation{American Astronomical Society, 1667 K~Street NW, Suite 800, Washington, DC 20006-1681, USA}}
\newcommand{\MMT}{\affiliation{MMT and Steward Observatories, University of Arizona, 933 North Cherry Avenue, Tucson, AZ 85721-0065, USA}}
\newcommand{\Geneva}{\affiliation{ISDC, Department of Astronomy, University of Geneva, Chemin d'\'Ecogia, 16 CH-1290 Versoix, Switzerland}}
\newcommand{\IUCAA}{\affiliation{Inter-University Center for Astronomy and Astrophysics, Post Bag 4, Ganeshkhind, Pune, Maharashtra 411007, India}}
\newcommand{\CMU}{\affiliation{Department of Physics, Carnegie Mellon University, 5000 Forbes Avenue, Pittsburgh, PA 15213-3815, USA}}
\newcommand{\NAOJ}{\affiliation{Division of Science, National Astronomical Observatory of Japan, 2-21-1 Osawa, Mitaka, Tokyo 181-8588, Japan}}
\newcommand{\IfA}{\affiliation{Institute for Astronomy, University of Hawai`i, 2680 Woodlawn Drive, Honolulu, HI 96822-1839, USA}}
\newcommand{\UCSC}{\affiliation{Department of Astronomy and Astrophysics, University of California, Santa Cruz, CA 95064-1077, USA}}
\newcommand{\Purdue}{\affiliation{Department of Physics and Astronomy, Purdue University, 525 Northwestern Avenue, West Lafayette, IN 47907-2036, USA}}
\newcommand{\Princeton}{\affiliation{Department of Astrophysical Sciences, Princeton University, 4 Ivy Lane, Princeton, NJ 08540-7219, USA}}
\newcommand{\Moore}{\affiliation{Gordon and Betty Moore Foundation, 1661 Page Mill Road, Palo Alto, CA 94304-1209, USA}}
\newcommand{\Durham}{\affiliation{Department of Physics, Durham University, South Road, Durham, DH1 3LE, UK}}
\newcommand{\JHU}{\affiliation{Department of Physics and Astronomy, The Johns Hopkins University, 3400 North Charles Street, Baltimore, MD 21218, USA}}
\newcommand{\Toronto}{\affiliation{David A.\ Dunlap Department of Astronomy and Astrophysics, University of Toronto,\\ 50 St.\ George Street, Toronto, Ontario, M5S 3H4 Canada}}
\newcommand{\Duke}{\affiliation{Department of Physics, Duke University, Campus Box 90305, Durham, NC 27708, USA}}
\newcommand{\NCU}{\affiliation{Graduate Institute of Astronomy, National Central University, 300 Jhongda Road, 32001 Jhongli, Taiwan}}
\newcommand{\Columbia}{\affiliation{Department of Physics and Columbia Astrophysics Laboratory, Columbia University, Pupin Hall, New York, NY 10027, USA}}
\newcommand{\Flatiron}{\affiliation{Center for Computational Astrophysics, Flatiron Institute, 162 5th Avenue, New York, NY 10010-5902, USA}}
\newcommand{\CIERA}{\affiliation{Center for Interdisciplinary Exploration and Research in Astrophysics and Department of Physics and Astronomy, \\Northwestern University, 1800 Sherman Avenue, 8th Floor, Evanston, IL 60201, USA}}
\newcommand{\GeminiNorth}{\affiliation{Gemini Observatory, 670 North A`ohoku Place, Hilo, HI 96720-2700, USA}}
\newcommand{\Keck}{\affiliation{W.~M.~Keck Observatory, 65-1120 M\=amalahoa Highway, Kamuela, HI 96743-8431, USA}}
\newcommand{\UW}{\affiliation{Department of Astronomy, University of Washington, 3910 15th Avenue NE, Seattle, WA 98195-0002, USA}}
\newcommand{\DiRAC}{\altaffiliation{DiRAC Fellow}}
\newcommand{\USask}{\affiliation{Department of Physics \& Engineering Physics, University of Saskatchewan, 116 Science Place, Saskatoon, SK S7N 5E2, Canada}}
\newcommand{\Thacher}{\affiliation{Thacher School, 5025 Thacher Road, Ojai, CA 93023-8304, USA}}
\newcommand{\Rutgers}{\affiliation{Department of Physics and Astronomy, Rutgers, the State University of New Jersey,\\136 Frelinghuysen Road, Piscataway, NJ 08854-8019, USA}}
\newcommand{\FSU}{\affiliation{Department of Physics, Florida State University, 77 Chieftan Way, Tallahassee, FL 32306-4350, USA}}
\newcommand{\Melbourne}{\affiliation{School of Physics, The University of Melbourne, Parkville, VIC 3010, Australia}}
\newcommand{\ASTROthreeD}{\affiliation{ARC Centre of Excellence for All Sky Astrophysics in 3 Dimensions (ASTRO 3D)}}
\newcommand{\Stromlo}{\affiliation{Mt.\ Stromlo Observatory, The Research School of Astronomy and Astrophysics, Australian National University, ACT 2601, Australia}}
\newcommand{\NCPAS}{\affiliation{National Centre for the Public Awareness of Science, Australian National University, Canberra, ACT 2611, Australia}}
\newcommand{\TAMU}{\affiliation{Department of Physics and Astronomy, Texas A\&M University, 4242 TAMU, College Station, TX 77843, USA}}
\newcommand{\Mitchell}{\affiliation{George P.\ and Cynthia Woods Mitchell Institute for Fundamental Physics \& Astronomy, College Station, TX 77843, USA}}
\newcommand{\ESO}{\affiliation{European Southern Observatory, Alonso de C\'ordova 3107, Casilla 19, Santiago, Chile}}
\newcommand{\MAS}{\affiliation{Millennium Institute of Astrophysics MAS, Nuncio Monsenor Sotero Sanz 100, Off.
104, Providencia, Santiago, Chile}}
\newcommand{\ICE}{\affiliation{Institute of Space Sciences (ICE, CSIC), Campus UAB, Carrer de Can Magrans, s/n, E-08193 Barcelona, Spain}}
\newcommand{\IEEC}{\affiliation{Institut d'Estudis Espacials de Catalunya, Gran Capit\`a, 2-4, Edifici Nexus, Desp.\ 201, E-08034 Barcelona, Spain}}
\newcommand{\Warwick}{\affiliation{Department of Physics, University of Warwick, Gibbet Hill Road, Coventry CV4 7AL, UK}}
\newcommand{\Macquarie}{\affiliation{School of Mathematical and Physical Sciences, Macquarie University, NSW 2109, Australia}}
\newcommand{\AAARC}{\affiliation{Astronomy, Astrophysics and Astrophotonics Research Centre, Macquarie University, Sydney, NSW 2109, Australia}}
\newcommand{\Capodimonte}{\affiliation{INAF - Capodimonte Astronomical Observatory, Salita Moiariello 16, I-80131 Napoli, Italy}}
\newcommand{\INFNNapoli}{\affiliation{INFN - Napoli, Strada Comunale Cinthia, I-80126 Napoli, Italy}}
\newcommand{\ICRANet}{\affiliation{ICRANet, Piazza della Repubblica 10, I-65122 Pescara, Italy}}
\newcommand{\MSU}{\affiliation{Center for Data Intensive and Time Domain Astronomy, Department of Physics and Astronomy,\\Michigan State University, East Lansing, MI 48824, USA}}
\newcommand{\SETI}{\affiliation{SETI Institute,
339 Bernardo Ave, Suite 200, Mountain View, CA 94043, USA}}
\newcommand{\IAIFI}{\affiliation{The NSF AI Institute for Artificial Intelligence and Fundamental Interactions}}
\newcommand{\ANUC}{\affiliation{Department of Astronomy, AlbaNova University Center, Stockholm University, SE-10691 Stockholm, Sweden}}
\newcommand{\UVA}{\affiliation{Department of Astronomy, University of Virginia, Charlottesville, VA 22904, USA}}
\newcommand{\THCA}{\affiliation{Physics Department and Tsinghua Center for Astrophysics (THCA), Tsinghua University, Beijing, 100084, People's Republic of China}}
\newcommand{\NARIT}{\affiliation{National Astronomical Research Institute of Thailand (NARIT), Don Kaeo, Mae Rim District, Chiang Mai 50180, Thailand}}
\newcommand{\HamObs}{\affiliation{Hamburger Sternwarte, Gojenbergsweg 112, 21029 Hamburg, Germany}}
\newcommand{\VaTech}{\affiliation{Department of Physics, Virginia Tech, 850 West Campus Drive, Blacksburg VA, 24061, USA}}
\newcommand{\Yunnan}{\affiliation{Yunnan Observatories, Chinese Academy of Sciences, Kunming 650216, P.R. China}}
\newcommand{\KLSECO}{\affiliation{Key Laboratory for the Structure and Evolution of Celestial Objects, Chinese Academy of Sciences, Kunming 650216, P.R. China}}
\newcommand{\ICS}{\affiliation{International Centre of Supernovae, Yunnan Key Laboratory, Kunming 650216, P.R. China}}
\newcommand{\FINCA}{\affiliation{Finnish Centre for Astronomy with ESO (FINCA), FI-20014 University of Turku, Finland}}
\newcommand{\Tuorla}{\affiliation{Tuorla Observatory, Department of Physics and Astronomy, FI-20014 University of Turku, Finland}}
\newcommand{\LAM}{\affiliation{Aix-Marseille Univ, CNRS, CNES, LAM, 13388 Marseille, France}}
\newcommand{\IAC}{\affiliation{Instituto de Astrof\'isica de Canarias, E-38205 La Laguna, Tenerife, Spain}}
\newcommand{\LPNHE}{\affiliation{LPNHE, (CNRS/IN2P3, Sorbonne Universit\'e, Universit\'e Paris Cit\'e), Laboratoire de Physique Nucl\'eaire et de Hautes \'Energies, 75005, Paris, France}}
\newcommand{\UniLag}{\affiliation{Universidad de La Laguna, Dept. Astrof\'isica, E-38206 La Laguna, Tenerife, Spain}}
\newcommand{\OU}{\affiliation{Homer L. Dodge Department of Physics and Astronomy, University of Oklahoma, 440 W. Brooks, Norman, OK 73019-2061, USA}}
\newcommand{\PSI}{\affiliation{Planetary Science Institute, 1700 East Fort Lowell Road, Suite 106, Tucson, AZ 85719-2395, USA}}
\newcommand{\TUM}{\affiliation{Technische Universit\"at M\"unchen, TUM School of Natural Sciences, Physik-Department, James-Franck-Stra\ss{}e 1, 85748 Garching, Germany}}
\newcommand{\UWarsaw}{\affiliation{Astronomical Observatory, University of Warsaw, Al. Ujazdowskie 4, 00-478 Warszawa, Poland}}
\newcommand{\Trinity}{\affiliation{School of Physics, Trinity College Dublin, The University of Dublin, Dublin
2, Ireland}}
\newcommand{\UNAB}{\affiliation{Departamento de Ciencias F\'isicas, Facultad de Ciencias Exactas, Universidad Andr\'es Bello, Fern\'andez Concha 700, Las Condes,
Santiago, Chile}}
\newcommand{\CCAPP}{\affiliation{Center for Cosmology and Astroparticle Physics, The Ohio State University, 191 West Woodruff Ave, Columbus, OH, USA}}
\newcommand{\OSU}{\affiliation{Department of Astronomy, The Ohio State University, 140 West 18th Avenue, Columbus, OH, USA}}
\newcommand{\LCOactual}{\affiliation{Las Campanas Observatory, Carnegie Observatories, Casilla 601, La
Serena, Chile}}

%% file: authors.tex
\author[0000-0002-0744-0047]{Jeniveve Pearson}
\UA
\author[0000-0003-4102-380X]{David J. Sand}
\UA
\author[0000-0002-3664-8082]{Peter Lundqvist}
\OKC
\author[0000-0002-1296-6887]{Llu\'is Galbany}
\ICE\IEEC
\author[0000-0003-0123-0062]{Jennifer E. Andrews}
\GeminiNorth
\author[0000-0002-4924-444X]{K. Azalee Bostroem}
\altaffiliation{LSSTC Catalyst Fellow} \UA
\author[0000-0002-7937-6371]{Yize Dong \begin{CJK*}{UTF8}{gbsn}(董一泽)\end{CJK*}}
\UCD
\author[0000-0003-2744-4755]{Emily Hoang}
\UCD
\author[0000-0002-0832-2974]{Griffin Hosseinzadeh}
\UA
\author[0000-0003-0549-3281]{Daryl Janzen}
\USask
\author[0000-0001-5754-4007]{Jacob E. Jencson}
\JHU
\author[0000-0001-9589-3793]{Michael J. Lundquist}
\Keck
\author{Darshana Mehta}
\UCD
\author[0000-0002-7015-3446]{Nicol\'as Meza Retamal}
\UCD
\author[0000-0002-4022-1874]{Manisha Shrestha}
\UA
\author[0000-0001-8818-0795]{Stefano Valenti}
\UCD
\author[0000-0003-2732-4956]{Samuel Wyatt}
\UW

\author[0000-0003-0227-3451]{Joseph P.\ Anderson}
\ESO\MAS
\author[0000-0002-5221-7557]{Chris Ashall}
\VaTech
\author[0000-0002-4449-9152]{Katie Auchettl}
\Melbourne \UCSC
\author[0000-0001-5393-1608]{Eddie Baron}
\PSI \HamObs \OU
\author[0000-0002-9388-2932]{St\'{e}phane Blondin}
\LAM
\author[0000-0003-4625-6629]{Christopher R. Burns}
\Carnegie
\author[0000-0002-7714-493X]{Yongzhi Cai \begin{CJK*}{UTF8}{gbsn}(蔡永志)\end{CJK*}}
\Yunnan\KLSECO\ICS
\author[0000-0002-1066-6098]{Ting-Wan Chen}
\NCU
\author[0000-0002-8400-3705]{Laura Chomiuk}
\MSU
\author[0000-0003-4263-2228]{David A.\ Coulter}
\UCSC
\author{Dane Cross}
\ICE
\author[0000-0002-5680-4660]{Kyle W. Davis}
\UCSC
\author[0000-0001-6069-1139]{Thomas de Jaeger}
\LPNHE
\author[0000-0002-7566-6080]{James M. DerKacy}
\VaTech
\author[0000-0002-2164-859X]{Dhvanil D. Desai}
\IfA
\author[0000-0001-9494-179X]{Georgios Dimitriadis}
\Trinity
\author[0000-0003-3429-7845]{Aaron Do}
\IfA
\author[0000-0003-4914-5625]{Joseph R. Farah}
\UCSB
\author[0000-0002-2445-5275]{Ryan J. Foley}
\UCSC
\author[0000-0002-1650-1518]{Mariusz Gromadzki}
\UWarsaw
\author[0000-0003-2375-2064]{Claudia P. Guti\'{e}rrez}
\IEEC\ICE
\author[0000-0002-6703-805X]{Joshua Haislip}
\UNC
\author[0000-0002-0264-7356]{Jonay I.\ Gonz\'alez Hern\'andez}
\IAC\UniLag
\author[0000-0001-9668-2920]{Jason T. Hinkle}
\IfA
\author[0000-0003-3953-9532]{Willem B. Hoogendam}
\altaffiliation{NSF Graduate Research Fellow}\IfA
\author[0000-0003-4253-656X]{D.\ Andrew Howell}
\LCO\UCSB
\author[0000-0002-4338-6586]{Peter Hoeflich}
\FSU
\author[0000-0003-1039-2928]{Eric Hsiao}
\FSU
\author[0000-0003-1059-9603]{Mark E. Huber}
\IfA
\author[0000-0001-8738-6011]{Saurabh W.\ Jha}
\Rutgers
\author[0000-0002-4374-0661]{Cristina Jim\'enez Palau}
\ICE\IEEC
\author[0000-0002-5740-7747]{Charles D.\ Kilpatrick}
\CIERA
\author[0000-0003-3642-5484]{Vladimir Kouprianov}
\UNC
\author[0000-0001-8367-7591]{Sahana Kumar}
\UVA
\author[0000-0003-3108-1328]{Lindsey A.\ Kwok}
\Rutgers
\author[0000-0003-2037-4619]{Conor Larison}
\Rutgers
\author[0000-0002-2249-0595]{Natalie LeBaron}
\UCB
\author[0009-0004-3242-282X]{Xavier Le Saux}
\UCSC
\author[0000-0002-3900-1452]{Jing Lu}
\MSU
\author[0000-0001-5807-7893]{Curtis McCully}
\LCO\UCSB
\author[0000-0001-5888-2542]{Tycho Mera Evans}
\FSU
\author[0000-0002-0370-157X]{Peter Milne}
\UA
\author[0000-0001-7132-0333]{Maryam Modjaz}
\UVA
\author[0000-0003-2535-3091]{Nidia Morrell}
\LCOactual
\author[0000-0003-3939-7167]{Tom\'as E. M\"{u}ller-Bravo}
\ICE\IEEC
\author[0000-0001-9570-0584]{Megan Newsome}
\LCO\UCSB
\author[0000-0002-2555-3192]{Matt Nicholl}
\QUB
\author[0000-0003-0209-9246]{Estefania Padilla Gonzalez}
\LCO\UCSB
\author[0000-0003-3490-3243]{Anna V. Payne}
\STScI\IfA
\author[0000-0002-7472-1279]{Craig Pellegrino}
\LCO\UVA
\author[0000-0001-6383-860X]{Kim Phan}
\IEEC\ICE
\author[0000-0003-0737-8463]{Jonathan Pineda-Garc\'ia}
\UNAB
\author[0000-0001-6806-0673]{Anthony L.\ Piro}
\Carnegie
\author[0009-0006-4637-4085]{Lara Piscarreta}
\ICE
\author[0000-0002-1633-6495]{Abigail Polin}
\Carnegie
\Caltech
\author[0000-0002-5060-3673]{Daniel E.\ Reichart}
\UNC
\author[0000-0002-7559-315X]{C\'esar Rojas-Bravo}
\UCSC
\author[0000-0003-4501-8100]{Stuart D.\ Ryder}
\Macquarie \AAARC
\author[0000-0003-1450-0869]{Irene Salmaso}
\OAPD\UniPd
\author[0009-0002-5096-1689]{Michaela Schwab}
\Rutgers
\author[0000-0002-9301-5302]{Melissa Shahbandeh}
\JHU\STScI
\author[0000-0003-4631-1149]{Benjamin J. Shappee}
\IfA
\author[0000-0003-2445-3891]{Matthew R. Siebert}
\STScI
\author[0000-0001-5510-2424]{Nathan Smith}
\UA
\author[0000-0002-1468-9668]{Jay Strader}
\MSU
\author[0000-0002-5748-4558]{Kirsty Taggart}
\UCSC
\author[0000-0003-0794-5982]{Giacomo Terreran}
\LCO 
\author[0000-0002-1481-4676]{Samaporn Tinyanont}
\NARIT
\author[0000-0002-2471-8442]{M. A. Tucker} 
\altaffiliation{CCAPP Fellow} \OSU \CCAPP
\author[0000-0002-3334-4585]{Giorgio Valerin}
\OAPD\UniPd
\author[0000-0002-1229-2499]{D. R. Young}
\QUB


%% file: spectra_log.tex
\startlongtable
\begin{deluxetable*}{ l c C c c c c c}\label{tab:specInst}
\tablecaption{Log of Spectroscopic Observations}
\tablehead{
\colhead{Date (UTC)} & \colhead{JD} & \colhead{Epoch (days)\tablenotemark{a}} & \colhead{Telescope} & \colhead{Instrument} & \colhead{Range (\AA)} & \colhead{Exp (s)} & \colhead{Slit ($''$)}}
\startdata
2022-10-13 & 2459865.938 & -13.6 & P60 & SEDM & 3780-9220 & 1800 & IFU\\
2022-10-13 & 2459866.496 & -13.0 & SALT & RSS & 3930-8600 & 1980 & 1.50\\
2022-10-15 & 2459867.833 & -11.7 & ESO-NTT & EFOSC2-NTT & 3670-9270 & 2700 & 1.0\\
2022-10-15 & 2459868.022 & -11.5 & FTN & FLOYDS & 3500-10000 & 3600 & 2\\
2022-10-15 & 2459868.041 & -11.5 & Keck & NIRES & 9410-24690 & 1000 & 0.55\\
2022-10-16 & 2459868.666 & -10.9 & ESO-NTT & EFOSC2-NTT & 3370-7490 & 1800 & 1.0\\
2022-10-16 & 2459868.692 & -10.9 & ESO-NTT & SOFI & 9380-16480 & 3240 & 1\\
2022-10-16 & 2459868.753 & -10.8 & Baade & IMACS & 4200-9390 & 900 & 0.7\\
2022-10-16 & 2459868.782 & -10.8 & ESO-NTT & EFOSC2-NTT & 6020-10010 & 1800 & 1.0\\
2022-10-17 & 2459869.571 & -10.0 & NOT & ALFOSC & 3400-9610 & 1800 & 1.3\\
2022-10-17 & 2459869.674 & -9.9 & SOAR & GHTS RED & 3650-7000 & 1800 & 1.0\\
2022-10-17 & 2459869.816 & -9.7 & P60 & SEDM & 3780-9220 & 1800 & IFU\\
2022-10-17 & 2459870.07 & -9.5 & FTN & FLOYDS & 3500-10000 & 3600 & 2\\
2022-10-18 & 2459870.669 & -8.9 & ESO-NTT & SOFI & 9380-16480 & 3240 & 1\\
2022-10-18 & 2459870.878 & -8.7 & Bok & B\&C & 4100-8000 & 4500 & 1.50\\
2022-10-18 & 2459871.484 & -8.1 & SALT & RSS & 3920-8600 & 1980 & 1.50\\
2022-10-19 & 2459871.8 & -7.8 & Bok & B\&C & 4870-6020 & 7500 & 1.50\\
2022-10-19 & 2459872.019 & -7.6 & UH88 & SNIFS & 3400-9100 & 1800 & IFU\\
2022-10-19 & 2459872.489 & -7.1 & SALT & RSS & 3920-8600 & 1980 & 1.50\\
2022-10-20 & 2459872.876 & -6.7 & Bok & B\&C & 4000-8000 & 6000 & 1.50\\
2022-10-20 & 2459873.071 & -6.5 & FTN & FLOYDS & 3500-10000 & 2100 & 2\\
2022-10-21 & 2459873.68 & -5.9 & SOAR & TripleSpec & 9400-24660 & 300 & 1.0\\
2022-10-21 & 2459874.093 & -5.5 & FTN & FLOYDS & 3800-10000 & 1800 & 2\\
2022-10-22 & 2459874.57 & -5.0 & NOT & ALFOSC & 3440-10260 & 900 & 1.0\\
2022-10-22 & 2459875.074 & -4.5 & UH88 & SNIFS & 3400-9100 & 1800 & IFU\\
2022-10-22 & 2459875.142 & -4.5 & Keck & LRIS & 3160-10150 & 250 & 1\\
2022-10-24 & 2459876.605 & -3.0 & NOT & ALFOSC & 3400-9660 & 900 & 1.3\\
2022-10-24 & 2459877.043 & -2.6 & FTN & FLOYDS & 3500-10000 & 1800 & 2\\
2022-10-26 & 2459878.718 & -0.9 & SOAR & GHTS RED & 4980-8990 & 600 & 1.0\\
2022-10-26 & 2459879.051 & -0.6 & UH88 & SNIFS & 3400-9100 & 1800 & IFU\\
2022-10-27 & 2459879.74 & 0.1 & SOAR & GHTS BLUE & 3700-7120 & 600 & 1.0\\
2022-10-27 & 2459879.743 & 0.1 & GTC & EMIR & 8900-23000 & 480 & 0.8\\
2022-10-27 & 2459879.779 & 0.1 & SOAR & TripleSpec & 9400-24660 & 180 & 1.1\\
2022-10-28 & 2459880.822 & 1.2 & LBT & MODS1 & 3630-10000 & 1800 & 1\\
2022-10-28 & 2459880.938 & 1.3 & Shane & Kast & 3250-10750 & 600 & 2.0\\
2022-10-29 & 2459881.928 & 2.3 & ARC & KOSMOS & 3900-9200 & 400 & 2.10\\
2022-10-29 & 2459882.46 & 2.8 & SALT & RSS & 3920-8600 & 1980 & 1.50\\
2022-10-30 & 2459882.781 & 3.1 & SOAR & GHTS RED & 4000-7910 & 1200 & 1.0\\
2022-10-30 & 2459883.141 & 3.5 & FTS & FLOYDS & 3500-10000 & 900 & 2\\
2022-10-31 & 2459883.55 & 3.9 & NOT & ALFOSC & 3420-9700 & 900 & 1.0\\
2022-10-31 & 2459883.787 & 4.1 & ESO-NTT & EFOSC & 3650-9250 & 1080 & 1.5\\
2022-10-31 & 2459884.019 & 4.3 & UH88 & SNIFS & 3400-9100 & 1800 & IFU\\
2022-11-01 & 2459884.788 & 5.1 & ESO-NTT & SOFI & 9380-16480 & 960 & 1\\
2022-11-02 & 2459886.047 & 6.4 & FTS & FLOYDS & 3500-10000 & 900 & 2\\
2022-11-04 & 2459887.526 & 7.8 & INT & IDS & 3800-9000 & 1500 & 0.974\\
2022-11-04 & 2459887.882 & 8.2 & Shane & Kast & 3250-10890 & 600 & 2.0\\
2022-11-05 & 2459888.501 & 8.8 & INT & IDS & 3800-9000 & 1200 & 0.974\\
2022-11-05 & 2459888.626 & 8.9 & GTC & EMIR & 8900-23000 & 480 & 0.8\\
2022-11-05 & 2459889.071 & 9.4 & FTS & FLOYDS & 3500-10000 & 900 & 2\\
2022-11-06 & 2459889.711 & 10.0 & INT & IDS & 3600-9000 & 1200 & 0.974\\
2022-11-06 & 2459890.005 & 10.3 & Keck & NIRES & 9410-24690 & 800 & 0.55\\
2022-11-07 & 2459890.586 & 10.9 & INT & IDS & 3800-9000 & 1500 & 1.488\\
2022-11-09 & 2459893.428 & 13.7 & SALT & RSS & 3920-8600 & 1980 & 1.50\\
2022-11-10 & 2459894.02 & 14.3 & FTS & FLOYDS & 3500-10000 & 899 & 2\\
2022-11-10 & 2459894.424 & 14.7 & SALT & RSS & 3920-8600 & 1980 & 1.50\\
2022-11-13 & 2459896.61 & 16.8 & NOT & ALFOSC & 3400-9630 & 900 & 1.3\\
2022-11-13 & 2459897.415 & 17.6 & SALT & RSS & 3920-8600 & 1980 & 1.50\\
2022-11-14 & 2459897.743 & 18.0 & SOAR & TripleSpec & 9400-24660 & 300 & 1.0\\
2022-11-15 & 2459899.029 & 19.2 & UH88 & SNIFS & 3400-9100 & 2200 & IFU\\
2022-11-16 & 2459899.874 & 20.1 & Bok & B\&C & 3940-8250 & 6000 & 1.50\\
2022-11-16 & 2459899.948 & 20.2 & UH88 & SNIFS & 3400-9100 & 1800 & IFU\\
2022-11-16 & 2459899.966 & 20.2 & FTS & FLOYDS & 3500-10000 & 900 & 2\\
2022-11-17 & 2459901.005 & 21.2 & Keck & NIRES & 9410-24690 & 480 & 0.55\\
2022-11-17 & 2459901.403 & 21.6 & SALT & RSS & 3920-8600 & 1980 & 1.50\\
2022-11-18 & 2459901.812 & 22.0 & ESO-NTT & SOFI & 9380-16480 & 960 & 1\\
2022-11-18 & 2459901.92 & 22.1 & Bok & B\&C & 4900-6000 & 7500 & 1.50\\
2022-11-21 & 2459904.841 & 25.0 & Shane & Kast & 3250-10690 & 600 & 2.0\\
2022-11-22 & 2459905.679 & 25.8 & SOAR & TripleSpec & 9400-24660 & 300 & 1.0\\
2022-11-23 & 2459907.019 & 27.2 & FTS & FLOYDS & 3500-10000 & 899 & 2\\
2022-11-27 & 2459910.994 & 31.1 & UH88 & SNIFS & 3400-9100 & 2400 & IFU\\
2022-11-28 & 2459911.545 & 31.7 & NOT & ALFOSC & 3400-9710 & 900 & 1.0\\
2022-11-28 & 2459911.686 & 31.8 & SOAR & GHTS RED & 4960-8970 & 600 & 1.0\\
2022-11-29 & 2459912.587 & 32.7 & SOAR & TripleSpec & 9400-24660 & 300 & 1.0\\
2022-11-29 & 2459912.683 & 32.8 & SOAR & GHTS BLUE & 3750-7110 & 600 & 1.0\\
2022-11-29 & 2459913.056 & 33.2 & FTS & FLOYDS & 3500-10000 & 900 & 2\\
2022-12-03 & 2459916.988 & 37.1 & FTS & FLOYDS & 3500-10000 & 900 & 2\\
2022-12-06 & 2459920.361 & 40.4 & SALT & RSS & 3920-8600 & 1910 & 1.50\\
2022-12-12 & 2459925.599 & 45.6 & SOAR & GHTS RED & 4950-8960 & 1800 & 1.0\\
2022-12-13 & 2459926.558 & 46.6 & INT & IDS & 3800-9000 & 1800 & 1.001\\
2022-12-14 & 2459927.902 & 47.9 & FTN & FLOYDS & 3500-10000 & 900 & 2\\
2022-12-15 & 2459928.761 & 48.7 & Bok & B\&C & 4000-8000 & 7500 & 1.50\\
2022-12-17 & 2459930.536 & 50.5 & GTC & EMIR & 8900-23000 & 1440 & 0.8\\
2022-12-19 & 2459932.772 & 52.7 & MMT & Binospec & 5260-7750 & 3600 & 1\\
2022-12-21 & 2459934.644 & 54.6 & SOAR & GHTS RED & 4960-8960 & 1800 & 1.0\\
2022-12-21 & 2459934.751 & 54.7 & ESO-NTT & SOFI & 9380-16480 & 2970 & 1\\
2022-12-21 & 2459934.806 & 54.7 & Shane & Kast & 3250-10890 & 1200 & 2.0\\
2022-12-24 & 2459937.838 & 57.8 & ARC & KOSMOS & 3780-9650 & 1800 & 2.10\\
2022-12-24 & 2459938.138 & 58.0 & FTS & FLOYDS & 5050-10000 & 2700 & 2\\
2022-12-29 & 2459942.691 & 62.6 & ESO-NTT & EFOSC & 3360-7480 & 1800 & 1.5\\
2022-12-29 & 2459942.691 & 62.6 & ESO-NTT & EFOSC & 6000-10010 & 1800 & 1.5\\
2022-12-31 & 2459944.971 & 64.8 & Keck & NIRES & 9410-24690 & 1200 & 0.55\\
2023-01-05 & 2459950.398 & 70.2 & GTC & EMIR & 8900-23000 & 1440 & 0.8\\
2023-01-07 & 2459952.451 & 72.3 & SALT & RSS & 3920-8600 & 1980 & 1.50\\
2023-01-11 & 2459955.72 & 75.5 & SOAR & GHTS RED & 4920-8930 & 1800 & 1.0\\
2023-01-13 & 2459957.646 & 77.4 & MMT & Binospec & 5260-7750 & 3600 & 1\\
2023-01-13 & 2459957.655 & 77.4 & ESO-NTT & EFOSC & 3360-7480 & 1800 & 1.5\\
2023-01-13 & 2459957.679 & 77.4 & ESO-NTT & EFOSC & 6000-10000 & 1800 & 1.5\\
2023-01-14 & 2459958.998 & 78.7 & FTS & FLOYDS & 3500-10000 & 2700 & 2\\
2023-01-25 & 2459970.452 & 90.1 & GTC & OSIRIS & 3630-7790 & 800 & 1.0\\
2023-01-25 & 2459970.452 & 90.1 & GTC & OSIRIS & 5100-10300 & 800 & 1.0\\
2023-01-26 & 2459970.592 & 90.3 & SOAR & GHTS RED & 4920-8930 & 1800 & 1.0\\
2023-01-27 & 2459971.757 & 91.4 & Shane & Kast & 3250-10740 & 2400 & 2.0\\
2023-01-31 & 2459976.388 & 96.0 & SALT & RSS & 3920-8600 & 1980 & 1.50\\
2023-02-01 & 2459977.001 & 96.6 & FTS & FLOYDS & 3500-10000 & 2700 & 2\\
2023-02-08 & 2459983.819 & 103.4 & Keck & NIRES & 9410-24690 & 2400 & 0.55\\
2023-02-11 & 2459987.346 & 106.9 & SALT & RSS & 6140-6940 & 2400 & 1.50\\
2023-02-23 & 2459998.551 & 118.0 & Clay & LDSS-3 & 3900-10630 & 900 & 1.0\\
2023-02-24 & 2459999.941 & 119.4 & FTS & FLOYDS & 3500-10000 & 3600 & 2\\
2023-03-05 & 2460008.582 & 128.0 & Baade & IMACS & 4200-9380 & 1200 & 0.7\\
2023-03-14 & 2460017.892 & 137.2 & FTS & FLOYDS & 3500-10000 & 3600 & 2\\
2023-03-30 & 2460033.507 & 152.7 & SOAR & GHTS RED & 4930-8930 & 3600 & 1.0\\
\enddata
\tablenotetext{a}{relative to B$_{\mathrm{max}}$}
\end{deluxetable*}